\documentclass[journal,two side,web]{ieeecolor}
\pdfoutput=1
\usepackage{etoolbox}
\makeatletter
\@ifundefined{color@begingroup}%
{\let\color@begingroup\relax
\let\color@endgroup\relax}{}%
\def\fix@ieeecolor@hbox#1{%
\hbox{\color@begingroup#1\color@endgroup}}
\patchcmd\@makecaption{\hbox}{\fix@ieeecolor@hbox}{}{\FAILED}
\patchcmd\@makecaption{\hbox}{\fix@ieeecolor@hbox}{}{\FAILED}

\usepackage{generic}

\usepackage{cite}
\usepackage{amsmath,amssymb,amsfonts}
\usepackage{algorithmic}
\usepackage{graphicx}
\usepackage{makecell}
\usepackage{float}
\graphicspath{ {pic/} }
\usepackage{textcomp}
\usepackage{multirow}
\usepackage{booktabs}
\usepackage{CJKutf8} 
\usepackage{mathrsfs} 
\usepackage{setspace} 
\usepackage{booktabs,multirow}
\usepackage[figuresright]{rotating}

\def\BibTeX{{\rm B\kern-.05em{\sc i\kern-.025em b}\kern-.08em
    T\kern-.1667em\lower.7ex\hbox{E}\kern-.125emX}}
\graphicspath{ {pic/} }
\begin{document}
\begin{CJK}{UTF8}{gbsn} 
\begin{spacing}{1} 
\title{Multimodal Identification of Alzheimer’s Disease:
A Review}
\author{Guian Fang, Mengsha Liu, Yi Zhong, Zhuolin Zhang, \\ Jiehui Huang\textsuperscript{\dag
}, Zhenchao Tang\textsuperscript{\dag
}, Calvin Yu-Chian Chen\textsuperscript{*}
\thanks{\textsuperscript{*}Corresponding author, \textsuperscript{\dag
}Project leader}
}

\maketitle

\begin{abstract}
Alzheimer's disease is a progressive neurological disorder characterized by cognitive impairment and memory loss. With the increasing aging population, the incidence of AD is continuously rising, making early diagnosis and intervention an urgent need. In recent years, a considerable number of teams have applied computer-aided diagnostic techniques to early classification research of AD. Most studies have utilized imaging modalities such as magnetic resonance imaging (MRI), positron emission tomography (PET), and electroencephalogram (EEG). However, there have also been studies that attempted to use other modalities as input features for the models, such as sound, posture, biomarkers, cognitive assessment scores, and their fusion. Experimental results have shown that the combination of multiple modalities often leads to better performance compared to a single modality. Therefore, this paper will focus on different modalities and their fusion, thoroughly elucidate the mechanisms of various modalities, explore which methods should be combined to better harness their utility, analyze and summarize the literature in the field of early classification of AD in recent years, in order to explore more possibilities of modality combinations.
\end{abstract}

\section{introduction}
\label{sec:introduction}

\quad As one of the most common neurodegenerative disorder, Alzheimer's disease (AD) is affecting the elderly around the world. According to the latest data from the World Health Organization (WHO), it is expected that the number of dementia patients will reach 55 million in 2019, and this number will increase to 139 million in the coming 2050. Among them, Alzheimer's disease is the most common cause of dementia, accounting for approximately 60-80\% of dementia cases. Alzheimer's disease is a degenerative neurological disease characterized by progressive loss of cognition and memory. Currently, the academic community usually believes that AD is related to the neurofibrillary tangles (NFT) and the extracellular Amyloid-β ($A\beta$) deposition, which cause neurons and synapses loss or damage, inflammation and brain tissue atrophy are other changes\cite{AD_report}. Alzheimer's disease can cause changes in brain structure and function, affecting patients from multiple aspects such as speech, emotion, and behavior. As the condition worsens, patients often become disconnected from society, lose their ability to take care of themselves, and burden their families and society.

\quad There is still no way to completely cure and reverse the progression of dementia. However,  early, accurate, and comprehensive diagnosis of Alzheimer's disease can provide timely intervention and slow down the progression of the disease. Experts are increasingly recognizing the importance of early diagnosis of Alzheimer's disease. At present, the clinical examination methods of Alzheimer's disease mainly include: Cognitive Assessment, Non-neuroimaging Biomarkers, Voice and Speech examination, Posture examination, Neuroimaging examination etc. 

\quad With the rapid development of large language models (LLMs) like ChatGPT, there has been an emergence of conversational systems based on natural language processing techniques. These systems, including HuatuoGPT\cite{huatuogpt}, BenTsao\cite{BenTsao}, and DoctorGLM\cite{DoctorGLM}, have shown promising performance in the field of medical diagnosis and consultation. Interestingly, we have found that this chatbot-based diagnostic model, utilizing ChatGPT, has exhibited relatively high intelligence. However, due to the complexity and multifactorial nature of Alzheimer's disease, the reliability of its inferences is often questionable, as they are solely based on textual input provided by the patient. The characteristics of a single modality may not be sufficient to support accurate early diagnosis. The changes caused by AD may also have similar manifestations in other diseases. A better approach would be to incorporate multiple modalities from the patient, including text, voice, images, and more, as diagnostic evidence. Multimodal diagnostic methods emerge as the times require. Based on the success of vision and language-based large models, multimodal diagnostic systems appear to be more robust and reliable. Multimodal diagnosis of AD poses a challenging problem with significant implications for the future. Therefore, this review will discuss the methods of multimodal AD diagnosis. Below, we will briefly introduce the diagnostic methods for each modality up to now, and finally discuss the methods for multimodal diagnosis of Alzheimer's disease (AD). More detailed content will be described in the main text.


\subsection{Neuroimaging}

\quad Clinical trials have shown that AD will bring key changes to the patient's brain, such as the accumulation of the protein fragment beta-amyloid into clumps (called beta-amyloid plaques) outside neurons and the accumulation of an abnormal form of the protein tau (called tau tangles) inside neurons\cite{AD_report}. Brain atrophy is another change, which is due to cell loss and decreased ability of cells to metabolize glucose (glucose is the main fuel for the brain). Thus, AD is associated with pathological amyloid deposition, structural brain atrophy, and altered brain metabolism\cite{image_progress}. In recent decades, major advances in neuroimaging techniques have made these techniques one of the most important biomarkers in the diagnosis of AD. Neuroimaging techniques offer valuable insights into the human brain. Structural magnetic resonance imaging (MRI) enables the detection of brain atrophy, while functional imaging modalities like positron emission tomography (PET) and functional MRI (fMRI) are capable of identifying hypometabolism\cite{PET-fMRI}. Furthermore, metrics such as mean diffusivity (MD) and fractional anisotropy (FA) measured by diffusion tensor imaging (DTI) provide indications of a person's cognitive status. Additionally, electroencephalography (EEG) allows for the assessment of communication activity between nerve cells, while magnetoencephalography (MEG) measures the magnetic fields generated by currents flowing within neurons, providing insights into brain activity\cite{god}. By employing these diverse techniques, researchers gain a comprehensive understanding of the brain's structure, function, and cognitive processes which can help them develop diagnostic methods. 

\quad However, multimodal imaging studies may offer various advantages over unimodal imaging studies. Multimodal imaging studies are able to study the temporal and topographical relations between many pathological variables, thus improving our understanding of pathophysiological interactions in the body. This approach allows direct comparison of the diagnostic capabilities of different imaging modalities in the same patient sample\cite{2015Multimodal}. Lu et al.\cite{muti_image_2} found that a network classifier constructed using a combination of FDG-PET and structural MRI images outperformed a network constructed using structural MRI or FDG-PET alone. Liu et al.\cite{muti_imge_3} used a zero-masking strategy for data fusion to extract complementary information from MR and PET to classify AD patients into four AD stages.

\quad Multimodality imaging studies are equally challenging because of the large number of imaging markers that are potential candidates for predicting disease transformation. This situation leads to two related problems. First, as the number of candidate features increases, the risk of data overfitting increases, and second, as the number of candidate features increases, covariance in the predictor variables becomes more severe\cite{2015Multimodal}. Whether these problems can be solved is the key to multimodal imaging research.


\subsection{Cognitive Assessment}

\quad Based on the patient's cognitive abilities, several tests are available to assess the level of AD (Alzheimer's disease) and MCI (Mild cognitive impairment). These include: the MMSE (Mini-Mental State)\cite{MMSE}, a simplified cognitive mental state test in the form of a score, which provides a quantitative assessment of cognitive state; The MoCA (The Montreal Cognitive Assessment)\cite{MoCA}, which provides a rapid assessment of different levels of cognitive impairment; and the ADAS-Cog (Alzheimer's Disease Assessment Scale - Cognitive)\cite{ADAS-Cog}, another commonly used clinical and experimental cognitive assessment tool; SCIP (Severe Cognitive Impairment Profile)\cite{SCIP}, SIB (Severe Impairment Battery)\cite{SIB} and so on. Roalf et al. compared the MMSE with the MoCA and concluded that the MoCA as a global assessment tool is superior to the MMSE and provides a reliable and simple conversion method from MoCA to MMSE scores\cite{MMSEvsMoCA}. 

\quad Most of these cognitive assessment methods are lengthy and complex and do not apply to all patients in all stages of dementia and do not perform well enough in terms of sensitivity\cite{cog_bad,god}. Although these cognitive assessment methods can provide a quantitative evaluation of cognitive status and help doctors understand the cognitive state of patients, they have limitations in terms of sensitivity. They may not capture subtle changes in certain cognitive domains. Some tests also require professional personnel for evaluation and interpretation, and the test results may be influenced by factors such as education and cultural background.

\subsection{Non-neuroimaging Biomarkers}

\quad Biomarkers are objective measurements of biological or pathogenic processes aimed at assessing disease risk or prognosis, guiding clinical diagnosis or monitoring therapeutic interventions\cite{Biomarker}. Changes in biomarkers can be obtained from neuroimaging on the one hand, and from changes in the composition of biofluids on the other.  Combined with clinical approaches and cognitive tests, biomarkers will be more useful to accomplish an accurate assessment of cognitive impairment and its causes, even allowing clinicians to identify and detect pathology caused by AD before it occurs\cite{2009Multimodal}. The section regarding neuroimaging-based biomarkers will be discussed in detail in the neuroimaging section. 

\quad Among the Non-neuroimaging Biomarkers and changes that have proven useful so far are hippocampal atrophy on decreased $A\beta_{42}$ in cerebrospinal fluid (CSF)\cite{a_beta42}, and increased tau protein and phosphorylated tau protein\cite{a_beta_tau}. In addition, blood-based biomarkers have likewise attracted the attention of experts, blood samples can be obtained in a less invasive and cheaper way\cite{blood_1}. Similar to CSF, blood-based biomarkers can measure $A\beta_{42}$ and other forms of $A\beta$  proteins, as well as different forms of tau. Additionally, Neurofilament light chain (NfL) shows promise as a blood-based biomarker for AD.  Moveover, in the process of diagnosing AD, genetic factors should not be overlooked. An example of a gene-based biomarker used to detect AD is the $\varepsilon4$ allele of the APOE gene. This genetic variation can be detected through blood samples or buccal swab samples. 

\quad These biomarkers can provide direct information about pathological processes such as abnormal protein deposition, inflammatory reactions, and neuronal damage, which help understand the development and progression of AD. However, they lack sufficient specificity, meaning that they may also be present in other neurological disorders or normal aging processes. Some biomarkers require the collection of specific samples, such as cerebrospinal fluid, which may involve a more specialized and technically demanding process, making it relatively less accessible for widespread use.

\subsection{Posture}

\quad In recent years, some progress has been made in research on posture (mainly face and gait) in the diagnosis of Alzheimer's disease (AD). Although postures are not currently the primary diagnostic criteria for AD, they play an important role as an aid in early diagnosis and monitoring.

\quad In terms of faces, studies have found that AD patients have problems with facial emotion expression and facial emotion comprehension. AD patients are impaired in facial expression recognition \cite{AD_face_ill} and have significant emotion recognition deficits \cite{AD_face_bar}, Fiona et al.\cite{AD_face_only} found that in Alzheimer's disease, emotion processing deficits were only found in complex and cognitively demanding emotion recognition tasks, while behavioral performance in simple face processing and emotion matching tasks was within the normal range. This offers the potential to use facial expression analysis as a tool for early AD diagnosis and monitoring.

\quad Gait is a complex cognitive task requiring coordination between a wide range of brain regions, and even in the milder stages of the disease, gait impairment may reflect dementia-induced neurodegeneration \cite{AD_gait_withCog}. There is growing evidence that cognitive, sensory, and motor changes may precede the clinical manifestations of AD by several years\cite{AD_gait_change}. Gait disturbances reported in early AD include slower gait, shorter stride length, lower cadence (longer stride time/gait cycle), and greater inter-stride variability\cite{AD_gait_bad}. Therefore, gait analysis is also considered to be an important tool for assessing motor function and cognitive status in AD patients. For the assessment of gait characteristics, Rosaria et al. suggested that gait characteristics can be divided into temporal, kinematic and kinetic characteristics\cite{AD_gait_chara}.

\quad In conclusion, facial and gait analysis has shown some potential in AD diagnosis. As technology continues to evolve and more research work is done, these aids are expected to be a useful addition to early AD diagnosis. However, facial and gait analysis is still in the research phase and more validation and standardization work is needed. Individual differences and other factors may have an impact on facial and gait performance, so further research is needed to determine their accuracy and reliability in AD diagnosis and monitoring.


\subsection{Sound}

\quad Voice and Speech problems are considered to be one of the most typical symptoms of AD, which is a direct and unavoidable consequence of cognitive impairment\cite{speech_result}. It has been demonstrated that AD patients perform poorly on different language tests\cite{speech_bad}, presenting naming and word-finding difficulties (anomia) leading to circumlocution, as well as difficulty accessing semantic information intentionally, leading to a general semantic deterioration\cite{speech_bad_1}, and behaving differently from normal in some acoustic and rhythmic features\cite{speech_ray}. This demonstrates the feasibility of using voice modality to diagnose Alzheimer's disease and that Voice and Speech can be used as an efficient, inexpensive, and easy-to-use tool to help in the diagnosis of AD.

\quad People with AD differ from normal people in semantics, syntax, and rhythm, and researchers have been able to diagnose AD by looking at a variety of features. The main conventional features used in Alzheimer's disease research are: Frequential Aspects (including interruptions, Voice periods, Fundamental frequency); Intensity (including Amplitude and Phonatory stability); Voice Quality (including noise); Biomechanical aspects (including Vocal fold body Movement Tongue movement)\cite{AD_speech_review}. Many studies have also identified acoustic measures that are highly correlated with pathological speech features or speech alterations \cite{voice_chara}. In recent years, with technological advances some new methods have been gradually invested in AD diagnosis, Fasih et al\cite{2020An} first used eGeMAPS\cite{[35]}, emobase\cite{2010Opensmile} and ComParE\cite{[37]} feature sets as Alzheimer's disease empirical attempts to introduce and evaluate a new method to represent these acoustic features ADR (active data representation). Liu et al. \cite{mh_1} divide a person's speech data into multiple segments and use the extracted spectrogram features from the speech data to identify AD.

\quad As a non-invasive and rapid diagnostic method, recognition technology based on patient voice data can effectively reduce medical costs compared to medical images that are difficult to obtain. Especially, natural language processing technology, signal processing technology, and deep learning technology have been developed significantly in recent years, and the technology based on automatic processing of voice signal records is gradually becoming mature.


\subsection{Multimodal}

\quad Currently, most studies on Alzheimer's disease (AD) use a single data model for prediction, which may have limitations. Psychological or Cognitive appraisal questionnaires may be too subjective and may lack sensitivity\cite{god}. Changes in both posture and voice may be influenced by factors unrelated to AD, such as normal aging. Neuroimaging also suffers from cost and availability issues (availability of PET and MRI scanning instruments varies widely between countries) and from patient bias (e.g., sensitivity to radiation exposure)\cite{2015Multimodal}.

\quad It has been shown that fusing complementary information from multiple modalities can improve the diagnostic performance of AD. Multimodal data contains the fusion of complementary information(e.g., Magnetic Resonance Imaging (MRI), Positron Emission Tomography (PET) and genetic data)\cite{mutimodal_1}. Landau et al. also found complementary information between acquired genetic, cerebrospinal fluid, neuroimaging, and cognitive measures\cite{hubu}. 

\quad  However, there are still challenges in fusing data from multiple modalities to diagnose AD. To begin with, it is important to recognize that different types of data are inherently diverse. Each modality, such as neuroimaging and genetic data, exhibits distinct data distributions, varying numbers of features, and differing levels of diagnostic discrimination for conditions like Alzheimer's disease (AD) \cite{zhou2019effective}. To fuse multimodal data, traditional approaches usually first perform feature selection for each modality separately, and then cascade the selected features used for diagnosis or prognosis. Nevertheless, this approach ignores the potential connection between different modal data \cite{mutimodal_1}. 

\quad The second challenge is the high dimensionality problem encountered in the fusion analysis of multimodal imaging for diagnostic AD. When combining data from multiple modalities, the resulting dataset tends to have a high number of dimensions. For instance, a single neuroimaging scan, such as MR or PET images, contains millions of voxels. Classical methods\cite{multimodal_c}, such as principal component analysis (PCA), independent component analysis (ICA), and linear discriminant analysis (LDA), are used in many studies to solve the high-dimensional problem of multimodal fusion analysis. These methods achieve attribute parsimony, but researchers need to put a lot of effort to analyze some important fusion features separately\cite{multimodal_c}.

\quad Lastly, there is the issue of incomplete data, where not all samples possess complete multimodal data. Typically, researchers opt to discard samples with missing data, thereby increasing the risk of sample loss. Alternatively, one approach involves interpolating the missing data using methods like zero interpolation, k-nearest neighbor (KNN), or Expectation Maximization. However, this interpolation approach may introduce unnecessary noise, subsequently compromising the model's performance.

\quad Many researchers have proposed solutions to the above challenges. 
Zhang et al.\cite{multimodal_d} proposed a general framework based on kernel methods, which can effectively combine MRI, PET, and CSF features and naturally embed them into traditional support vector machines to effectively solve, achieving high accuracy in AD classification. Zhou et al.\cite{zhou2019effective} proposed a three-stage deep feature learning and fusion framework, which utilizes multimodal neural image data (i.e. MRI and PET) and genetic data (i.e. SNP) to learn potential representations for each individual modality and joint potential representations for each pair of modalities in the first two stages. In the third stage, the classification model is learned using joint potential representations from all modality pairs. Janani et al.\cite{multimodal_e}used the stack denoising automatic encoder to process EHR and SNP data, used 3D Convolutional neural network (CNNs) to train MRI imaging data, cascaded these intermediate features and transferred them to the classification, indicating that the multi-modal data analysis using DL is better than the single-mode DL model.

\subsection{The main highlights of this literature survey}


\quad(1) We have conducted a comprehensive review of the current mainstream AD diagnostic methods based on various modalities, summarizing the research progress and recent advancements in these modalities over the past five years.

\quad(2) We have analyzed the latest research on the application of multimodal techniques in AD diagnosis and discussed the current challenges in multimodal fusion. We also present different solutions proposed by researchers in this field.

\quad(3) We provide possible directions and suggestions for future multimodal AD diagnostic technologies.



\section{neuroimaging} 
Currently, the consensus regarding Alzheimer's disease (AD) is that individuals first enter the preclinical stage (such as being asymptomatic), then progress to mild cognitive impairment (MCI), and eventually develop AD dementia. MCI patients are further classified into early-stage MCI (EMCI) or late-stage MCI (LMCI) based on their performance on cognitive screening tools, with LMCI patients having the highest risk of AD conversion, while EMCI has a relatively lower conversion rate \cite{2010Clinical}. However, relying solely on cognitive deficits to determine whether a patient is in the early or late stage of MCI is quite challenging, as traditional cognitive assessment tools often fail to provide accurate judgments. This is where neuroimaging methods prove valuable, as they can detect substantial changes within the brain and provide crucial clues by visualizing and measuring gross alterations in the brains of AD subjects\cite{2016Preclinical}. Neuroimaging is not only applicable to individuals already diagnosed with AD dementia but also to MCI patients and even cognitively normal individuals in the preclinical stage, allowing for accurate assessment of the extent of the disease in patients. The Alzheimer's Disease Neuroimaging Initiative (ADNI) database (http://adni.loni.usc.edu) collects and analyzes a large-scale dataset consisting of neuroimaging, biomarkers, cognitive assessments, and clinical data, including clinical evaluations, brain imaging modalities (such as MRI, PET), and biological samples (such as cerebrospinal fluid, plasma). These data are utilized to investigate the pathological mechanisms of AD, biomarkers for disease progression, the interplay between genetics and environmental factors, and the evaluation of novel treatment approaches.

In this chapter, we extensively discussed several mainstream neuroimaging techniques (Table \ref{sum}) and explored their interrelationships. Additionally, we analyzed computer-based approaches for diagnosing AD, aiming to provide more accurate prediction and diagnostic tools to anticipate cognitive decline and the progression from MCI to AD dementia in patients.

\begin{table*}[htbp]
  \caption{Neuroimaging Techniques in AD Diagnosis }
   \label{sum}
   \centering
  \begin{tabular}{llll}
    \toprule
    \thead{Technique} & \thead{Function} & \thead{Advantages} & \thead{Disadvantages} \\
    \midrule
    \thead{MRI} & \thead{Provides detailed brain structural images, \\ including brain volume, atrophy, etc.} & \thead{Non-invasive, high resolution} & \thead{Cannot 
    directly observe metabolism\\ and functional activity} \\
    
    \thead{DTI} & \thead{Assesses integrity and connectivity \\of white matter fiber bundles} & \thead{Provides white matter microstructure \\information, detects fiber bundle damage} & \thead{Limited in fiber crossing\\ and challenging areas} \\
    
    \thead{sMRI} & \thead{Analyzes changes in brain\\ tissue morphology and density} & \thead{Non-invasive, provides\\ structural information} & \thead{Does not provide functional \\and metabolic information} \\
    
    \thead{fMRI} & \thead{Detects brain activity and \\functional connectivity} & \thead{Observes brain functional activity, \\provides functional connectivity maps} & \thead{Limited sensitivity, affected \\by motion and noise} \\
    
   \thead{PET} & \thead{Evaluates brain metabolism \\and function} & \thead{Observes brain metabolism 
   \\and functional activity, widely\\ used in biomarker research} & \thead{Requires radioactive tracers, expensive} \\
   
    \thead{FDG-PET} & \thead{Assesses brain glucose metabolism} & \thead{Detects abnormal brain\\ glucose metabolism, \\associated with AD} & \thead{Requires radioactive tracers,\\ expensive, cannot directly observe \\other metabolic activities} \\
    \bottomrule
  \end{tabular}
\end{table*}

\subsection{Magnetic Resonance Imaging (MRI)}
In AD research, amyloid-beta (A$\beta$) is the most commonly studied biomarker, and there is substantial evidence indicating the presence of A$\beta$ deposition in a large proportion of individuals at the preclinical stage\cite{1989Morphology}. $\beta$-amyloid protein (A$\beta$) is generated through the cleavage of the amyloid precursor protein by the $\beta$-secretase and γ-secretase complex. Once released as monomers, A$\beta$ can form oligomers with cellular toxicity or neuroregulatory properties\cite{CAPPAI1999885}, leading to neurodegeneration and cognitive impairment. Therefore, it is crucial to detect the presence of A$\beta$ before its deposition reaches a significant level.

Magnetic Resonance Imaging (MRI) can effectively measure the levels of A$\beta$ in various brain regions. Changes associated with A$\beta$ deposition are closely related to increased gray matter atrophy, particularly in the hippocampus. Thus, the hippocampus is one of the key regions affected in AD dementia. By scanning brain regions such as the hippocampus, entorhinal cortex, amygdala, medial and lateral temporal lobes, lateral ventricles, medial temporal gyri, and cortical gray matter, MRI can detect brain atrophy caused by severe neuronal loss\cite{2010TheAL}.

Currently, MRI techniques are well-established. For example, the team at the University of California, San Francisco (UCSF) Medical Center utilizes FreeSurfer for cortical reconstruction and volume segmentation\cite{Michael2013The}. Detailed information regarding MRI data acquisition, preprocessing, and quality control can be found at (https://ida.loni.usc.edu).

\subsection{Diffusion Tensor Imaging (DTI)}
DTI is a magnetic resonance imaging (MRI) technique used to assess microstructural changes within the white matter fiber tracts of the brain\cite{2010Update}. It provides information by measuring the diffusion of water molecules in tissues. The most commonly used metric in DTI is Fractional Anisotropy (FA), which reflects the degree of diffusion anisotropy in different directions. A decrease in FA is often considered an indicator of axonal degradation and demyelination in the white matter of the brain\cite{2009The}.

Using DTI imaging to evaluate white matter fibers, studies have found significant fiber damage in the connections between the hippocampus and the posterior cingulate gyrus in AD patients. This suggests that white matter damage may be associated with gray matter atrophy in the temporo-parietal brain network implicated in AD. A recent meta-analysis of DTI studies in AD patients confirmed the significant role of fiber damage in the posterior cingulate gyrus and the major white matter tracts connecting the prefrontal cortex with the medial temporal or parietal cortex\cite{2013Effectiveness}.

\subsection{Structural MRI}
\begin{figure}[h]
    \centering
    \includegraphics[width=0.5\textwidth]{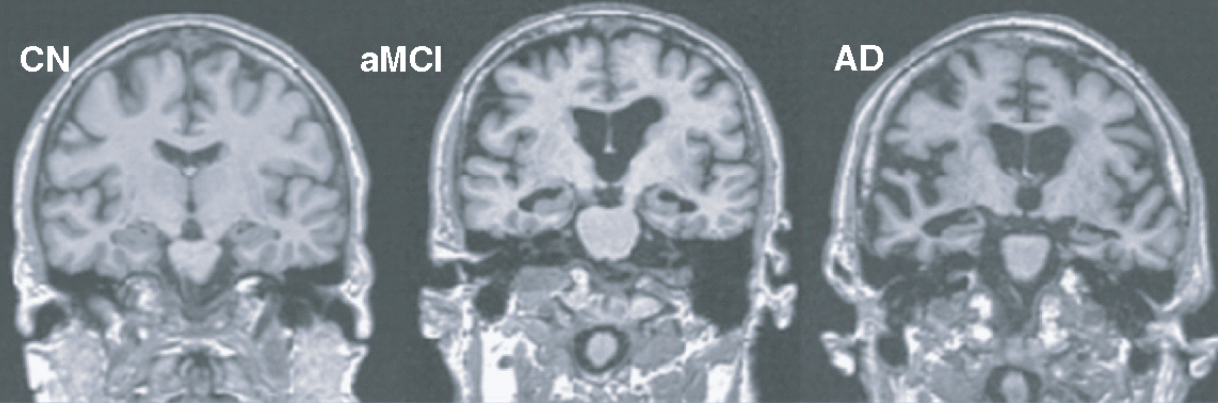}
    \caption{MRI imaging of hippocampal volume structure in normal control group (CN), MCI and AD patients }
    \label{SMRI}
\end{figure}

Structural MRI is a technique used to assess brain changes in AD patients by detecting alterations in the gray matter and white matter. This technique is particularly sensitive to changes in gray matter volume due to neuronal loss and atrophy\cite{2010Correlates}. Brain atrophy is widely recognized as a hallmark of AD and is associated with the severity of the disease\cite{2010Use}. In Alzheimer's disease, neurons in the medial temporal lobe and hippocampus are particularly vulnerable to loss, which is consistent with the occurrence of significant cognitive impairment in clinical presentations. Studies have found that in the early stages of AD dementia, the volume of the hippocampus has already decreased by 20\%\cite{Kehoe2014AdvancesIM} (Figure \ref{SMRI}). Other brain regions such as the lateral temporal lobe, parietal lobe, and frontal lobe have also been found to be adversely affected\cite{2010Relative}.

\subsection{Functional Magnetic Resonance Imaging (fMRI)}
\begin{figure}[h]
    \centering
    \includegraphics[width=0.4\textwidth]{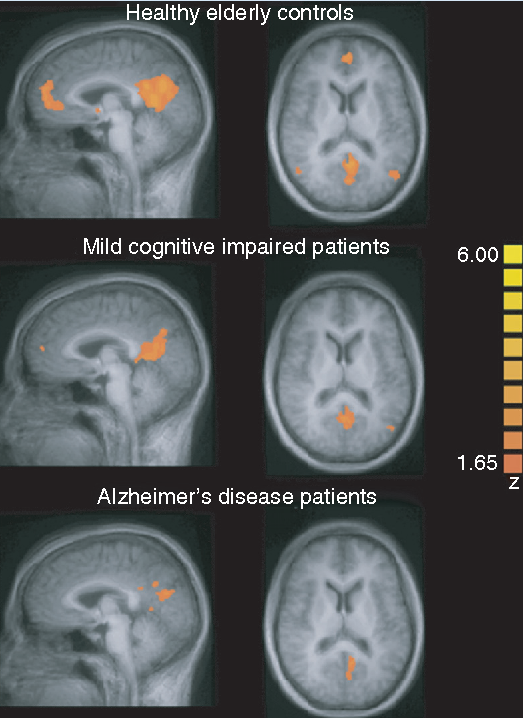}
    \caption{Schematic diagram of fMRI default network function connections for normal control group, MCI, and AD patients}
    \label{FMRI}
\end{figure}

Functional Magnetic Resonance Imaging (fMRI) measures synaptic activity and assesses changes in brain during resting state and cognitive tasks. Studies have found reduced hippocampal activity in individuals with AD during episodic memory tasks, consistent with early memory deficits\cite{Kehoe2014AdvancesIM} (Figure \ref{FMRI}).

In healthy individuals, the default mode network is active during resting state but deactivated during cognitive tasks. AD patients exhibit impaired intrinsic functional connectivity of the default mode network during resting state\cite{2004Default}. fMRI evaluation has shown that in the early stages of MCI, there is overactivation in the hippocampus during memory tasks, but a reduction in activity is observed in the preclinical stage, potentially indicating a sign of clinical decline. However, fMRI has limitations, such as the inability to directly measure neuronal activity, observe subcortical functional activity, and susceptibility to motion artifacts, which may result in false-negative and false-positive outcomes.

\subsection{Positron Emission Tomography (PET)}
\begin{figure}[!h]
    \centering
    \includegraphics[width=0.4\textwidth]{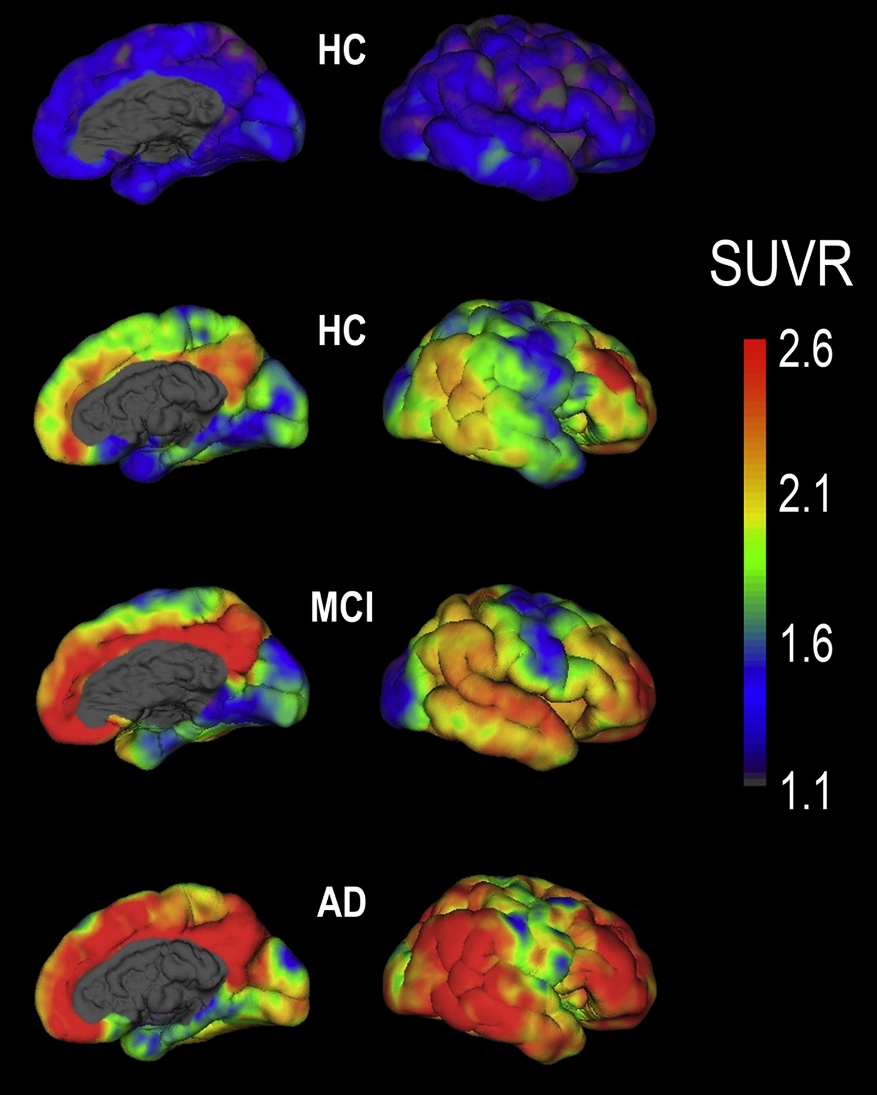}
    \caption{PET images of AD patients and healthy controls.It can be observed that AD patients have significant A$\beta$ deposition in the brain.}
    \label{pet1}
\end{figure}

In AD, the deposition of $\beta$-amyloid (A$\beta$) is a prominent feature. As the disease progresses, A$\beta$ leads to the pathological accumulation of tau protein (Figure \ref{pet1})\cite{2016Neuroimaging}, resulting in the formation of 'senile' plaques and neurofibrillary tangles\cite{2008Rapid}. These abnormal accumulations are associated with an increased risk of developing AD\cite{2010Clin}. Among the amyloid-binding compounds, ${ }^{11}$C-PIB and tau have been widely used, and their uptake can be measured using Positron Emission Tomography (PET).

Longitudinal PET scans have shown the accumulation of tau protein in AD patients. In the early stages, tau primarily accumulates in the medial temporal lobe, and later spreads to the lateral temporal lobe, as well as the superior and medial regions of the parietal lobe, before severe cognitive impairment occurs. Various radiotracers have been developed that selectively bind and highlight pathological structures in the central nervous system, including amyloid plaques, neurofibrillary tangles, activated microglia, and reactive astrocytes. Examples of compounds that enable amyloid plaque imaging include [${ }^{18}$F]FDDNP, ${ }^{18}$F-BAY94-9172, ${ }^{11}$C-S B-13, ${ }^{11}$C-B F-227, and ${ }^{11}$C-PIB.

Using different PET tracers, the deposition of A$\beta$ can be visualized and quantitatively measured throughout the brain. Current imaging agents include 11C-labeled radioactive tracer PiB or 18F-labeled tracers such as [${ } ^{18}$F]florbetaben, [${ } ^{18}$]florbetapir (also called [${ } ^{18}$]AV-45), and [${ } ^{18}$]flutemetamol.

\subsection{18F-Fluorodeoxyglucose PET (FDG-PET)}
\begin{figure}[!h]
    \centering
    \includegraphics[width=0.4\textwidth]{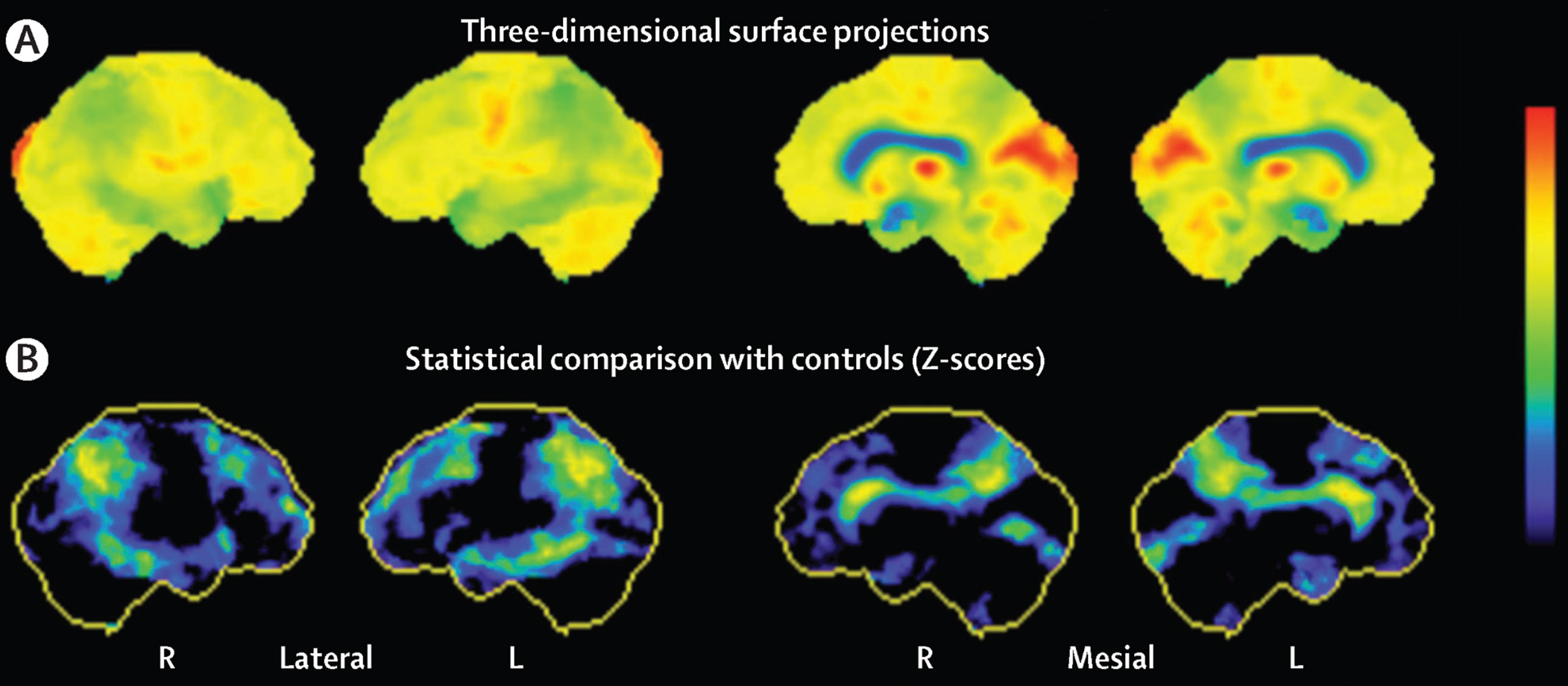}
    \caption{Figure A shows an FDG-PET image of an AD patient, with yellow or red indicating hyperglycemia and green indicating hypometabolism. Figure B shows the comparison between the FDG-PET scan of this patient and the healthy population, with green or yellow indicating high deviation, and black or blue indicating no deviation or low deviation.}
    \label{FDG}
\end{figure}

In dementia patients, 18F-FDG-PET (18F-labeled fluorodeoxyglucose positron emission tomography) can detect the reduction of brain metabolism, which is a hallmark of neurodegeneration. 18F-FDG-PET measures the regional glucose consumption directly associated with the local intensity of glutamatergic synaptic and astrocytic activity in the brain. It can assess the degree and location of decreased brain metabolism, reflecting impaired neuronal function \cite{2016CSF} (Figure \ref{FDG}).

18F-FDG-PET is particularly useful in early diagnosis as it can reveal the characteristic pattern of neurodegeneration in individuals with mild cognitive impairment associated with Alzheimer's disease earlier than MRI. It detects early changes related to AD pathology more sensitively and can be used to predict the transition from cognitive normality to mild cognitive impairment.

\subsection{Neuroimaging Analysis Methods}
Several classical machine learning techniques have been used to explore Alzheimer's disease (AD), ranging from image decomposition techniques like principal component analysis to more complex nonlinear decomposition algorithms. With the advent of deep learning paradigms, it is now possible to directly extract high-level abstract features from MRI images \cite{2019Deepl}.

The methods for diagnosing AD based on neuroimaging can be broadly categorized into four types\cite{2019Deepl}: slice-based, patch-based, voxel-based, and ROI-based. Slice-based methods assume that useful features are contained within 2D image slices, reducing the number of learnable parameters. Voxel-based methods are the most direct, using voxel intensity values from the entire 3D brain scan. Patch-based methods extract features related to AD patterns by extracting small 3D cubes (called "patches") from the brain. ROI (Region of Interest) methods emphasize specific regions in the brain that are related to AD rather than the entire brain. The diagnostic process often requires prior knowledge of abnormal regions associated with AD, such as the hippocampus.

Slice-based methods simplify certain important regions into 2D images, reducing the number of hyperparameters. Farooq et al. \cite{2017Ad} performed slice-based axial scanning of GM volume in the images, discarding slices that did not contain informative information. Convolutional neural networks (CNNs) have been shown to have a revolutionary impact in the field of neuroimaging. Recent studies have utilized 2D CNNs for Alzheimer's disease (AD) detection on magnetic resonance imaging (MRI) scans, where each MRI scan is divided into multiple 2D image slices. The CNN is trained by computing the loss function between the labels of each subject and the predicted outputs of each image slice. Wang et al. \cite{2021ADVIAN} propose a VGG-inspired network that integrates convolutional block attention modules into the VIN backbone. However, a major drawback of slice-based methods is that 2D CNNs cannot understand the dependencies between voxels in the images. Converting 3D MRI scans into 2D image slices results in data loss. This is because brain regions span multiple 2D slices of the MRI scan, and the features related to the size and shape of the brain region will be lost after slicing. Ebrahimi et al. \cite{WOS:000674484500001} introduced Temporal Convolutional Networks (TCNs) to address this issue. TCNs model the MRI feature sequences generated by CNNs to capture time-based dependencies, which are used for AD detection. TCNs can understand the temporal dependencies between 3D image slices within the 2D MRI volume.

Voxel-based methods capture all 3D information within a single brain scan, but they often treat all brain regions uniformly without considering specific anatomical structures. At the voxel level, tissue density (e.g., GM/WM) is commonly used as a feature for classification algorithms. 3D CNNs are typically employed in voxel-based methods \cite{2021Multi}. Murcia et al. \cite{2020Studying} modeled the dataset by applying Convolutional Autoencoders (CAEs). CAEs can directly extract data-driven features from three-dimensional images without the need for feature extraction beforehand. However, voxel-based methods are structurally complex and require a large number of training parameters, which can lead to overfitting. Additionally, voxel-level feature representations have high dimensionality \cite{2017Ar}. Therefore, dimensionality reduction of features becomes a primary challenge in improving the classification performance of voxel-based methods.

Patch-based methods offer an intermediate scale between voxel-level and slice-level approaches, allowing for more effective capturing of local structural changes in images. By extracting features from patches (also known as patches) sampled randomly from MR images, many weak classifiers can be built and combined to make the final decision for AD diagnosis \cite{0Development}. This approach involves training a fully convolutional network (FCN) using randomly sampled patches from the entire MRI volume. With the trained FCN, high-risk voxels can be selected and fed into a multilayer perceptron (MLP) for individual-level AD classification. Lian et al. \cite{2018Hierarchical} proposed a hierarchical fully convolutional network (H-FCN) that learns multiscale feature representations (e.g., patch-level, region-level, and subject-level) from sMRI scans, constructing a hierarchical classification model.

In recent years, patch-based methods have gained wide application in deep learning. In AD diagnosis, gray matter (GM) images of the brain are typically divided into 3D patches based on regions defined by the Automated Anatomical Labeling (AAL) atlas. These patches are then used to train various deep learning models. Through this approach, the models can learn to extract high-level features and patterns from the patches for AD classification and diagnosis. Dense networks \cite{2020Studying}, 2D and 3D convolutional neural networks (CNNs)\cite{payan2015predicting}\cite{2017Deepo}\cite{2018Convolutional}\cite{Ruoxuan2018Hippocampus}, and residual networks \cite{Geert2017A} have been used to learn features from sMRI for AD diagnosis. Zhu et al. \cite{2021Dual} proposed a dual-attention multiple instance deep learning network (DA-MIDL), which consists of patch networks with spatial attention blocks to extract discriminative features from each sMRI patch. The method also employs attention-based multiple instance learning (MIL) pooling operations to balance the relative contributions of each patch. Patch-based methods capture disease-related patterns in the brain by extracting features from small image patches. The main challenge in this approach is selecting image patches with the most informative content to capture both local (patch-level) and global (image-level) features.

The ROI-based methods, which rely on pre-segmented regions of interest (ROIs), have also been widely used in AD diagnosis. This process typically involves three main components: 1) predefining the ROIs of interest, 2) extracting imaging features, and 3) constructing a classification model. With the emergence of machine learning (ML) methods, automatic feature extraction techniques have been employed\cite{2018Assessing}\cite{2014Spatial}\cite{2009Robustness}. Heckemann et al.\cite{Rolf2010Improving} proposed a multi-atlas propagation with enhanced registration (MAPER), the first automatic whole-brain multi-region segmentation method that enables automatic segmentation of acquired MRI images. Liu et al.\cite{0Neuroimaging} used the MAPER method for feature extraction and selection, segmenting 15 brain structures in each MRI image and calculating the gray matter (GM) volume of brain regions as candidate predictive factors. Finally, by utilizing classifiers such as support vector machines (SVM), the impact of predictive factors on the AD progression can be evaluated. Li et al.\cite{2021An} employed principal component analysis (PCA) for dimensionality reduction to obtain highly discriminative features by removing redundant features. However, ROI-based methods still have some limitations. The definition of ROIs requires researchers to accumulate extensive experience, and the segmentation of ROIs can be influenced by individual differences and subjective factors among experts. Moreover, morphological abnormalities caused by brain diseases may involve multiple ROIs or partially predefined ROIs, which may result in unstable performance.

Lee et al. \cite{lee2019toward}proposed a feature representation method that combines voxel-based, region-based, and patch-based approaches, known as a hybrid method. Specifically, they employed a deep neural network (DNN) to learn complex voxel relationships within each region. In this approach, the brain was divided into predefined regions (region-based method), and the intricate nonlinear relationships between voxels (voxel-based method) were determined based on the anatomical shape of each region (patch-based method). By integrating these three approaches, the proposed method aimed to capture comprehensive information from different spatial scales, improving the representation of brain features for AD diagnosis.

\section{Clinical Data of Biomarkers}

Although neuroimaging plays a crucial role in the diagnosis and research of Alzheimer's disease (AD), it may not provide sufficient sensitivity and specificity, especially in early diagnosis, due to the complex nature of AD. Therefore, there is a need to identify more specific biomarkers to detect pathological changes in AD at an earlier stage. In the field of AD research, the identification and detection of biomarkers play a vital role in early diagnosis, monitoring disease progression, and evaluating treatment efficacy. Various methods have been developed for biomarker detection, providing valuable insights into the underlying pathological processes of AD. This section provides an overview of four commonly used biomarker detection methods: cerebrospinal fluid (CSF) analysis, plasma analysis, APOE gene variants, and assessment of oxidative stress and inflammation(Table \ref{biomarker}). To facilitate AD research and early diagnosis, the Alzheimer's Disease Neuroimaging Initiative (ADNI) database (http://adni.loni.usc.edu) was launched in 2003, aiming to develop and validate biomarkers for early detection and treatment of AD.

\begin{table*}[htbp]
  \caption{Biomarker Detection Methods }
  \label{biomarker}
  \centering
  \begin{tabular}{lll}
    \toprule
    \thead{Biomarker Method} & \thead{Characteristics} & \thead{Pros and Cons} \\
    \midrule
   \thead{ Cerebrospinal Fluid (CSF)} & \thead{- Provides direct information \\about neurological diseases} & \thead{- Detection method is relatively complex, \\requiring lumbar puncture or intrathecal\\ injection for sample collection} \\ & \thead{- Allows detection of various biomarkers\\ such as amyloid-beta and tau proteins} & \thead{- Invasive sampling may cause \\discomfort and risks} \\
    \midrule
  \thead{Blood Plasma} & \thead{- Non-invasive and easy to collect} & \thead{- Biomarker concentrations in \\plasma are relatively low, \\limited detection sensitivity} \\
                 & \thead{- Allows detection of certain biomarkers \\such as proteins and metabolic products} & \thead{- Susceptible to interference from \\other factors such as diet and medication} \\
    \midrule
    \thead{APOE Allelic Variants} &\thead{- Genetic biomarker for \\assessing individual genetic risk} & \thead{- Genetic variants associated \\with AD risk exhibit diversity} \\
                          & \thead{- Detected through genetic typing or sequencing} & \thead{- Genetic factors only partially explain AD risk} \\
    \midrule
   \thead{- Oxidative Stress and Inflammation} & \thead{- Reflects the activity and \\extent of cellular oxidative damage\\ and inflammatory response} & \thead{- Oxidative stress and inflammation \\levels are influenced by multiple factors} \\
    & \thead{- Evaluation can be done \\by detecting related biomarkers} & \thead{- Lack of specificity, potential overlap with other diseases} \\
     & \thead{- Requires further validation and\\ replication of research findings} &\thead{ - Consistency and reliability of results need confirmation} \\
    \bottomrule
  \end{tabular}
\end{table*}

\subsection{Cerebrospinal Fluid and Blood Plasma Sampling}

Sampling of cerebrospinal fluid (CSF) and plasma is one of the most direct and convenient methods for studying biochemical changes occurring in the central nervous system. Compared to brain biopsy or direct insertion of microdialysis probes into the brain, CSF and plasma sampling methods are less invasive and easier to obtain \cite{BLENNOW200718}. Through the collection of CSF and plasma samples, valuable information about biomarkers related to the central nervous system can be obtained. These biomarkers include amyloid-$\beta$ (A$\beta$₁₋₄₂), total tau (Tau), and phosphorylated p-TAU\_181p (pTau). CSF samples can be collected through lumbar puncture, and in some cases, from the fourth ventricle or lateral ventricle puncture. Studies have found a significant decrease in the average concentration of A$\beta$42 in the CSF of Alzheimer's disease (AD) patients \cite{1995Reduction}, which is believed to be associated with the deposition of A$\beta$ protein in plaques. According to the "amyloid cascade hypothesis," these deposits prevent A$\beta$42 from entering the CSF, leading to a decrease in its concentration. Low levels of CSF A$\beta$42 may also be indicative of amyloid deposition \cite{Mawuenyega2010Decreased}.

The advantages of CSF collection include relatively low cost and providing insights into neurodegeneration, tau protein, and amyloid pathology \cite{2016CSF}. However, CSF biomarkers alone cannot provide detailed information about the extent and progression of pathology or neurodegeneration over time. Additionally, CSF analysis cannot directly determine the location and severity of pathology. Therefore, when using CSF biomarkers for diagnosis and disease progression monitoring, it is often necessary to combine them with other clinical and imaging examinations to obtain a more comprehensive evaluation.

Compared to cerebrospinal fluid (CSF), plasma sampling is less invasive and more cost-effective. However, the utility of plasma A$\beta$42 as a biomarker for Alzheimer's disease (AD) is still unclear, as it may not accurately reflect A$\beta$ changes in the brain. Some studies have shown that plasma A$\beta$42 and A$\beta$40 do not reflect A$\beta$ accumulation in the brains of AD patients \cite{2006Inverse}\cite{2007Plasma}. Most studies have found no significant changes in plasma A$\beta$ levels in sporadic AD patients \cite{2000Standardization}\cite{Tomasz2005Plasma}. However, there are also studies that have found increased \cite{2003Plasma} or decreased \cite{2006Plasma} plasma A$\beta$42 levels in AD patients. Furthermore, some studies have suggested that the ratio of A$\beta$40/A$\beta$42 in plasma can predict the transition from normal cognition to mild cognitive impairment and AD \cite{Richard1999Plasma}. Nakamura et al. \cite{2018Highp} efficiently measured plasma amyloid-$\beta$ biomarkers using immunoprecipitation and mass spectrometry combined methods, demonstrating the potential clinical utility in predicting brain amyloid-$\beta$ burden at an individual level.
While plasma sampling is less invasive and more easily accessible, the use of plasma biomarkers for AD diagnosis and progression monitoring is still under investigation. Further research is needed to better understand the relationship between plasma biomarkers and the underlying pathophysiological changes in AD.

In addition to A$\beta$ protein, other plasma biomarkers such as pTau forms, Neurofilament light chain, and Glial fibrillary acidic protein have also been found to be associated with AD \cite{Prof2022Blood}. Researchers have analyzed the correlation between plasma biomarkers and AD and used statistical methods to assess the strength of the association between plasma biomarkers and A$\beta$-PET burden. They found a significant correlation between plasma biomarkers and regions of high A$\beta$ deposition in the brain \cite{2018Highp}. These findings suggest that plasma biomarkers may provide valuable insights into the pathological processes associated with AD and have the potential to serve as non-invasive tools for AD diagnosis and monitoring. However, further research is still needed to fully understand the relationship between plasma biomarkers and AD and to validate their clinical utility.

Hansson et al. \cite{2008Evaluation} conducted statistical analysis using Spearman's correlation coefficient to evaluate the predictive value of plasma A$\beta$ as a factor for the subsequent development of AD. Verberk et al. \cite{2018pla} compared baseline demographic and clinical characteristics using t-tests, Mann-Whitney U tests, and chi-square tests. In addition, they used logistic regression analysis and receiver operating characteristic (ROC) curve analysis to assess the association between plasma biomarkers and abnormal amyloid status based on CSF and PET \cite{2021Ab}. These analyses provide valuable insights into the relationship between plasma biomarkers and AD and contribute to the understanding of their diagnostic and prognostic value.

\subsection{Genetics}
In addition to the methods mentioned above, there are genetic factors associated with AD that need to be considered. The APOE gene has three common alleles: APOE ε2, APOE ε3, and APOE ε4. Among them, the APOE ε4 allele is considered a major genetic risk factor for AD \cite{0Apolipoprotein}.

    In AD research, the APOE gene and its polygenic hazard score (PHS) are of great importance. The polygenic hazard score (PHS) is based on 31 single nucleotide polymorphisms (SNPs) and can reliably identify AD risk in individuals of any age \cite{2017Genetic1}. It has been found that the effect of APOE*ε4 on AD risk is mediated through the inhibition of amyloid-beta (A$\beta$) clearance and promotion of A$\beta$ aggregation \cite{2010Accumulation}\cite{2010APOE}. Furthermore, carrying the ApoE ε4 allele is associated with accelerated A$\beta$ deposition, hippocampal atrophy, and disruption of the default mode network, and these effects can manifest from the preclinical stages of AD. APOE4 promotes the pathogenic mechanisms of AD by impairing microglial reactivity, lipid transport \cite{2017APOE}, synaptic integrity and plasticity \cite{0Plastic}, glucose metabolism \cite{2017Apolipoprotein}, and cerebrovascular integrity and function \cite{2015Genetics}. Although all isoforms of apolipoprotein E (APOE) are associated with A$\beta$ deposition in the brain, the APOE ε4 allele confers a significantly higher risk for AD compared to the presence of APOE ε2 or APOE ε3. Therefore, incorporating APOE genotyping into clinical decision-making and treatment planning can enable more personalized and precise management and treatment of AD.

\subsection{Oxidative Stress and Inflammation}
There is growing evidence that oxidative damage plays a significant role in the pathological processes of Alzheimer's disease (AD). Inflammatory processes associated with AD pathology, including microglia and astrocytes surrounding plaques, are also implicated in the disease \cite{2006The}. In addition to their potential direct involvement, the secretions of these inflammatory cells, such as acute-phase proteins, alpha-1 antichymotrypsin (ACT, also known as SERPINA3 protein), alpha-2 macroglobulin (α2M), as well as activators of the classical pathway of complement and cytokines like interleukin-1$\beta$ (IL-1$\beta$) and tumor necrosis factor-alpha (TNF-α, also known as TNF), persist within plaques \cite{2018Highp}.

In the pathogenesis of Alzheimer's disease, free radicals also play a significant role in neuronal damage. One important consequence of free radical damage is lipid peroxidation, leading to the generation of F2-isoprostanes, which serve as biomarkers of this pathological mechanism. Multiple studies have shown that levels of F2-isoprostanes in the cerebrospinal fluid of AD patients are higher compared to healthy elderly individuals or non-AD dementia patients \cite{2007F2}.

\section{Posture}
As research on Alzheimer's disease (AD) advances, it has become apparent that relying solely on traditional neuroimaging and biomarker-based methods may not fully capture the early changes and subtle pathological features of AD. Therefore, finding a complementary approach to improve the accuracy and sensitivity of AD diagnosis has become increasingly important. In recent years, there has been growing interest in the use of posture assessment (including facial and gait analysis) in the diagnosis and study of AD. Posture, as a comprehensive expression of movement and emotion, can reflect changes in multiple aspects such as motor control, balance, and cognitive function. It can be utilized to identify early pathological changes in AD and evaluate disease progression.

\subsection{Facial Analysis}
In interpersonal communication, the recognition and production of facial expressions play a vital role in conveying emotional experiences and are closely linked to emotions, cognition, and behavioral adaptation. However, as AD progresses, patients may experience difficulties in recognizing emotional facial expressions. Through experiments, researchers have found that AD patients have impaired facial emotion expression abilities\cite{gressie_kumfor_teng_foxe_devenney_ahmed_piguet_2023}. They lose some of their capacity to understand and express emotional rhythms\cite{2022emo}. Significant differences have been observed between AD patients and healthy older adults in emotional and cognitive tasks. When unable to discern the emotions of others, patients may become more confused and agitated, leading to abnormal behaviors such as sadness, reluctance to engage in conversations, or throwing objects. Therefore, studying the emotional recognition abilities of AD patients is of great importance for their treatment and care\cite{Shirley2000Emotion}. Additionally, communication difficulties in the early stages of AD may arise from an inability to correctly identify others' emotions, leading to impaired early emotional attention. Therefore, studying facial emotion recognition in AD patients can contribute to early AD diagnosis.

In addition to the decline in cognitive abilities associated with AD, problems can also arise in the production of facial expressions. The amygdala, which is known to be interconnected with the frontal cortex, plays a crucial role in rapid responses to emotional stimuli and regulates emotion-related autonomic reactions\cite{2013Heightened}. However, some studies have shown that AD patients perform poorly on emotional tasks and exhibit deficits in emotional processing\cite{keith}. This can result in their facial expressions appearing somewhat rigid. Therefore, analyzing specific facial expressions made by AD patients can contribute to AD diagnosis to some extent.

\subsection{Gait Behavior}

In normal aging, there is a decline in executive attention. AD patients, who have neurodegenerative disorders, exhibit more pronounced deficits in the nervous system, leading to impairments in gait speed and stride length, resulting in gait abnormalities such as slower walking speed and increased gait variability (i.e., fluctuations in stride length or timing) that occur in the pre-dementia stage. Gait has been shown to be strongly associated with cognition\cite{Rosie2016Gait}. Verghese et al. studied three gait parameters - pace, rhythm, and variability - and proposed that quantitative gait impairments serve as markers of preclinical dementia\cite{2007Quantitative}. Ríona analyzed gait features such as step, swing time, step velocity, step length variability, and stance time asymmetry, suggesting that the association between gait variability and cognitive impairment reflects neurophysiological changes\cite{2019Do}. Therefore, gait detection can be used to predict dementia diagnosis\cite{2016Poor}. As gait is a complex cognitive task that requires coordination between widespread brain regions to select and organize motor programs suitable for the desired action\cite{2013Higher}, even in the early stages of the disease, gait abnormalities may reflect neurodegeneration due to dementia. Therefore, gait analysis may serve as a useful screening tool to differentiate mild dementia from normal aging.

\subsection{Assessment Methods}

\subsubsection{Facial Emotion Understanding}
Gressie et al.\cite{gressie_kumfor_teng_foxe_devenney_ahmed_piguet_2023} used the Facial Affect Selection Task (FAST)\cite{Fiona2014Degradation} to assess facial emotion recognition in AD patients and compared their performance with a healthy control group. Participants were shown arrays of seven faces from the same person expressing six basic emotions (happiness, anger, sadness, surprise, fear, and disgust) and a neutral expression, and were asked to indicate the target indicated by verbal prompts.

Similar to the FAST task, Jiskoot et al.\cite{2021Emotion} developed the Emotion Recognition Task (ERT) to investigate emotion recognition deficits in AD patients. The ERT is a computerized neuropsychological test that can be obtained through the DiagnoseIS neuropsychological assessment system (www.diagnoseis.com). Participants are required to select one of six options (i.e., anger, disgust, fear, happiness, sadness, and surprise) to label facial emotional expressions.

Shirley et al.\cite{Shirley2000Emotion} combined facial emotion recognition with a rhythm task through seven subtasks to study the abilities of AD patients in emotional prosody and emotion recognition. In their study, individuals with AD showed comparable emotional rhythm processing abilities to the healthy control group, but their facial emotion processing abilities were impaired, especially in recognizing negative emotions, with AD patients demonstrating poorer results.

\subsubsection{Facial Emotion Production}
Facial emotion expression can be elicited voluntarily, and many researchers induce spontaneous facial expressions by having participants view movies, video clips, or read books. Keith et al.\cite{keith} constructed two sets of stimuli using the International Affective Picture System (IAPS) and recorded participants' facial expression data using facial electromyography (EMG) before and during the presentation of the images. Mograbi et al.\cite{2012Emotional} developed two novel experimental success-failure manipulation (SFM) paradigms, in which participants were instructed to perform tasks arranged in the experiment, and researchers observed the responses of the normal and AD groups during successful and failed tasks to analyze differences in facial emotion production between AD patients and healthy individuals. The results revealed that while AD patients did not show significant differences in emotion compared to normal individuals when the tasks were successfully completed, they were more likely to exhibit negative emotions when the tasks failed.

\subsubsection{Gait Data}
Different types of sensors are used to collect real-time statistical data on human gait, which can be broadly categorized into two main classes: wearable sensors and environmental sensors. Wearable sensors are typically placed on various parts of the patient's body, and the captured data is either transmitted via wireless connection or collected on onboard storage devices \cite{2015Accuracy}\cite{2016The2}. Environmental sensors, on the other hand, are installed in the environment and do not require older adults to wear them. For example, Magnus et al. \cite{2022Components} measured objective spatiotemporal gait parameters using an electronic walkway, where participants were instructed to walk on the walkway at a comfortable, self-selected pace. Rosaria et al. \cite{2017Spatio} collected data using a stereo-photogrammetric system with eight infrared cameras. If the combination of the first two categories is considered, a third category can also be obtained. In this case, wearable and environmental sensors are used together to form a hybrid system.

\subsubsection{Facial Feature Extraction}

Static facial feature extraction methods can be categorized into geometric-based or appearance-based approaches. In geometric-based methods, the shape and location of facial organs are extracted to form feature vectors representing the facial geometry. Appearance-based methods, on the other hand, utilize image filters (such as Gabor wavelets) applied to the entire face or parts of the face to detect variations in facial appearance. Keith et al. \cite{keith} used EMG electrodes connected to the face to extract features for facial emotion recognition. By analyzing the electromyographic activity of the zygomaticus major and corrugator muscles, the facial expressions of the participants were analyzed. In addition to using facial electrode sensors, many studies analyze facial expressions by directly capturing images of the person's face. Ekman and Friesen proposed the Facial Action Coding System (FACS) \cite{facer}, in which facial expressions are composed of 46 action units (AUs) that correspond to specific sets of facial muscles, and the combination of AUs can form different emotions. Mograbi et al. encoded 29 facial behaviors involved in emotional expression \cite{2012Emotional}, and facial feature extraction can be used to classify captured expressions.

In addition to capturing static photos, some researchers diagnose AD by analyzing facial feature videos. Static methods typically extract facial features from a single image or frame within an image sequence, while dynamic methods extract facial features from a series of images with time information, utilizing multiple image sequences as input to classifiers that leverage temporal information \cite{2017Facial1}. Dynamic methods are sensitive to subtle changes in the face and can effectively utilize temporal information.

\subsubsection{Gait Feature Extraction}
Gait detection features can be categorized into three main types: spatiotemporal features, kinematic features, and kinetic features.

Spatiotemporal features are the most commonly used features in gait analysis and have been widely researched and tested over the years \cite{2022Components}. Examples of spatiotemporal features include stride length, step width, stance time, swing time, single support time, double support time, step count, step duration, heel strike time, toe strike time, heel-off time, and toe-off time. For instance, Bg et al. \cite{ghoraani2021detection} used the ProtoKinetics Movement Analysis Software (PKMAS) to extract gait features from the Zenomat and GAITRite systems. They extracted features such as stride time and single support time. For each trial, the mean, standard deviation (SD), and asymmetry of eight gait features, including stride time, step time, single support time, swing time, double support time, stance time, stride length, and step length, were extracted.

Kinematic analysis of gait involves studying joint angle deviations, typically calculated using wearable IMU sensors placed on the upper and lower body or motion capture systems (MCS) that allow for three-dimensional motion analysis. From a gait analysis perspective, the most significant kinematic features are the angle values at the ankle, knee, hip, and trunk, which represent the actual changes in joint function when they undergo maximum flexion/extension. For example, Silvia et al. \cite{2015Validation} utilized accelerometer-based monitors to measure kinematic gait features. Another study \cite{2017Spatio} calculated lower limb joint deviations kinematically by analyzing joint deviations in the sagittal plane at the thigh, knee, and ankle joints to obtain the range of motion.

Kinetic features involve the extraction of joint moments and powers, and data are often evaluated using force and pressure sensors embedded in platforms, instrumented walkways, shoes, or insoles. One study \cite{0A} extracted features from the vertical ground reaction force (VGRF) signal to describe the amplitude distribution of foot forces throughout the complete gait cycle, characterizing abnormal tremor movements in neurodegenerative diseases.

\section{Sound}
In addition to traditional modalities such as neuroimaging, biomarkers, and posture, sound has emerged as a novel non-invasive biological signal that is widely studied for early diagnosis and monitoring of Alzheimer's disease (AD). Sound, as a medium for information transmission, offers unique advantages and relevance in AD research. It is easy to collect, cost-effective, and can be captured using common audio devices such as smartphones and microphones, without the need for specialized equipment or expensive imaging instruments. This makes sound a feasible tool for large-scale screening, with potential applications in various settings including communities, clinical settings, and homes.

\subsection{Characteristics of Language Impairment in AD}

In Alzheimer's disease (AD), the language networks and their underlying connectivity fibers in the cerebral cortex and subcortical regions are extensively damaged, resulting in language impairments. The severity of language impairments is positively correlated with the severity of dementia \cite{2017Voice}. In the early stages of AD, individuals often experience noticeable difficulties in word-finding and a reduction in vocabulary. Their speech becomes meaningless due to a lack of substantive words, and they may compensate by using excessive explanations to compensate for their inability to express certain words. As the disease progresses, the difficulties in word-finding become more pronounced \cite{Mirheidari2017Toward}. In the middle to late stages, individuals with AD exhibit reduced spontaneous speech, a decrease in vocabulary, and an inability to name common objects or recall the names of family members. In the advanced stages of Alzheimer's disease, language impairments manifest as output disorders, with individuals often giving unrelated responses, ultimately leading to an inability to communicate effectively.

\subsection{Criteria for Assessing AD through Speech}

Various language tests have been utilized to identify and quantify cognitive impairments in individuals with Alzheimer's disease (AD) \cite{2015Speaking}. Samrah et al. \cite{2013Connected} analyzed the speech of individuals with AD across four dimensions: speech production, syntactic complexity, lexical content, and fluency errors. M.F. et al. \cite{2003Speech} developed an assessment tool consisting of 37 subscales to measure spontaneous speech, auditory comprehension, repetition, naming, reading comprehension, and other abilities in individuals with AD. Each subscale is assigned a score ranging from 0 (normal) to 6 (most abnormal). Ulla et al. \cite{2020AB} categorized tasks into three main types: spontaneous speech (SS), verbal fluency tasks (VF), and other tasks (OT).

The purpose of the SS task is to elicit participants' spontaneous speech. This is typically achieved by asking participants to describe pictures or engage in conversations. The spontaneous speech task can be used to analyze various linguistic attributes such as word retrieval processes, syntax, semantic and acoustic impairments, and communication errors.

Verbal fluency tasks (VF) consist of two types: phonemic verbal fluency (PVF) and semantic verbal fluency (SVF). In PVF tasks, participants are asked to generate as many words as possible within two minutes that begin with a specific letter, such as words starting with the letter "F." In SVF tasks, participants are instructed to produce as many words as possible within one minute belonging to a specific semantic category, such as animal names. Recently, natural language processing (NLP) techniques have been employed for automated analysis of semantic clusters, while speech processing techniques are utilized to analyze timing and acoustic measures.

Other tasks (OT) include tests unrelated to spontaneous speech and semantic clusters, such as sentence repetition, paragraph reading, story writing, counting backward, pronunciation, or naming tests. These tasks allow examination of different aspects of memory, semantic processing, as well as acoustic and speech measurements.

\subsection{Extraction of Speaker-specific Data and Dataset Creation}

VBSD Dataset \cite{mh_1}: This dataset collects voice data from 36 participants, including 254 voice samples from individuals with AD and 250 voice samples from healthy controls (HC). A total of 504 spectrogram features were extracted from these voice data. The purpose of this dataset is to identify common features among multiple AD patients (https://github.com/LinLLiu/AD).

Dem@Care Dataset \cite{2016The}: This dataset consists of voice data from 24 AD patients and 8 healthy controls. The dataset covers four tasks, including picture description, description of pictures from memory, sentence repetition, and rapid pronunciation repetition. The aim of this dataset is to study the performance of AD patients' speech across different tasks.

Pitt Dataset \cite{1994The}: The Pitt Corpus collects voice data from participants in Alzheimer's and related dementia research at the University of Pittsburgh School of Medicine. The dataset includes description tasks, word fluency tasks, story recall tasks, and sentence construction tasks. The dataset aims to explore the characteristics of voice data across different tasks.

ADReSS Dataset \cite{luz2020alzheimer}: This dataset was created for the ADReSS challenge, where participants were matched for age and gender to minimize bias risk in the prediction task. The dataset includes speech recordings and textual oral picture descriptions from participants based on the Cookie Theft picture from the Boston Diagnostic Aphasia Examination.

\begin{table*}[ht]
\centering
\caption{Voice Datasets }
  \begin{tabular}{ll}
\toprule
\textbf{Dataset Name} & \textbf{Description} \\
\midrule
VBSD Dataset\cite{mh_1} & Contains voice data from AD patients and healthy controls, extracting spectral features. \\

Dem@Care Dataset\cite{2016The} & Includes voice data from AD patients and healthy controls across multiple tasks. \\

Pitt Dataset\cite{1994The} & Collects voice data from participants in Alzheimer's and related dementia studies. \\

ADReSS Dataset\cite{luz2020alzheimer} & Includes voice recordings and descriptions based on pictures. \\
\bottomrule
\end{tabular}
\end{table*}

\subsection{Extraction and Introduction of Voice Features}

As the disease progresses, AD patients exhibit semantic, syntactic, and phonetic abnormalities compared to normal individuals. Therefore, traditional machine learning methods often utilize lexical and syntactic structures as classification features. For instance, individuals with AD may experience word-finding difficulties, leading to temporal variations in their speech, such as noticeable pauses and slow tempo. During oral reading, speakers with AD dementia demonstrate decreased speech rate, reduced articulation rate, decreased phonation time, as well as increased number and proportion of pauses. In addition to simple syntactic analysis, researchers have incorporated other analyses, such as additional prosodic analysis on the Pitt dataset, extracting 42 Mel Frequency Cepstral Coefficients (MFCC) features \cite{Fraser2016Linguistic}.

Chen et al. introduced a novel feature called Multi-Resolution Cochleagram (MRCG), which combines cochleagrams of four different temporal resolutions to capture local and contextual information. It is capable of extracting acoustic features from mixtures even under very low signal-to-noise ratio (SNR) conditions \cite{2014A}. Furthermore, researchers have diagnosed Alzheimer's disease by extracting spectrogram-based voice features, including speech rate, number of pauses, and response time. Objective acoustic measurements reveal a significant reduction in pitch modulation in the Alzheimer's group, with individuals with AD often exhibiting a higher proportion of speech interruptions, a decreased ratio of noise to harmonic content, and a general decrease in speech intensity and energy. Their speech sounds highly dysarthric.

\begin{figure}[!h]
    \centering
    \includegraphics[width=0.4\textwidth]{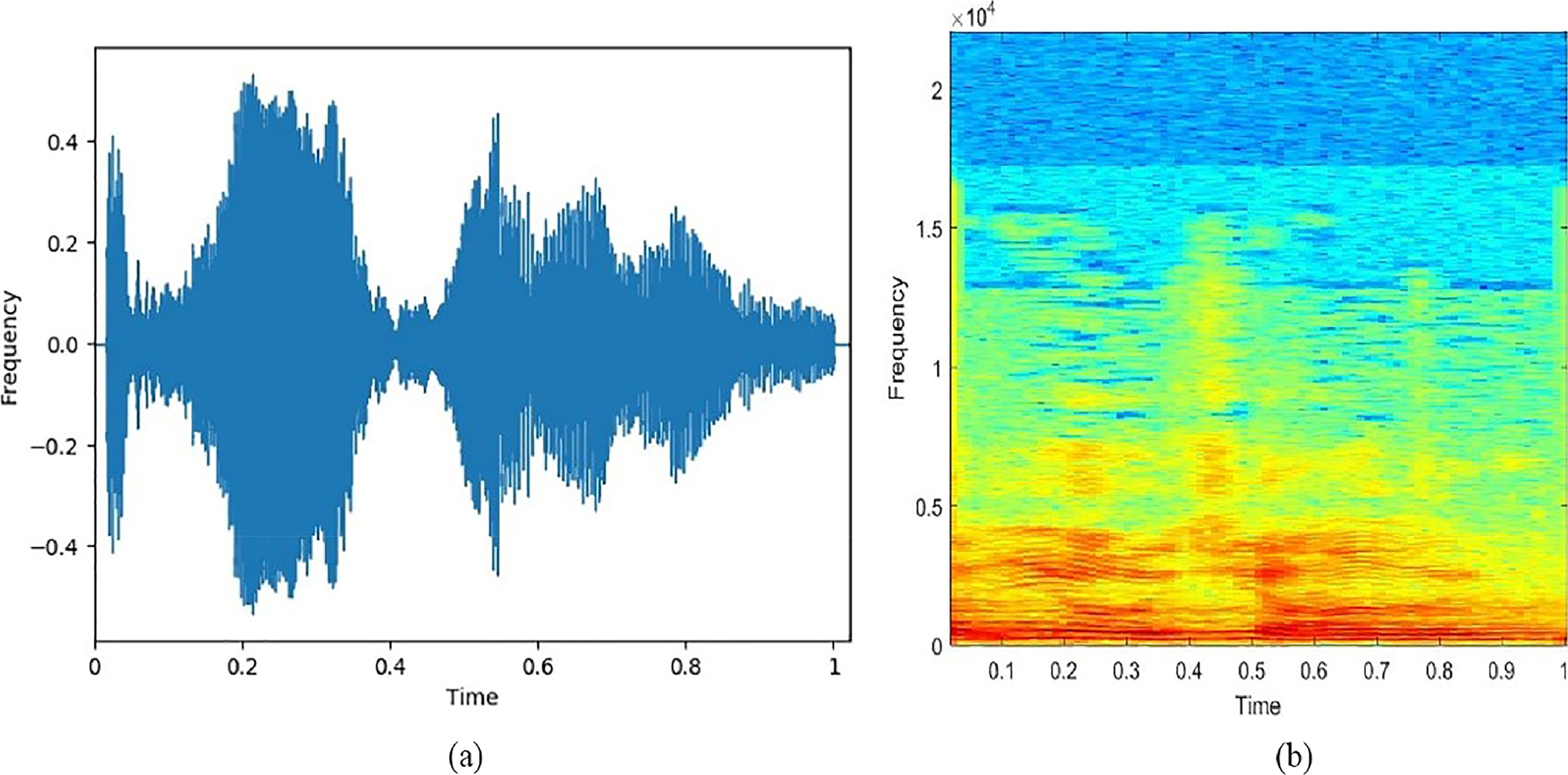}
    \caption{The spectrum of AD patients' speech, the darker the color, the stronger the speech energy}
    \label{sound}
\end{figure}

For example, Liu et al. \cite{mh_1} recorded participants' voices using wearable Internet devices and obtained a series of speech waveforms. After collecting the data, spectrogram features were extracted from the audio segments (Figure \ref{sound}), and the resulting spectrogram data were used to train a model. Finally, this trained model was used to identify whether a new set of data belonged to AD. Florian et al. proposed the openSMILE feature extraction toolkit \cite{2010Opensmile}, which can extract acoustic features from speech segments. This toolkit combines feature extraction algorithms from speech processing and can extract various features from speech. In addition to mainstream rhythmic features, Jesús B et al. \cite{2015New} extracted paralinguistic features, which are related to pitch and spectral energy balance. These features place appropriate linguistic information within the context.

\section{Research methods}
The research methods for early diagnosis of AD using computer-assisted techniques mainly include traditional machine learning methods and deep learning methods. Traditional machine learning methods usually use a series of manually engineered features to classify patient data through classification algorithms. Deep learning methods construct deep neural network models to automatically learn features from raw data and classify them.

Although traditional machine learning methods have achieved some success in the early diagnosis of AD, deep learning methods perform better in processing high-dimensional data and have achieved many important results in other medical fields. Therefore, more and more researchers are applying deep learning methods to the early diagnosis of AD and have achieved some very promising results in this field.

This article will describe research on early classification of AD based on computer-assisted diagnosis from both traditional machine learning and deep learning perspectives.
\subsection{Traditional machine learning methods}
\subsubsection{Feature Extraction}
Feature extraction is a common application in pattern recognition and image processing in machine learning. Its goal is to remove the redundant parts caused by too many features in the original data, in order to save memory space and improve computational power, improve classification performance, and belongs to the dimensionality reduction process.

Feature extraction focuses on the size, shape, and volume of the extracted features for classification research. Early AD diagnosis research mainly involves the hippocampus and the structure of the entorhinal cortex. The common feature extraction objects mainly have the following forms:
\begin{itemize}
    \item \textbf{Volume-based features}: Previous studies have shown that the hippocampus has atrophied in the early stages of AD, and the research shows that about 50\% of AD patients have hippocampal atrophy. Therefore, the hippocampal volume can be used as a biomarker feature for AD diagnosis.
    \item \textbf{Thickness-based features}: Lerch et al. found that AD patients have significant differences in the thickness of the entire brain cortex compared to the normal control group NC, reflecting a decrease in the thickness of the cortex in multiple parts of the brain. Clinical studies have found that changes in brain volume caused by brain atrophy are an important biomarker in the early stages of AD.
    \item \textbf{Morphological features}: Changes in brain volume caused by hippocampal and medial temporal lobe atrophy have been used as biomarker features for early detection of AD. Gerardin et al. found in their research that morphological changes were already present in the brains of early-stage AD patients before hippocampal atrophy occurred. Using morphological features as biomarker features for early diagnosis of AD has more sensitive characteristics.
    \item \textbf{Texture feature analysis}: Texture is a visual feature that reflects the homogeneous phenomenon of an image or pattern, reflecting the characterizing attributes of the tissue structure that presents slow or periodic changes on the surface of an object. Freeborough et al. found significant differences between AD patients and normal control group (NC) through texture analysis.
    \item \textbf{Voxel Morphology Analysis}: Voxel (Volume Element) is the minimum unit of digital data in three-dimensional space segmentation, relative to the minimum unit of two-dimensional space, pixel. Voxel-Based Morphometry (VBM) is a comprehensive, objective and automatic analysis method based on voxel units for MRI medical images, which can be used for morphological research on living human brain tissue. By quantitatively calculating the relative changes in gray and white matter density and volume corresponding to different voxels in MRI images through VBM, it can be used for early diagnosis research of AD. Karas et al. found that there are morphological differences in GM (Gray Matter) voxels between AD patients and normal control group through VBM technology.
\end{itemize}
\subsubsection{Classification Algorithm Based}
The traditional machine learning methods used for early AD diagnosis and classification research have become increasingly systematic and mature. The algorithms for AD early classification using machine learning assistance mainly include the following categories:
\begin{itemize}
    \item \textbf{Support Vector Machine (SVM)}: In the AD early diagnosis and classification research, SVM is the most frequently used algorithm for AD classification. Kim et al. \cite{2020Slice} used Boundary Equilibrium Generative Adversarial Networks (BEGAN) to extract features of Alzheimer's disease (AD) and normal cognition (NC) states for stable convergence. Then, the authors trained a support vector machine (SVM) classifier on these features to distinguish AD from NC.
    \item \textbf{Logistic Regression}: Logistic regression can achieve good classification results when learning a large number of training samples. Ieracitano et al. \cite{2020A} used the logistic regression algorithm based on the brainwave (EEG) CWT and BiS features to classify patients with mild cognitive impairment (MCI) or Alzheimer's disease (AD) in the neurological system.
    \item \textbf{Linear Discriminant Analysis (LDA)}: The LDA algorithm is commonly used to evaluate the severity of patients' condition and predict the disease, and can achieve better classification results than logistic regression when the sample data is small. Ahmad et al. \cite{2019Classification} used biomarker patterns and resting-state functional magnetic resonance imaging (fMRI) to label the hippocampus (HP), middle temporal gyrus (MTG), olfactory cortex, and posterior cingulate cortex (PCC), and then used LDA for AD classification research.
    \item \textbf{Bayesian Classifier}: The Bayesian classifier has a smaller classification error rate and is commonly used in pattern recognition and other fields. Yubraj et al. \cite{2019Alzheimer} proposed using cortical thickness and subcortical volume as biological markers for AD classification, using principal component analysis (PCA) for dimensionality reduction, and using Bayesian classifier for classification research.
    \item \textbf{Random Forest}: Random Forest has higher accuracy than decision tree algorithms and can handle more input data. Dimitriadis et al. \cite{2017Random} proposed using the random forest algorithm to select multiple biomarkers from a subset of feature sets based on preprocessed MRI images, and fusing these biological markers to classify using the random forest algorithm.
    \item \textbf{K-Nearest Neighbor (KNN)}: The classification of the KNN algorithm is based on whether the K nearest samples in the feature space of the sample data point belong to the same category. Farouk et al. \cite{2019Supervised} extracted biological marker information from structural magnetic resonance images (SMRI), fused texture features based on gray-level co-occurrence matrix and voxel-based morphological features, and used the KNN algorithm for early AD diagnosis and classification.
\end{itemize}
\subsection{Deep Learning Methods}
Deep learning is a machine learning technique based on neural networks, which has achieved remarkable results in many fields. In the diagnosis of Alzheimer's disease, deep learning technology has begun to play an increasingly important role. By analyzing a large amount of medical data of patients, such as neuroimaging, voiceprints, posture, and biomarkers, deep learning technology can improve the accuracy and precision of Alzheimer's disease diagnosis.

\subsubsection{Neural Network Classification Models}
\begin{itemize}
    \item \textbf{LetNet-5}: Sarraf et al. \cite{2016DeepAD} conducted classification studies of AD and NC based on MRI images using the LeNet model. LetNet-5 can obtain effective representations of original images, and can identify the rules of original images with very few preprocessing, but due to the characteristics of its model, the results are not ideal for processing large-scale data and complex problems.
    \item \textbf{AlexNet}: Afzal et al. \cite{afzal2019data} used the pre-trained AlexNet network model to select 199 subjects from the Open Access Series of Imaging Studies (OASIS) dataset for classification studies of AD and NC. The classification accuracy reached 98.41\%.
    \item \textbf{VGG}: Jain et al. \cite{2019Convolutional} used the VGG-16 network model and MRI single-mode imaging to conduct a three-classification study of NC, MCI, and AD. Experimental data came from the Alzheimer's Disease Neuroimaging Initiative (ADNI) dataset, and 50 samples were selected for the experiment. The classification accuracy can reach 99.14\%.
    \item \textbf{GoogLeNet}: Sarraf et al. \cite{2016DeepAD} respectively obtained 144 fMRI and 302 MRI sample data from the ADNI dataset based on MRI and fMRI modal imaging data, and used the GoogLeNet network model to extract low to high features from a large number of training images. The experimental training set and test set ratio was 3:1. The average accuracy of GoogLeNet model in AD and NC classification under fMRI images was 94.24\%. The average accuracy of GoogLeNet and LeNet models based on MRI images were 98.74\% and 97.88\%, respectively. GoogLeNet model performed better.
    \item \textbf{ResNet}: Ji et al. \cite{2019Early} conducted AD classification studies based on ResNet network model in single-mode MRI, multi-mode MRI, and FDG-PET images. The classification accuracy reached 97.65\% in AD/mild cognitive impairment and 88.37\% in mild cognitive impairment/normal control.
    \item \textbf{3D-CNN}: Kompanek et al. \cite{2019Volumetrie} used 3D-CNN based on MRI images to classify NC and AD, and used Volumetric Data Augmentation (VDA) to enhance the classification accuracy and generalization ability of the network model.
\end{itemize}
\subsubsection{Based on unimodal deep learning methods}
Modality refers to the form in which things happen or exist, which can be information such as sound, images, and text. For the early diagnosis of AD research, it can be data such as brain imaging, voiceprints, and postures. 

The application of deep learning methods to biomedical research is one of the main tasks to use different modal biomarkers for early AD classification research. Barbaroux et al. \cite{2020Encoding} explored AD classification tasks by analyzing the human brain cortex using the SCNN (Spherical CNN) model. SCNN extends the regular CNN from planar data to spherical data by defining convolutional operators that are equidistant to the same 3D rotation on spherical input. The experiment is based on the morphological measurement method of T1-weighted MRI images. The classification accuracy between AD and NC can reach 92.16\%, indicating the feasibility and superiority of using the SCNN model directly for discriminating analysis of human brain cortex.

Kim et al. \cite{2020Slice}used an improved deep learning method with a GAP layer based on FDG-PET imaging data, and conducted experiments based on the ADNI and Severance dataset imaging data from the nuclear medicine department of the Westfalenus Hospital. They achieved accuracy rates of 91.02\% and 86.09\% in the classification results of AD and NC, respectively. The experiment used slice-selective learning to reduce computation, and at the same time, used transfer learning and a GAP layer instead of a fully connected layer to avoid overfitting and improve the model's generalization ability. The experimental results on different datasets showed that the performance of the GAP layer was statistically significant compared with the fully connected layer in terms of model accuracy, sensitivity, and specificity (p$<$0.01), and the model's performance on accuracy, sensitivity, and specificity did not show statistically significant differences on different datasets (p$>$0.05).

Bin et al. \cite{2018Binary} proposed using multiple deep 2D-CNNs constructed by separable convolution layers to learn various features of brain regions. Based on the SMRI imaging in the OASIS cross-sectional image dataset for AD classification research, the Xception and Inception version 3 structures were used as transfer learning models. The experimental results showed that the model had superior performance in AD classification.

Suganthe et al. \cite{2020Diagnosis} proposed to establish a deep convolutional neural network (DCNN) and a VGG-16-based convolutional neural network (VCNN) model, using MRI images from the ADNI database for classification research on 4 stages: AD, NC, EMCI, and LMCI. By changing the number of convolutional layers, convolutional kernel sizes and numbers, dropout layers, pooling layers, and other parameters to compare the model classification results, the results showed that the DCNN model with 7 convolutional layers had much better classification performance than the DCNN model with 4 convolutional layers. The classification accuracy of AD and LMCI reached 93.76\%. The comparative experiments of the 2 models showed that the accuracy of the VCNN model was higher than that of the DCNN model, but its number of trainable parameters was higher than that of the DCNN model.

Böhle et al. \cite{Moritz2019Layer} used layer-wise relevance propagation (LRP) to visualize the decisions of an MRI-based Alzheimer's disease (AD) convolutional neural network. LRP produces a heatmap in the input space indicating the relevance/importance of each voxel to the final classification result, which can help clinicians interpret neural network decisions for diagnosis of AD (and other diseases) based on structural MRI data.

Zhu et al. \cite{2021Dual} proposed a dual-attention multiple instance deep learning network (DA-MIDL) for early diagnosis of AD and mild cognitive impairment (MCI), its prodromal stage. DA-MIDL consists of three main components: 1) Patch-Nets based on spatial attention blocks to extract discriminative features within each sMRI patch, while enhancing abnormal microstructural features of the brain, 2) attention-based multiple instance (MIL) pooling to balance relative contributions of each patch and generate global differentially-weighted representations for the entire brain structure, and 3) attention-based global classifier to further learn holistic features and make AD-related classification decisions. The model was evaluated on 1689 subjects from the ADNI and AIBL datasets, achieving state-of-the-art results.

Liu et al. \cite{muti_imge_3} proposed a deep learning embedded AD diagnosis framework with less prior clinical knowledge to differentiate four stages of AD progression. Due to unsupervised feature representation embedded in the workflow, it has the potential to be extended to more unlabeled data for feature engineering in practice. In the unsupervised pretraining stage, they used SAEs to obtain high-level features. When using multiple neural imaging modalities, they employed a zero-mask strategy to extract synergistic effects between different modalities, processed in a denoising manner. After unsupervised feature engineering, the authors used a softmax regression model. They also adopted a novel method to visualize high-level brain biomarkers for analyzing extracted high-level features. The proposed framework was evaluated in AD stages 2-4 classification. Based on MR and PET ADNI datasets, the authors' framework outperformed state-of-the-art SVM-based methods and other deep learning frameworks.

Lian et al. \cite{2018Hierarchical} proposed a hierarchical fully convolutional network (H-FCN) for automated identification of discriminative local regions in the entire brain sMRI. Then, joint learning and fusion of multiscale feature representations were conducted to construct a hierarchical classification model for AD diagnosis. Their proposed H-FCN method was evaluated in a large number of subjects from two independent datasets, i.e., ADNI-1 and ADNI-2, demonstrating good performance in joint discriminative atrophy localization and brain disease diagnosis.

Ebrahimi \cite{WOS:000674484500001} used a ResNet-18 pretrained on the ImageNet dataset and employed serialized models, time convolutional networks (TCN), and various types of recurrent neural networks. Their proposed TCN model obtained the best classification performance using MRI, with an accuracy of 91.78\%, sensitivity of 91.56\%, and specificity of 92\%. Their results indicate that adopting sequence-based models can improve classification accuracy by 10\% compared to 2D and 3D CNNs.

Westman et al. \cite{WOS:000285486000043} used orthogonal partial least squares (OPLS) to differentiate Alzheimer's disease (AD), mild cognitive impairment (MCI), and elderly control groups, using local and global magnetic resonance imaging (MRI) volume measurements. The study included 117 AD patients, 122 MCI patients and 112 control subjects (from the AddNeuroMed study). High-resolution sagittal 3D MP-RAGE data sets were obtained from each subject, with automatic region segmentation and manual contouring of the hippocampus for each image. A total of 24 different anatomical structures were used for sample MVP analysis. Seventeen randomly selected AD patients, 12 randomly selected control subjects and 22 MCI patients who converted to AD in the 1-year follow-up were excluded from the initial OPLS analysis to provide a small external test set for model validation. Using only hippocampal measurements compared to controls, the authors found a sensitivity of 87\% and a specificity of 90\% for AD. Combining global and regional measurements resulted in a sensitivity of 90\% and a specificity of 94\%. This increase in sensitivity and specificity led to a positive likelihood ratio from 9 to 15. From the external test set, the model correctly predicted 82\% of AD patients and 83\% of control subjects. Finally, 73\% of MCI subjects who converted to AD in 1 year of follow-up were found to be closer to AD patients than control subjects.

Morra et al. \cite{WOS:000259927400007} introduced a new brain MRI segmentation method called automatic context model (ACM) for automatic segmentation of the hippocampus in images of elderly dementia tissues. During the training stage, the algorithm used 21 manually labeled segmentations to learn the classification of hippocampal and non-hippocampal regions using an improved AdaBoost method based on approximately 18,000 features, including image intensity, location, image curvature, image gradient, tissue classification map of gray/white matter and CSF, mean, standard deviation, and Haar filters of sizes ranging from 1x1x1 to 7x7x7. All brain images were linearly registered to a standard template, and a basic shape prior was designed to capture the global shape of the hippocampus, defined as the point-wise sum of all training masks. They also incorporated shape priors and features of curvature, gradient, mean, standard deviation, and Haar filters of the tissue classification image as features. In each iteration of the ACM, the Bayesian posterior distribution and the surrounding features were fed to AdaBoost as new features. In the validation study, the results were compared with the manual annotations of two experts. Using the leave-one-out method and standard overlap and distance error metrics, the automated segmentation results were highly consistent with those of human evaluators, with any differences comparable to those between trained human evaluators.

In another study\cite{2021An}, a groundbreaking deep belief network (DBN)-based multi-task learning algorithm is proposed for solving classification issues. Specifically, the overfitting problem is addressed by utilizing the dropout technology and zero-masking strategy. These techniques also improve the model's generalization ability and robustness. Subsequently, a novel framework based on DBN-based multi-task learning is constructed for precise diagnosis of AD. After MRI preprocessing, principal component analysis reduces the feature dimension, and multi-task feature selection approach selects the feature set related to all tasks based on the internal relevancy among multiple related tasks. Using data from the ADNI dataset, their method achieved satisfactory results in six tasks: health control (HC) vs. AD, HC vs. pMCI, HC vs. sMCI, pMCI vs. AD, sMCI vs. AD, and sMCI vs. pMCI, with accuracies of 98.62\%, 96.67\%, 92.31\%, 91.89\%, 99.62\%, and 87.78\%, respectively. Experimental findings demonstrate that the DBN-based MTL algorithm developed in this study is a remarkably efficient, superior, and practical method for AD diagnosis.

Wang et al.\cite{2021ADVIAN} firstly propose a VIN, a VGG-inspired network as the backbone, and investigate the utilization of attention mechanisms. They introduce the Alzheimer's Disease VGG-Inspired Attention Network (ADVIAN), where convolutional block attention modules are integrated into the VIN backbone. Additionally, they propose 18-way data augmentation to avoid overfitting. They perform ten runs of 10-fold cross-validation to report unbiased performance. Their results show that sensitivity and specificity reach 97.65 +/- 1.36 and 97.86 +/- 1.55, respectively. ADVIAN achieves a precision and accuracy of 97.87 +/- 1.53 and 97.76 +/- 1.13, respectively. Furthermore, the F1 score, MCC, and FMI obtain 97.75 +/- 1.13, 95.53 +/- 2.27, and 97.76 +/- 1.13, respectively. The AUC is 0.9852. In conclusion, they proposed ADVIAN outperforms 11 state-of-the-art methods. Additionally, their experimental results demonstrate the effectiveness of 18-way data augmentation.

In another study, Liu et al.\cite{liu2019using}  utilize a deep-learning framework that employs Siamese neural networks, which are trained on paired lateral inter-hemispheric regions. The approach aims to harness the discriminative power of whole-brain volumetric asymmetry. To achieve this objective, they use the MRICloud pipeline to obtain volumetric features of pre-defined atlas brain structures, with a novel non-linear kernel trick applied to normalize these features. This normalization process works to reduce batch effects across various datasets and populations. They work with the low-dimensional features of the brain and show that Siamese networks achieve comparable performance to studies that use whole-brain MR images. This is an advantage, as they reduce complexity and computational time while still preserving the biological information density. Their experimental results demonstrate that Siamese networks outperform traditional prediction methods that do not incorporate the asymmetry in brain volumes on the ADNI and BIOCARD datasets in certain metrics.

In another paper, Lee et al.\cite{WOS:000491861000105} propose a new method for diagnosing Alzheimer's disease (AD) or mild cognitive impairment (MCI) using magnetic resonance imaging (MRI). Their approach integrates voxel-based, region-based, and patch-based methods into a single framework. They parcellate the brain into predefined anatomical regions using templates and derive complex nonlinear relationships among the voxel intensities, which provide volumetric measurements, within each region. Unlike other approaches that use cubical or rectangular shapes, they focus on the unique, atypical shapes of the regions. Using deep neural networks, they learn the complex nonlinear relationships among the voxels in each region and extract a "regional abnormality representation." Finally, they make a clinical decision by integrating the regional abnormality representations across the whole brain. Their method has an advantage in allowing us to interpret and visualize symptomatic observations of subjects with AD or MCI in the brain space. On the baseline MRI dataset from the Alzheimer's Disease Neuroimaging Initiative (ADNI) cohort, their method achieves state-of-the-art performance for four binary classification tasks and one three-class classification task.

Cui et al.\cite{WOS:000467664000001} present a classification framework that combines convolutional and recurrent neural networks for longitudinal analysis of structural MR images in AD diagnosis. First, they construct a Convolutional Neural Network (CNN) to learn the spatial features of MR images for the classification task. Then, they build a recurrent Neural Network (RNN) with three cascaded bidirectional gated recurrent unit (BGRU) layers on the outputs of CNN at multiple time points to extract longitudinal features for AD classification. Their proposed method jointly learns the spatial and longitudinal features along with the disease classifier, achieving optimal performance. Furthermore, their approach models longitudinal analysis using RNN from the imaging data at various time points. They evaluate the proposed method with longitudinal TI-weighted MR images of 830 participants, including 198 AD, 403 mild cognitive impairment (MCI), and 229 normal control (NC) subjects from the Alzheimer's Disease Neuroimaging Initiative (ADNI) database. Experimental results show the proposed method achieves a classification accuracy of 91.33\% for AD vs. NC and 71.71\% for pMCI vs. sMCI. These results demonstrate the promising performance of their method for longitudinal MR image analysis.

In another paper, Ge et al.\cite{ge2019multi} propose a deep learning method for multi-scale feature learning that is based on segmented tissue areas. They propose a novel deep 3D multi-scale convolutional network scheme that generates multi-resolution features for AD detection. The proposed scheme employs several parallel 3D multi-scale convolutional networks, each of which applies to individual tissue regions (GM, WM and CSF), followed by feature fusions. The proposed fusion is applied in two separate levels: the first level fusion is applied on different scales within the same tissue region, and the second level is applied on different tissue regions. To further reduce the dimensions of features and mitigate overfitting, a feature boosting and dimension reduction method, XGBoost, is utilized before classification. The proposed deep learning scheme has been tested on a moderate open dataset of ADNI (consisting of 1198 scans from 337 subjects) and has shown excellent performance on randomly partitioned datasets (with a top score of 99.67\% and an average score of 98.29\%) and good performance on subject-separated partitioned datasets (with a top score of 94.74\% and an average score of 89.51\%).

Cui et al.\cite{cui2018hippocampus} propose a new method for analyzing the hippocampus that combines global and local features using three-dimensional densely connected convolutional networks and shape analysis for diagnosing Alzheimer's Disease. Their method effectively incorporates local visual features and global shape characteristics to improve classification without requiring tissue segmentation or nonlinear registration. They evaluated their method using T1-weighted structural MRIs from 811 subjects, which included 192 AD patients, 396 MCI patients (of which 231 were stable MCI and 165 were progressive MCI), and 223 normal control subjects from the Alzheimer's disease neuroimaging initiative database. Their experimental results demonstrate that their proposed method achieved a classification accuracy of 92.29\% and area under the ROC curve of 96.95\% for AD diagnosis.

The methodology proposed by Ieracitano et al.\cite{ieracitano2019convolutional} involves evaluating the power spectral density (PSD) of EEG traces from 19 channels and using this information to generate 2D grayscale images (PSD-images) representing the spectral profiles. A customized Convolutional Neural Network, which includes a processing module comprising of convolution, Rectified Linear Units (ReLu), and pooling layer (CNN1), is designed to extract suitable features from the PSD-images, thereby facilitating two and three-way classification tasks. Compared to more conventional learning machines, the resulting CNN provides better classification performance, achieving an average accuracy of 89.8\% for binary classification and 83.3\% for three-way classification.

In another paper, Li et al.\cite{li2020detecting} present a 4D deep learning model, called C3d-LSTM, for AD discrimination. The model is capable of utilizing spatial and time-varying information simultaneously by processing 4D fMRI data directly. The proposed C3d-LSTM combines 3D convolutional neural networks (CNNs) to extract spatial features from each volume of the 3D static image in an fMRI image sequence. Subsequently, the feature maps obtained are fed to the long short-term memory (LSTM) network to capture the time-varying information contained within the data. Because of its design, the C3d-LSTM is an end-to-end data-driven model, making it more convenient for processing 4D fMRI data. The proposed model was evaluated on the AD Neuroimaging Initiative dataset against controlled experiments. This evaluation produced better results in AD detection than the functional connectivity, 2D, or 3D fMRI data methods when utilizing 4D fMRI data directly. The results demonstrate that the proposed method using the C3d-LSTM model provided a far better result in AD detection.

In another study, Chitradevi et al.\cite{chitradevi2020analysis} considered various brain sub-regions to diagnose Alzheimer's disease using four optimization algorithms: Genetic Algorithm (GA), Particle Swarm Optimization algorithm (PSO), Grey Wolf Optimization (GWO), and Cuckoo Search (CS). Among these optimization techniques, GWO produced promising results due to its proper selection of global optimal solution. The segmented regions were classified using a deep learning classifier and validated with Ground Truth (GT) images. The results demonstrate that GWO can segment brain sub-regions with a high accuracy of 98\% similarity between the ground truth and segmented regions. The segmented regions were then classified using a deep learning classifier, providing a high accuracy of 95\%. The HC sub-region showed better classification accuracy (95\%), sensitivity (95\%), and specificity (94\%) compared to all other regions. Based on the segmentation and classification measures, it is clear that the proposed method performed better than other methods. Finally, clinical validation was performed on normal and AD subjects using mini-mental state examination (MMSE) scores. The proposed pipeline correlated strongly with MMSE scores. Therefore, the results suggest that the HC region is a major factor in diagnosing AD.

The aim of Puente et al.\cite{puente2020automatic}'s study is to develop an automated system for detecting disease presence in sagittal MRI scans, which are underutilized. They utilized sagittal MRIs from ADNI and OASIS datasets and employed Transfer Learning techniques for better accuracy. Their findings indicate that damages related to AD can be identified in sagittal MRI scans and that the results are similar to those from state-of-the-art DL models that use horizontal-plane MRI scans.

In another paper, Chen et al.\cite{chen2021iterative} propose an iterative model for sparse and deep learning (ISDL) that jointly extracts deep features and identifies critical cortical regions for diagnosing AD and MCI. Their first step is to design a Deep Feature Extraction (DFE) module that captures structural information ranging from local to global that is derived from 62 cortical regions. They then incorporate a sparse regression module to identify the critical cortical regions and integrate it into the DFE module. This step helps to exclude irrelevant cortical regions from the diagnostic process. They update the parameters of the two modules alternatively and iteratively in an end-to-end manner. Their experimental results suggest that the ISDL model provides a state-of-the-art solution for both AD-CN classification and MCI-to-AD prediction.

Hazarika et al.\cite{hazarika2022experimental} believe that, amongst all the models presented, the DenseNet-121 model achieves a convincing result with an average performance rate of 88.78\%. However, one of the downsides of the DenseNet model is that it utilizes a lot of convolutional operations, which make the model computationally slower than several other models discussed. Depth-wise convolution is a popular method to make convolutional operation faster and better. Therefore, to enhance the execution time, they propose replacing the convolution layers in the original DenseNet-121 architecture with depth-wise convolution layers. The new architecture not only improved the execution time, but it also improved the performance of the model, with an average rate of 90.22\%.

Bae et al.\cite{bae2021transfer} have developed a deep learning model for predicting conversion from MCI (Mild Cognitive Impairment) to DAT (Dementia of Alzheimer's Type) using structural magnetic resonance imaging scans as input to a 3-dimensional convolutional neural network. This neural network was trained using transfer learning, with normal control and DAT scans used as input in the source task to pretrain the model. Later, they retrained the model on a target task of classifying which MCI patients eventually progressed to DAT. Their model had an outstanding classification accuracy of 82.4\% at the target task, surpassing current models in the field. They also employed an occlusion map approach to visualize the brain regions that significantly contribute to the prediction. Surprisingly, they found that among the contributory regions were the pons, amygdala, and hippocampus.

In another study, Zhang et al.\cite{zhang2021explainable} propose a novel computer-aided approach for early diagnosis of AD. They introduce an explainable 3D Residual Attention Deep Neural Network (3D ResAttNet) for end-to-end learning from sMRI scans. Their approach is distinct from existing methods in three ways: Firstly, they propose a Residual Self-Attention Deep Neural Network to capture local, global, and spatial information of MR images and improve diagnostic performance. Secondly, they introduce an explainable method using Gradient-based Localization Class Activation mapping (Grad-CAM) to improve the interpretability of the proposed method. Thirdly, their work provides a complete end-to-end learning solution for automated disease diagnosis. They evaluate the proposed 3D ResAttNet method on a large cohort of subjects from real datasets for two challenging classification tasks, namely Alzheimer's disease (AD) vs. Normal cohort (NC) and progressive MCI (pMCI) vs. stable MCI (sMCI). The experimental results demonstrate that their approach has a competitive advantage over state-of-the-art models in terms of accuracy performance and generalizability.

In another paper, Huang et al.\cite{2019Diagnosis} propose the use of a convolutional neural network (CNN) to integrate all the multi-modality information contained in both T1-MR and FDG-PET images of the hippocampal area for the diagnosis of Alzheimer's disease (AD). Unlike traditional machine learning algorithms, their method does not require manually extracted features - instead, it utilizes 3D image-processing CNNs to learn features for the diagnosis or prognosis of AD. To evaluate the performance of the proposed network, they trained the classifier with paired T1-MR and FDG-PET images in the ADNI dataset. The dataset includes 731 cognitively unimpaired subjects labeled as CN, 647 subjects with AD, 441 subjects with stable mild cognitive impairment (sMCI), and 326 subjects with progressive mild cognitive impairment (pMCI). They obtained higher accuracies of 90.10\% for the CN vs. AD task, 87.46\% for the CN vs. pMCI task, and 76.90\% for the sMCI vs. pMCI task. The proposed framework yielded a state-of-the-art performance.

Ghoraani et al.\cite{ghoraani2021detection} collected data on both "single-tasking" gait (walking) and "dual-tasking" gait (walking with cognitive tasks) from 32 healthy participants, 26 participants with mild cognitive impairment (MCI), and 20 participants with Alzheimer's disease (AD) using a computerized walkway. Each participant was assessed with the Montreal Cognitive Assessment (MoCA), and a set of gait features (such as mean, variance, and asymmetry) was extracted. Significant gait features for three classifications (MCI/healthy, AD/healthy, and AD/MCI) were identified, and a support vector machine model was trained in a one-vs.-one manner for each classification. The majority vote of the three models was assigned as either healthy, MCI, or AD. Regarding the results, the average accuracy of the classification was 78\% (F1-score of 77\%), which was reasonable when compared to the MoCA score with an accuracy of 83\% (F1-score of 84\%). Additionally, the performance of healthy vs. MCI or AD was 86\% (F1-score of 88\%), which was also comparable to MoCA's accuracy of 88\% (F1-score of 90\%).

Rucco et al.\cite{2017Spatio}'s study explored the gait patterns in Alzheimer’s disease (AD) and behavioral variant of Frontotemporal Dementia (bvFTD) patients. The bvFTD group showed more instability and slowness in a single task, and worsened further in a dual motor task. AD patients had lower speed and stride length than healthy subjects. During a cognitive dual task, both groups showed impaired velocity and stability, with AD patients showing a significant deterioration in gait performance.

Emotion processing and social interaction difficulties are present in dementia, particularly in frontotemporal dementia. Corticobasal syndrome is related to frontotemporal dementia, with emotion processing changes found in frontoparietal cortical regions and the basal ganglia, but has not been systematically explored. A study was conducted by Kumfor et al.\cite{Fiona2014Degradation} to examine emotion processing in corticobasal syndrome and Alzheimer’s disease. Results showed that emotional processing deficits are present in patients with corticobasal syndrome and are related to changes in brain regions crucial for emotion processing.

These studies demonstrate the potential of unimodal deep learning methods in leveraging specific modalities of data, such as brain imaging, for accurate AD classification. However, these approaches are limited to a single modality, and the fusion of information from multiple modalities can provide more comprehensive insights.

\subsubsection{Based on Multimodal Deep Learning Methods}
Multimodal deep learning refers to the use of multiple single-modal information in deep learning methods to achieve the fusion of information between different modalities.

Many researchers use multimodal data for early classification research of Alzheimer's disease (AD). Huang et al.\cite{2019Diagnosis} used various multimodal feature information under the hippocampal TI-weighted magnetic resonance imaging and FDG-PET imaging, using convolutional neural networks for AD classification research. The classification accuracies obtained for NC and PMCI, SMCI and PMCI, AD and NC were 87.46\%, 76.90\%, and 90.10\% respectively. The results showed that the classification results combined with multi-modal imaging data were better than those in the single-modal form.

Forouzannzhad et al.\cite{2018A} used the multimodal imaging technology of positron emission tomography (PET) and magnetic resonance imaging (MRI) and the results of standard neuropsychological testing scores, to classify early diagnosis of AD using deep neural networks (DNN). The classification accuracies for normal control group NC and early mild cognitive impairment EMCI were as high as 84.0\%. The classification accuracies for NC and late mild cognitive impairment LMCI, CN and AD, EMCI and LMCI, EMCI and AD, LMCI and AD were 84.1\%, 96.8\%, 69.5\%, 90.3\% and 80.2\%, respectively. The classification accuracy for NC and EMCI was only 68.0\% on MRI alone. The study showed that the multimodal method was better than single-modal imaging analysis.

Kang et al.\cite{2020Identifying} constructed a transfer learning method of the VGG16 model based on SMRI and diffusion tensor imaging DTI (Diffusion Tensor Imaging) bimodal data for the classification research of EMCI and NC. The data came from the ADNI dataset. The experiment used a multimodal fusion strategy to merge slices with the same index into RGB slices, and input them into the model for training. LASSO (Least Absolute Shrinkage and Selection Operator) algorithm was used to extract the features related to EMCI disease, and the experimental obtained a classification accuracy of 94.2\%, with a high sensitivity of 97.3\%. The experimental results showed that multimodal data can provide more useful information for distinguishing EMCI and NC, and from a clinical perspective, DTI is an important biomarker for EMCI.

Khvostikov et al.\cite{20183D} fused the SMRI and DTI imaging modes in the hippocampal region of interest, and compared the use of single-modal data for the experiment with the AD classification algorithm based on the 3D-CNN model. In the AD and MCI classifications, a high accuracy of 93.3\% was obtained under the multimodal situation, which had a significant advantage over the accuracy of 65.8\% under the SMRI single-mode. In the experiment, data augmentation was used to balance the sizes of different classes, in order to eliminate the impact of different amounts of data on the network training process.

Lin et al. \cite{2021Bidirectional} tackled the problem of expensive missing PET data by first using a reversible generative adversarial network (RevGAN) model to reconstruct missing data. Then, the team proposed a 3D convolutional neural network (CNN) classification model with multimodal inputs for AD diagnosis. Finally, their method was evaluated on the Alzheimer's Disease Neuroimaging Initiative (ADNI) database. The experimental results showed that the proposed framework can effectively map the structure and functional information of brain tissue to images. The synthesized images were very close to real images, and the framework significantly improved the performance of AD diagnosis and MCI conversion prediction.

Martinez-Murcia et al. \cite{2020Studying} attempted to explore AD through deep convolutional autoencoders. By fusing the information of neuropsychological test results, diagnosis, and other clinical data with imaging features extracted only through MRI, the study aimed to find the relationship between cognitive symptoms and potential neurodegenerative processes. Then, the extracted features in different combinations were analyzed and visualized, and regression and classification analyses were used to estimate the impact of each coordinate of the autoencoder's manifold on the brain. In terms of AD diagnosis, the study achieved over 80\% classification accuracy.

In the study by Shi \cite{shi2017multimodal} et al., a multimodal stacked dense pyramid network (MM-SDPN) algorithm was proposed, which consists of two levels of SDPN for feature representation and fusion of multimodal neuroimaging data for AD diagnosis. Specifically, the high-level features of MRI and PET were learned by two SDPNs, and then the features were fed into another SDPN to fuse multimodal neuroimaging information. The proposed MM-SDPN algorithm was applied to the ADNI dataset for binary and multi-class classification tasks. The experimental results showed that MM-SDPN outperformed the state-of-the-art algorithms based on multimodal feature learning for AD diagnosis.

El-Sappagh et al. \cite{WOS:000616803100043} developed an accurate and interpretable diagnostic and progression detection model for Alzheimer's disease (AD), providing doctors with accurate decisions and a set of explanations for each decision. Specifically, the model integrates 11 modalities from 1048 subjects from the ADNI dataset: 294 cognitively normal, 254 stable mild cognitive impairment (MCI), 232 progressive MCI, and 268 AD. This is actually a two-layer model that uses random forest (RF) as the classifier algorithm. In the first layer, the model performs multi-class classification aimed at early diagnosis of AD patients. In the second layer, the model applies binary classification to detect MCI to AD progression that may occur within three years after baseline diagnosis. The performance of the model was optimized using key markers selected from a large number of biological and clinical indicators. For interpretability, the authors provide RF classifier global and instance-based explanations for each layer using the SHapley Additive exPlanations (SHAP) feature attribution framework. Additionally, they implemented 22 interpreters based on decision trees and fuzzy rules to provide supplementary explanations for each RF decision in each layer. Moreover, these explanations are presented in natural language to help doctors understand the prediction results. The designed model achieved a cross-validation accuracy of 93.95\% and F1 score of 93.94\% in the first layer, and a cross-validation accuracy of 87.08\% and F1 score of 87.09\% in the second layer. Due to the wide consistency of the explanations provided with AD medical literature, the resulting system is not only accurate but also trustworthy, traceable, and medically applicable. The system can help enhance the clinical understanding of AD diagnosis and progression by providing detailed insights into the influence of different modalities on disease risk.

AD is a progressive neurodegenerative disease, and biomarkers based on pathological physiology may provide objective indicators for disease diagnosis and staging. Neuroimaging scans obtained from MRI and metabolic imaging obtained through FDG-PET provide in vivo measurements of brain structure and function (glucose metabolism) in brain life activities. There is currently a hypothesis that combining multiple different image modalities providing complementary information can help improve the ability to diagnose AD early. Lu et al. \cite{muti_image_2} proposed a novel framework based on deep learning that uses multi-modality and multi-scale deep neural networks to identify individuals with AD. Their method provided 82.4\% accuracy in identifying patients with mild cognitive impairment (MCI) who will transition to AD within 3 years (86.4\% comprehensive accuracy in transitioning within 1-3 years), 94.23\% sensitivity in classifying individuals who may be clinically diagnosed with AD, and 86.3\% specificity in classifying non-dementia control groups, which is better than results in previous literature.

Qiu's team \cite{0Development} reported an interpretable deep learning strategy that uses multi-modal input of MRI, age, gender, and minimum mental state examination scores to describe unique Alzheimer's disease markers. Their framework links a fully convolutional network with a multi-layer perceptron, which can construct high-resolution maps of disease probability from local brain structures to generate accurate, intuitive individual AD risk visualization for accurate diagnosis. The model was trained using patients diagnosed with Alzheimer's disease and cognitively normal subjects from the Alzheimer's Disease Neuroimaging Initiative (ADNI) dataset (n=417), and validated on three independent cohorts: the Australian Imaging, Biomarkers and Lifestyle representational study (AIBL) (n=382), the Framingham Heart Study (n=102), and the National Alzheimer's Coordinating Center (NACC) (n=582). The performance of the model with multi-modal input remained consistent across datasets, with average area under the curve values of 0.996, 0.974, 0.876, and 0.954 for the ADNI study, AIBL, Framingham Heart Study, and NACC data sets, respectively. In addition, their method outperformed the diagnostic performance of a multi-institutional team of eleven practicing neurologists, and the high-risk brain areas predicted by the model closely tracked post-mortem pathological findings.

Shangran et al.\cite{WOS:000813768100006} present a deep learning framework that conducts several diagnostic steps consecutively to identify individuals with normal cognition (NC), mild cognitive impairment (MCI), Alzheimer’s disease (AD), and non-AD dementias (nADD). Their framework uses multiple models that can accept flexible combinations of routinely collected clinical information, including demographics, medical history, neuropsychological testing, neuroimaging, and functional assessments. They then demonstrate that their models perform comparably to the diagnostic accuracy of practicing neurologists and neuroradiologists. Additionally, they utilize interpretability methods in computer vision to reveal that disease-specific patterns detected by their models correspond closely with the presence of neuropathological lesions on autopsy and track distinct patterns of degenerative changes throughout the brain. Their work demonstrates methodologies for validating computational predictions against established standards of medical diagnosis.

In another study, Martinez et al.\cite{2020Studying} have implemented a novel approach for exploratory data analysis of Alzheimer's Disease (AD) which utilizes deep convolutional autoencoders. Their main objective is to establish connections between cognitive symptoms and the underlying neurodegeneration process by integrating information from neuropsychological test outcomes, clinical diagnoses, and other related data with imaging features extracted using a data-driven decomposition of MRI scans. To accomplish this, they analyze and visualize the distribution of extracted features in various combinations through regression and classification analysis. Furthermore, they estimate the influence of each coordinate of the autoencoder manifold on the brain. Their results show that the imaging-derived markers are able to predict clinical variables with correlations greater than 0.6, particularly in the case of neuropsychological evaluation variables like the MMSE or the ADAS11 scores. Moreover, their approach achieves a classification accuracy above 80 for the diagnosis of AD.

Liu et al.\cite{liu2020multi} introduce a multi-model deep learning framework based on a convolutional neural network (CNN) for automatic hippocampal segmentation and Alzheimer's disease (AD) classification through structural MRI data. Firstly, they construct a multi-task deep CNN model to learn jointly about hippocampal segmentation and disease classification. Using the segmentation results, they then construct a 3D Densely Connected Convolutional Network (3D DenseNet) to learn the features of 3D patches. Their method utilizes a combination of the features learned from the multi-task CNN and DenseNet models to classify disease status. Ther proposed method is evaluated using baseline T1-weighted structural MRI data that they have collected from 97 AD patients, 233 patients with mild cognitive impairment (MCI), and 119 normal control (NC) subjects from the ADNI database. They have achieved excellent results, with a dice similarity coefficient of 87.0\% for hippocampal segmentation. Furthermore, their proposed method has attained an accuracy of 88.9\% and an AUC (area under the ROC curve) of 92.5\% for classifying AD versus NC subjects, along with an accuracy of 76.2\% and an AUC of 77.5\% to classify MCI versus NC subjects. 

These findings emphasize the advantages of multimodal deep learning methods in leveraging diverse data sources to improve AD classification accuracy. By integrating information from multiple modalities, these approaches provide more comprehensive and robust insights for early AD diagnosis.

\section{Experiment}
\quad In this chapter, we will discuss the relevant researches on
the early identification of Alzheimer’s disease (AD) patients
using various modalities in recent years, focusing on three aspects: 1) Dataset and feature selection, 2) Modality and method,
3) Evaluation index and performance.

\subsection{Dataset and feature selection}

\quad The early classification of AD is a complex task that
requires specific characteristics in the data samples, including:
1) The dataset needs to have a certain scale; 2) The samples
should be labeled by professional doctors; 3) Acquiring a
large amount of experimental data and annotations often
requires significant human and material resources. 


\quad Currently, the most widely used general datasets for
AD research include ADNI, OASIS, AIBL, FHS, NACC and
BIOCARD. These datasets cover a wide range of data types
and can meet the needs of most AD-related studies.

\subsubsection{ADNI}
\begin{figure*}[!h]
    \includegraphics[width=1.0\textwidth]{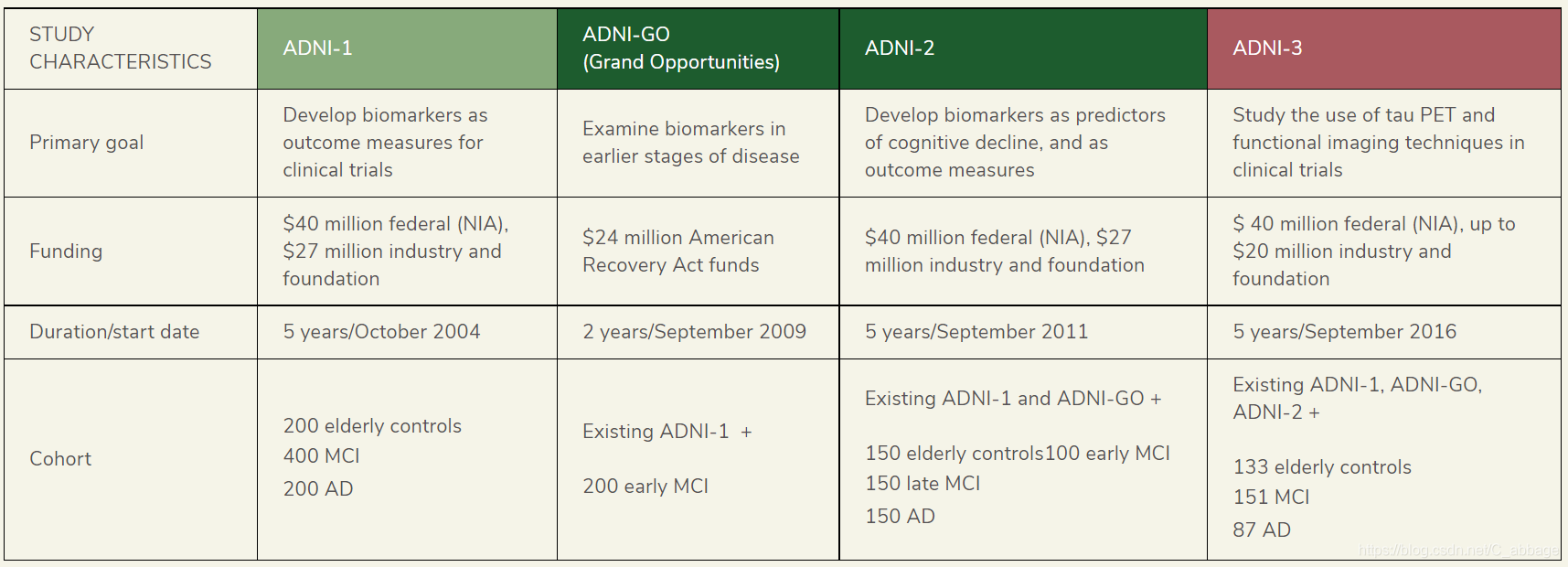}
    \centering
    \caption{ADNI dataset. The ADNI series includes ADNI-1, ADNI-GO, ADNI-2, and ADNI-3. The figure summarizes them in the following terms: primary goal, funding, duration/start date, and cohort}
    \label{ADNI}
\end{figure*}

\quad The Alzheimer's Disease Neuroimaging Initiative (ADNI) is renowned for its longitudinal and multi-center research and is the most common dataset used in AD research. ADNI researchers collect various types of data from patients, including clinical, genetic, MRI, PET images, and biomarkers, with the aim of investigating whether they can measure the progression of Mild Cognitive Impairment (MCI) and early AD. The ADNI series, including ADNI-1, ADNIGO, ADNI-2, and ADNI-3, are supplements and improvements to previous series, as shown in Fig. \ref{ADNI}.(Source: https://adni.loni.usc.edu/)

\subsubsection{OASIS}

\quad The Open Access Series of Imaging Studies (OASIS) aims to share brain neuroimaging datasets with researchers in related fields. OASIS is available in the following versions: OASIS-1 includes 434 cross-sectional MRI scans from 416 subjects aged 18 to 96 years. OASIS-2 consists of 373 longitudinal MRI scans from 150 subjects. OASIS-3 includes 2,842 longitudinal MRI scans, 2,157 PET scans, and 1,472 CT scans from 1,379 subjects. Each subject typically has two or more scans with an interval of at least one year. Currently, OASIS is the second largest core data resource after ADNI. (Source: https://www.oasis-brains.org/\#about)

\subsubsection{AIBL}

\quad The Australian Imaging, Biomarker \& Lifestyle (AIBL) study was initiated in 2006 to conduct long-term assessments (over 4.5 years) on more than 1,100 patients in order to determine which biomarkers, cognitive features, health, and lifestyle factors contribute to the subsequent development of symptomatic Alzheimer's disease (AD). AIBL collects imaging and medical data from 211 AD patients, 133 patients with mild cognitive impairment (MCI), and 768 cognitively normal healthy individuals, making it the largest study of its kind in Australia. (Source: https://aibl.csiro.au/about/)

\subsubsection{FHS}

\quad The Framingham Heart Study (FHS)\cite{massaro2004managing} is a longitudinal community-based cohort study that has collected extensive clinical data on three generations. Since 1976, the FHS has expanded to include the evaluation of factors leading to cognitive decline, dementia, and Alzheimer's disease. In March 1999, the offspring cohort began undergoing brain MRI scans and neuropsychological (NP) testing. The MRI and NP studies allow for quantitative measurements of brain structure, such as white matter hyperintensities (WMH) volume, and cognitive function, respectively. Combined with its comprehensive clinical data, the FHS has become one of the largest databases in its field.

\subsubsection{NACC}

\quad The National Alzheimer's Coordinating Center (NACC) is responsible for developing and maintaining a database that collects participant information from 29 Alzheimer's Disease Centers (ADCs) funded by the National Institute on Aging (NIA). The database includes standardized clinical and neuropathological research data. The goal is to provide researchers with a longitudinally collected set of standardized assessment procedures to better characterize participants with mild Alzheimer's disease and mild cognitive impairment compared to non-dementia control groups.

\subsubsection{BIOCARD}

\begin{figure*}[!h]
    \centering
    \includegraphics[width=0.8\textwidth]{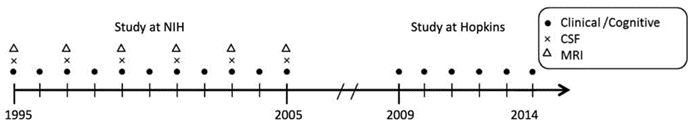}
    \caption{BIOCARD：the timeline of data collection}
    \label{BIOCARD}
\end{figure*}

\quad The overall objective of this study\cite{soldan2015relationship} was to identify variables that can predict the subsequent development of mild to moderate Alzheimer's disease symptoms in cognitively normal individuals. From 1995 to 2005, during the tenure of the National Institutes of Health (NIH) in the United States, participants underwent comprehensive neuropsychological testing annually. MRI scans, cerebrospinal fluid (CSF) samples, and blood specimens were collected approximately every 2 years. In 2009, the research team at Johns Hopkins University continued with annual clinical and cognitive evaluations, collected blood samples, and assessed previously obtained data. Due to funding constraints, CSF and MRI scans were not collected since the study moved to Johns Hopkins University. Please refer to Figure \ref{BIOCARD} for the timeline of data collection.

\quad However, the aforementioned datasets do not include audio attribute data such as pause counts, reaction time, speech rate, etc. Therefore, in a series of studies using the sound modality for AD patient diagnosis, research teams have to recruit participants and collect voice data autonomously to meet specific research needs. These self-built datasets include the previously mentioned PITT Corpus, VBSD, Dem@Care, ADReSS, and others. Similarly, these datasets also do not contain posture attribute data from various body parts such as chest, hips, wrists, and ankles, resulting from actions such as sitting, standing, walking, and lying down. These data are primarily obtained from sensors worn by patients.

\subsubsection{Smartphone-based single-sensor dataset}

\quad The dataset\cite{raza2019diagnosis} consists of six Activities of Daily Living (ADLs), including sitting, standing, walking, lying, going upstairs, and going downstairs. These ADLs were performed by 30 participants ranging in age from 19 to 49 years. The participants wore a smartphone attached to their waist and followed a structured protocol to perform the aforementioned ADLs. The dataset includes three-dimensional acceleration signal data (from the accelerometer) and three-dimensional angular velocity data (from the gyroscope). The sampling frequency of the smartphone sensors was 50Hz. For more details, please refer to Fig.\ref{tab:zitai}.

\subsubsection{Multi-sensor dataset}
\quad The dataset \cite{raza2019diagnosis} was developed using four inertial sensors placed on the chest, right hip, right wrist, and ankle. A total of 19 participants (26±8 years old) took part in the experiment. Data collection was performed using wireless sensors based on blink sensing technology. The sensors captured three-dimensional acceleration and three-dimensional angular velocity signals, with a sampling frequency of 204.8Hz. For more detailed information, please refer to Fig. \ref{tab:zitai}.

    \begin{table*}[htbp]
      \centering
      \caption{Two pose datasets based on sensors}
        \begin{tabular}{rrrp{9.78em}rp{10.78em}}
        \toprule
        \multicolumn{1}{p{7em}}{Author} &
          \multicolumn{1}{p{4em}}{Sampling} &
          \multicolumn{1}{p{7em}}{Sensor Location} &
          Experiment Setting &
          \multicolumn{1}{p{9.78em}}{Recorded Signals} &
          Activities Analyzed
          \\
         &
          \multicolumn{1}{p{4em}}{Frequency} &
           &
          (Population) &
          \multicolumn{1}{p{9.78em}}{} &
          \multicolumn{1}{r}{}
          \\
        \midrule
        \multicolumn{1}{p{7em}}{Anguita et al.} &
          \multicolumn{1}{p{4em}}{50 Hz} &
          \multicolumn{1}{p{7em}}{Waist-mounted} &
          Semi-naturalistic &
          \multicolumn{1}{p{9.78em}}{3D Acceleration Signal,} &
          Standing, sitting, lying,
          \\
         &
           &
          \multicolumn{1}{p{7em}}{Smartphone} &
          conditions(30 subjects) &
          \multicolumn{1}{p{9.78em}}{3D Angular Velcovity} &
          walking, walking upstairs,
          \\
         &
           &
           &
          (19 to 48 years) &
           &
          Walking downstairs.
          \\
        \multicolumn{1}{p{7em}}{Leutheuser et al.} &
          \multicolumn{1}{p{4em}}{204.8 Hz} &
          \multicolumn{1}{p{7em}}{Wrist, hip, chest,} &
          Laborary settings (23 &
          \multicolumn{1}{p{9.78em}}{3D Acceleration Signal,} &
          Walking, sitting, lying,
          \\
         &
           &
          \multicolumn{1}{p{7em}}{ankle} &
          young adults) (27±7 &
          \multicolumn{1}{p{9.78em}}{3D Angular Velcovity} &
          Standing, stairs up, stairs
          \\
         &
           &
           &
          years) &
           &
          down
          \\
        \bottomrule
        \end{tabular}%
      \label{tab:zitai}%
    \end{table*}%


\quad At the gene-level granularity, previous studies have utilized gene expression profiles and DNA methylation profiles as datasets for research purposes \cite{park2020prediction}. The specific utilization of the dataset is depicted in Fig.\ref{tab:gene dataset}.

    \begin{table*}[htbp]
      \centering
      \caption{Datasets of gene expression profiles and DNA methylation profiles}
        \begin{tabular}{p{12.5em}p{9em}rp{18.5em}}
        \toprule
        Dataset &
          \multicolumn{2}{p{18em}}{Gene expression} &
          DNA methylation
          \\
        \multicolumn{1}{r}{} &
          \multicolumn{2}{p{18em}}{Rosetta/Merck Human 44k 1.1 microarray} &
          Illumina Human Methylation 450 Bead Chip
          \\
        \midrule
        GEO ID &
          GSE33000 &
          \multicolumn{1}{p{9em}}{GSE44770} &
          GSE80970
          \\
        Number of normal samples &
          \multicolumn{1}{l}{157} &
          \multicolumn{1}{l}{100} &
          \multicolumn{1}{l}{68}
          \\
        Number of AD samples &
          \multicolumn{1}{l}{310} &
          \multicolumn{1}{l}{129} &
          \multicolumn{1}{l}{74}
          \\
        Number of features &
          19,488 genes &
           &
          485,577 probes for CpG site
          \\
        \bottomrule
        \end{tabular}%
      \label{tab:gene dataset}%
    \end{table*}%


\quad Apart from the aforementioned datasets, there are also some niche datasets collected independently by universities, research centers, or hospitals to meet specific research needs. These datasets can be seen in the "Dataset" column in the table below.


\quad While general-purpose datasets can fulfill the needs of the majority of research studies, different studies often select subsets from these datasets based on factors such as the number of participants, their distribution, and the types of features available. It is ideal to have a balanced distribution of participants across gender, educational background, and age, as it reduces sampling bias and enhances the persuasiveness of experimental results, allowing for generalization. Therefore, this paper has compiled the "Datasets and Feature Selection" information from various studies in this field in recent years, as shown in Table \ref{tab:11}-\ref{tab:14}.


\quad MRI is the most commonly used imaging modality. Based on the size of the selected region of interest (ROI), the most commonly used feature extraction methods can be roughly classified into four categories.

1) Voxel-based methods \cite{2020Studying}\cite{2019Volumetrie}\cite{wang2019ensemble} treat voxel intensity as features. However, analyzing features only on isolated voxels ignores the high correlation between voxels and has limitations such as high data dimensionality, computational complexity and overfitting.

2) Region-based methods \cite{shi2017multimodal}\cite{colliot2008discrimination}\cite{2021An}\cite{li2015robust}\cite{zhou2019effective}\cite{liu2019using}\cite{lee2019toward}\cite{cui2019rnn}\cite{ge2019multi}\cite{li2019deep}\cite{chitradevi2020analysis}\cite{spasov2019parameter}\cite{huang2019diagnosis}\cite{lei2022predicting}\cite{ghazi2019training}\cite{kang2021multi} extract features from predefined brain regions. Commonly selected regions include the hippocampus (HC), gray matter (GM), white matter (WM), corpus callosum (CC), cerebrospinal fluid (CSF) and cortical thickness (typically represented as vertex-level scales on the cortical surface \cite{chen2021iterative}\cite{barbaroux2020encoding}\cite{huang2019diagnosis}). The dimensionality of the data is significantly reduced compared to voxel-level features, making it the most commonly used scale for features.

3) Patch-based methods \cite{0Development}\cite{2018Hierarchical}\cite{2021Dual}\cite{cui2018hippocampus}\cite{liu2020multi}\cite{liu2019weakly}\cite{mendoza2020single} learn new feature representations from local patches, providing a scale between voxel-level and region-level features that can better capture local features in the images.

4) Slice-based methods.
Depending on the different directions of the two-dimensional slices, they can be divided into axial \cite{2021ADVIAN}\cite{tufail2020binary}\cite{multimodal_e}, sagittal \cite{cui2019rnn}\cite{puente2020automatic}, and coronal \cite{ebrahimi2021deep}\cite{raza2019diagnosis}\cite{kang2021multi} slices. Some studies, such as \cite{bae2021transfer}\cite{zhang2019multi}, use slices from all three directions, but most studies only use axial slices.

Preprocessing steps for MRI often include skull-stripping \cite{hazarika2022experimental}\cite{zhang2021explainable} to optimize the results by removing the skull.


\quad The scores obtained from cognitive assessment tests are commonly used features in cognitive modalities. Such as MMSE, MoCA, CDR-SOB, CDR-GLOB, ADAS-cog, RAVLT, FAQ, CERAD-NB, and DSRS\cite{MMSEvsMoCA}\cite{2010Clin}\cite{0Development}\cite{2021An}\cite{li2015robust}\cite{li2019deep}\cite{spasov2019parameter}\cite{liu2019weakly}\cite{abuhmed2021robust}\cite{lei2022predicting}\cite{lei2020deep}.These measures are interrelated, and under certain conditions, they can be converted or transformed into one another. For example, the conversion from MMSE to MoCA is discussed in \cite{MMSEvsMoCA}. However, each measure focuses on different cognitive aspects, highlighting the need for their combination to achieve complement, thereby providing a more comprehensive description of cognitive abilities.


\quad Gait analysis is a less commonly used modality in cognitive research. It involves having participants wear sensors that detect the motion information of joints and limbs over time using gyroscopes, accelerometers, and other sensors. This information is then used as features in the analysis \cite{ghoraani2021detection}\cite{matsuura2019statistical}\cite{raza2019diagnosis}. Voice modality, on the other hand, often directly utilizes existing feature sets such as eGeMAPS, emobase, ComParE, and MRCG as features \cite{luz2020alzheimer}\cite{haider2019assessment}. Some studies also use the spectral information of voice segments as features \cite{mh_1}. These modalities and their features have been discussed in detail in the previous text.


\quad Gender, age, education level, and ethnic background are commonly used features in demographic modality \cite{0Development}\cite{multimodal_e}\cite{puente2020automatic}. These features provide a more detailed and comprehensive characterization of the participants' profiles. APOE4, SNPs, amyloid-beta protein, and cerebrospinal fluid (CSF) are commonly used features in biomarker modality \cite{2010Clin}\cite{multimodal_e}\cite{bi2020multimodal}\cite{li2015robust}\cite{li2019deep}\cite{abuhmed2021robust}. Incorporating genetic and molecular biomarkers related to disease progression significantly improves the prediction results and facilitates pathological research on Alzheimer's disease (AD). This helps address the limited interpretability of deep learning models. Some biomarkers are often represented in the form of images, hence intersecting with the image modality.

    \begin{table*}[htbp]
      \centering
      \caption{Dataset and feature selection}
        \begin{tabular}{c|c|p{11.5em}|p{16.8em}|p{17em}}
        \toprule
        \multicolumn{1}{c|}{\textbf{Ref.}} &
          \multicolumn{1}{c|}{\textbf{Year}} &
          \multicolumn{1}{c|}{\textbf{Dataset}} &
          \multicolumn{1}{c|}{\textbf{Sample}} &
          \multicolumn{1}{c}{\textbf{Feature}}
          \\
        \midrule
        \multicolumn{1}{c|}{\cite{MMSEvsMoCA}} &
          2020 &
          Penn Memory Center + Clinical Core of the University of Pennsylvania’s Alzheimer’s Disease Center &
          \multirow{3}[2]{*}{587 subjects(321 AD,126 MCI,140 CN)} &
          \multicolumn{1}{p{17em}}{\multirow{3}[2]{*}{MMSE + MoCA + CERAD-NB + DSRS}}
          \\
        \midrule
        \multicolumn{1}{c|}{\multirow{2}[2]{*}{\cite{2010Clin}}} &
          \multirow{2}[2]{*}{2022} &
          \multicolumn{1}{c|}{\multirow{2}[2]{*}{ADNI dataset}} &
          \multirow{2}[2]{*}{490 subjects (56 AD, 287 MCI, 147 CN)} &
          \multicolumn{1}{p{17em}}{MRI + Amyloid-β PET + CSF + Cognitive sub-scores + APOE + SNP}
          \\
        \midrule
        \multicolumn{1}{c|}{\cite{ghoraani2021detection}} &
          2021 &
           A retrospective cohort of community-dwelling older adults &
          \multicolumn{1}{c|}{\multirow{3}[2]{*}{78 subjects (20 AD, 26 MCI, 32 CN)}} &
          \multicolumn{1}{p{17em}}{Stride time + Step time + Single Support time + Swing time + Double support time + Stance time + Stride length + Step length}
          \\
        \midrule
        \multicolumn{1}{c|}{\multirow{2}[2]{*}{\cite{mh_1}}} &
          \multirow{2}[2]{*}{2020} &
          \multicolumn{1}{c|}{VBSD dataset} &
          VBSD:36 subjects (23 AD, 13 CN) &
          \multicolumn{1}{p{17em}}{Spectrogram features extracted from }
          \\
         &
           &
           \multicolumn{1}{c|}{Dem@Care dataset} &
          Dem@Care:32 subjects (24 AD, 8 CN) &
          \multicolumn{1}{p{17em}}{speech segments}
          \\
        \midrule
        \multicolumn{1}{c|}{\multirow{2}[2]{*}{\cite{haider2019assessment}}} &
          \multirow{2}[2]{*}{2021} &
          \multicolumn{1}{c|}{\multirow{2}[2]{*}{Pitt corpus}} &
          \multicolumn{1}{c|}{\multirow{2}[2]{*}{164 subjects (82 AD, 82 non-AD)}} &
          \multicolumn{1}{p{17em}}{Feature sets: eGeMAPS + emobase + ComParE + MRCG}
          \\
        \midrule
        \multicolumn{1}{c|}{\multirow{3}[2]{*}{\cite{luz2020alzheimer}}} &
          \multirow{3}[2]{*}{2020} &
          \multirow{3}[2]{*}{ADReSS Challenge Dataset} &
          \multicolumn{1}{c|}{156 subjects (78 AD, 78 non-AD)} &
          \multicolumn{1}{p{17em}}{Feature sets: eGeMAPS + emobase + }
          \\
         &
           &
          \multicolumn{1}{c|}{} &
          \multicolumn{1}{c|}{4077 speech segments} &
          \multicolumn{1}{p{17em}}{ComParE + MRCG + Minimal + }
          \\
         &
           &
          \multicolumn{1}{c|}{} &
          \multicolumn{1}{c|}{(2122 AD, 1955 non-AD)} &
          \multicolumn{1}{p{17em}}{Linguistic}
          \\
        \midrule
        \multicolumn{1}{c|}{\multirow{2}[2]{*}{\cite{shi2017multimodal}}} &
          \multirow{2}[2]{*}{2018} &
          \multicolumn{1}{c|}{\multirow{2}[2]{*}{ADNI dataset}} &
          202 subjects (51 AD, 99 MCI, 52 CN) &
          \multicolumn{1}{p{17em}}{\multirow{2}[2]{*}{MRI (Region-level: GM volume) + PET}}
          \\
         &
           &
          \multicolumn{1}{c|}{} &
          99 MCI (43 MCI-C, 56 MCI-NC) &
          
          \\
        \midrule
        \multicolumn{1}{c|}{\multirow{2}[2]{*}{\cite{muti_image_2}}} &
          \multirow{2}[2]{*}{2018} &
          \multicolumn{1}{c|}{\multirow{2}[2]{*}{ADNI dataset}} &
          1242 subjects (360 sCN, 409 sMCI, 18  &
          \multicolumn{1}{c}{\multirow{2}[2]{*}{FDG-PET + MRI}}
          \\
         &
           &
          \multicolumn{1}{c|}{} &
          pCN, 217 pMCI, 238 AD) &
          
          \\
        \midrule
        \multicolumn{1}{c|}{\multirow{4}[2]{*}{\cite{qiu2020development}}} &
          \multirow{4}[2]{*}{2020} &
          \multicolumn{1}{c|}{ADNI dataset} &
          ADNI: 417 subjects (188 AD, 229 CN) &
          \multicolumn{1}{p{17em}}{MRI (Patch-level) + Age + Gender +}
          \\
         &
           &
          \multicolumn{1}{c|}{AIBL dataset} &
          AIBL: 382 subjects (62 AD, 320 CN) &
          \multicolumn{1}{p{17em}}{MMSE}
          \\
         &
           &
          \multicolumn{1}{c|}{FHS dataset} &
          FHS: 102 subjects (29 AD, 73 CN) &
          
          \\
         &
           &
          \multicolumn{1}{c|}{NACC dataset} &
          NACC: 565 subjects (209 AD, 356 CN) &
          
          \\
        \midrule
        \multicolumn{1}{c|}{\multirow{2}[2]{*}{\cite{2018Hierarchical}}} &
          \multirow{2}[2]{*}{2020} &
          \multicolumn{1}{c|}{\multirow{2}[2]{*}{ADNI dataset}} &
          \multicolumn{1}{c|}{1457 subjects} &
          \multicolumn{1}{c}{\multirow{2}[2]{*}{MRI (Patch-level)}}
          \\
         &
           &
          \multicolumn{1}{c|}{} &
          (429 CN, 465 sMCI, 205 pMCI, 358 AD) &
          
          \\
        \midrule
        \multicolumn{1}{c|}{\multirow{2}[2]{*}{\cite{ebrahimi2021deep}}} &
          \multirow{2}[2]{*}{2021} &
          \multicolumn{1}{c|}{ImageNet dataset} &
          ImageNet: 1000 object categories &
          \multicolumn{1}{c}{\multirow{2}[2]{*}{MRI (Coronal slice)}}
          \\
         &
           &
          \multicolumn{1}{c|}{ADNI dataset} &
          ADNI: 450 subjects (225 AD, 225 CN) &
          
          \\
        \midrule
        \multicolumn{1}{c|}{\cite{colliot2008discrimination}} &
          2008 &
          Database of the Centre Hospitalo-Universitaire of Caen, France &
          \multicolumn{1}{c|}{\multirow{2}[2]{*}{74 subjects (25 AD, 24 MCI, 25 CN)}} &
          \multicolumn{1}{c}{\multirow{2}[2]{*}{MRI (Region-level: HC)}}
          \\
        \midrule
        \multicolumn{1}{c|}{\multirow{6}[2]{*}{\cite{multimodal_e}}} &
          \multirow{6}[2]{*}{2022} &
          \multicolumn{1}{c|}{\multirow{6}[2]{*}{ADNI dataset}} &
          \multicolumn{1}{c|}{Clinical data: 2004 subjects} &
          \multicolumn{1}{p{17em}}{Clinical data: Demographics + }
          \\
         &
           &
          \multicolumn{1}{c|}{} &
          \multicolumn{1}{c|}{(598 CN, 699 MCI, 707 AD)} &
          \multicolumn{1}{p{17em}}{Neurological exams + Cognitive sub-}
          \\
         &
           &
          \multicolumn{1}{c|}{} &
          \multicolumn{1}{c|}{Imaging: 502 subjects}  &
          \multicolumn{1}{p{17em}}{ scores + Biomarkers + Medication +}
          \\
         &
           &
          \multicolumn{1}{c|}{} &
          \multicolumn{1}{c|}{(132 CN, 104 MCI, 266 AD)} &
          \multicolumn{1}{p{17em}}{Imaging summary scores}
          \\
         &
           &
          \multicolumn{1}{c|}{} &
          \multicolumn{1}{c|}{Genetic: 809 subjects} &
          \multicolumn{1}{p{17em}}{Imaging: Cross-sectional MRI}
          \\
         &
           &
          \multicolumn{1}{c|}{} &
           \multicolumn{1}{c|}{(245 CN, 338 MCI, 226 AD)} &
          \multicolumn{1}{p{17em}}{Genetics: Whole genome sequencing}
          \\
        \midrule
        \multicolumn{1}{c|}{\multirow{2}[2]{*}{\cite{zeng2021new}}} &
          \multirow{2}[2]{*}{2021} &
          \multicolumn{1}{c|}{\multirow{2}[2]{*}{ADNI dataset}} &
          \multicolumn{1}{c|}{361 subjects} &
          \multicolumn{1}{p{17em}}{MRI (Region-level: GM volume) + }
          \\
         &
           &
          \multicolumn{1}{c|}{} &
          （92 CN, 82 sMCI, 95 pMCI, 92 AD） &
          \multicolumn{1}{p{17em}}{MMSE + ADAS-Cog}
          \\
        \midrule
        \multicolumn{1}{c|}{\multirow{2}[2]{*}{\cite{2021ADVIAN}}} &
          \multirow{2}[2]{*}{2021} &
          \multicolumn{1}{c|}{OASIS-1 dataset} &
          OASIS-1：126 subjects (28 AD,98 CN) &
          \multicolumn{1}{c}{\multirow{2}[2]{*}{MRI (Axial key slices)}}
          \\
         &
           &
          \multicolumn{1}{c|}{Local hospitals} &
          Local hospitals: 70 subjects (70 AD) &
          
          \\
        \bottomrule
        \end{tabular}%
      \label{tab:11}%
    \end{table*}%


    \begin{table*}[htbp]
      \centering
      \caption{(Continued)}
        \begin{tabular}{c|c|p{10em}|p{17.5em}|p{16.8em}}
        \toprule
        \multicolumn{1}{c|}{\textbf{Ref.}} &
          \multicolumn{1}{c|}{\textbf{Year}} &
          \multicolumn{1}{c|}{\textbf{Dataset}} &
          \multicolumn{1}{c|}{\textbf{Sample}} &
          \multicolumn{1}{c}{\textbf{Feature}}
          \\
        \midrule
        \multicolumn{1}{c|}{\multirow{2}[2]{*}{\cite{2021Dual}}} &
          \multirow{2}[2]{*}{2021} &
          \multicolumn{1}{c|}{\multirow{2}[2]{*}{ADNI dataset}} &
          ADNI: 1193 subjects (389 AD, 172 pMCI, &
          \multicolumn{1}{c}{\multirow{2}[2]{*}{MRI (Patch-level)}}
          \\
         &
           &
          \multicolumn{1}{c|}{\multirow{2}[2]{*}{AIBL dataset}} &
           232 sMCI, 400 CN); AIBL: 496 subjects &
          
          \\
         &
           &
           &
           (79 AD, 17 pMCI, 93 sMCI, 307 CN) &
          
          \\
        \midrule
        \multicolumn{1}{c|}{\multirow{2}[2]{*}{\cite{martinez2019studying}}} &
          \multirow{2}[2]{*}{2020} &
          \multicolumn{1}{c|}{\multirow{2}[2]{*}{ADNI dataset}} &
          \multicolumn{1}{c|}{479 subjects} &
          \multicolumn{1}{c}{MRI (Voxel-level:}
          \\
         &
           &
           &
          \multicolumn{1}{c|}{(168 CN, 212 MCI, 99 AD)} &
          \multicolumn{1}{c}{voxel intensity of GM and WM)}
          \\
        \midrule
        \multicolumn{1}{c|}{\multirow{2}[2]{*}{\cite{lin2021bidirectional}}} &
          \multirow{2}[2]{*}{2021} &
          \multicolumn{1}{c|}{\multirow{2}[2]{*}{ADNI dataset}} &
          \multicolumn{1}{c|}{1086 subjects} &
          \multicolumn{1}{c}{\multirow{2}[2]{*}{MRI (Region-level: HC) + PET}}
          \\
         &
           &
           &
          (362 AD, 308 CN, 183 pMCI, 233 sMCI) &
          
          \\
        \midrule
        \multicolumn{1}{c|}{\cite{bi2020multimodal}} &
          2020 &
          \multicolumn{1}{c|}{ADNI dataset} &
          \multicolumn{1}{c|}{72 subjects (37 AD, 35 CN)} &
          \multicolumn{1}{c}{fMRI + SNP}
          \\
        \midrule
        \multicolumn{1}{c|}{\multirow{7}[2]{*}{\cite{matsuura2019statistical}}} &
          \multirow{7}[2]{*}{2019} &
          \multicolumn{1}{p{10em}|}{\multirow{3}[2]{*}{Fujidera Assisted Living}} &
          \multicolumn{1}{r|}{} &
          \multicolumn{1}{p{16.8em}}{Average speed of stepping +}
          \\
         &
           &
          \multicolumn{1}{p{11.5em}|}{\multirow{3}[2]{*}{Facility}} &
          \multicolumn{1}{c|}{\multirow{4}[2]{*}{10,833 behavior data from 90 subjects}} &
          \multicolumn{1}{p{16.8em}}{The standard deviation of stepping }
          \\
         &
           &
          \multicolumn{1}{p{10em}|}{\multirow{3}[2]{*}{Tsudou Elderly Care Center}} &
          \multicolumn{1}{c|}{\multirow{4}[2]{*}{(Male: 32, Female: 68)}} &
          \multicolumn{1}{p{16.8em}}{speed + The ratio of correct answers +}
          \\
         &
           &
           \multicolumn{1}{p{10em}|}{\multirow{3}[2]{*}{Daisen Elderly Care Center}} &
          \multicolumn{1}{r|}{} &
          \multicolumn{1}{p{16.8em}}{ Average time of answering the calculus}
          \\
         &
           &
           &
          \multicolumn{1}{r|}{} &
          \multicolumn{1}{p{16.8em}}{ questions + The average height of }
          \\
         &
           &
           &
          \multicolumn{1}{r|}{} &
          \multicolumn{1}{p{16.8em}}{knee joint + The standard deviation of }
          \\
         &
           &
           &
          \multicolumn{1}{r|}{} &
          \multicolumn{1}{p{16.8em}}{height for the knee joint.}
          \\
        \midrule
        \multicolumn{1}{c|}{\multirow{2}[2]{*}{\cite{li2015robust}}} &
          \multirow{2}[2]{*}{2015} &
          \multicolumn{1}{c|}{\multirow{2}[2]{*}{ADNI dataset}} &
          \multicolumn{1}{c|}{202 subjects} &
          \multicolumn{1}{p{16.8em}}{(MRI + PET) (Region-level: ROI volume)}
          \\
         &
           &
           &
          \multicolumn{1}{c|}{(51 AD, 43 pMCI, 56 sMCI, 52 CN)} &
          \multicolumn{1}{p{16.8em}}{ + CSF + MMSE + ADAS-Cog}
          \\
        \midrule
        \multicolumn{1}{c|}{\cite{park2020prediction}} &
          2019 &
          \multicolumn{1}{p{11.5em}|}{Rosetta/Merck Human 44k} &
          \multicolumn{1}{c|}{\multirow{2}[2]{*}{GSE33000：157 CN，310 AD}} &
          \multicolumn{1}{c}{\multirow{2}[2]{*}{Gene expression}}
          \\
         &
           &
          \multicolumn{1}{p{11.5em}|}{1.1 microarray + Illumina} &
          \multicolumn{1}{c|}{\multirow{2}[2]{*}{GSE44770：100 CN，129 AD}} &
          \multicolumn{1}{c}{\multirow{2}[2]{*}{(GEO ID: GSE33000，GSE44770) +}}
          \\
         &
           &
            \multicolumn{1}{p{11.5em}|}{Human Methylation 450 Bead Chip} &
          \multicolumn{1}{c|}{\multirow{2}[2]{*}{GSE80970：68 CN，74 AD}} &
          \multicolumn{1}{c}{\multirow{2}[2]{*}{DNA methylation (GEO ID: GSE80970）}}
          \\
        \midrule
        \multicolumn{1}{c|}{\multirow{2}[2]{*}{\cite{zhou2019effective}}} &
          \multirow{2}[2]{*}{2019} &
          \multicolumn{1}{c|}{\multirow{2}[2]{*}{ADNI dataset}} &
          \multicolumn{1}{c|}{805 subjects} &
          \multicolumn{1}{p{16.8em}}{MRI (Region-level: GM volume) + PET}
          \\
         &
           &
           &
          \multicolumn{1}{c|}{(226 CN，389 MCI，190 AD)} &
          \multicolumn{1}{p{16.8em}}{ (Region-level: ROI intensity) + SNP}
          \\
        \midrule
        \multicolumn{1}{c|}{\multirow{2}[2]{*}{\cite{liu2019using}}} &
          \multirow{2}[2]{*}{2019} &
          \multicolumn{1}{c|}{ADNI dataset} &
          \multicolumn{1}{c|}{1143 subjects} &
          \multicolumn{1}{c}{MRI (Region-level: volumes of brain}
          \\
         &
           &
          \multicolumn{1}{c|}{BIOCARD database} &
          \multicolumn{1}{c|}{(324 from BIOCARD, 819 from ADNI)} &
          \multicolumn{1}{p{16.8em}}{structures)}
          \\
        \midrule
        \multicolumn{1}{c|}{\multirow{2}[2]{*}{\cite{lee2019toward}}} &
          \multirow{2}[2]{*}{2019} &
          \multicolumn{1}{c|}{\multirow{2}[2]{*}{ADNI dataset}} &
          \multicolumn{1}{c|}{801 subjects} &
          \multicolumn{1}{c}{\multirow{2}[2]{*}{MRI (Region-level: GM volume)}}
          \\
         &
           &
           &
          \multicolumn{1}{c|}{(198 AD,160 pMCI,214 sMCI,229 CN)} &
          
          \\
        \midrule
        \multicolumn{1}{c|}{\multirow{2}[2]{*}{\cite{cui2019rnn}}} &
          \multirow{2}[2]{*}{2019} &
          \multicolumn{1}{c|}{\multirow{2}[2]{*}{ADNI dataset}} &
          \multicolumn{1}{c|}{830 subjects} &
          \multicolumn{1}{c}{\multirow{2}[2]{*}{Sagittal MRI(Region-level: GM volume)}}
          \\
         &
           &
           &
          (198 AD,167 pMCI,236 sMCI,229 CN) &
          \\
        \midrule
        \multicolumn{1}{c|}{\cite{ge2019multi}} &
          2019 &
          \multicolumn{1}{c|}{ADNI dataset} &
          \multicolumn{1}{c|}{337 subjects (198 AD,139 CN)} &
          \multicolumn{1}{c}{MRI(Region-level: GM + WM + CSF)}
          \\
        \midrule
        \multicolumn{1}{c|}{\multirow{3}[2]{*}{\cite{li2019deep}}} &
          \multirow{3}[2]{*}{2019} &
          \multicolumn{1}{c|}{\multirow{2}[2]{*}{ADNI dataset}} &
          \multicolumn{1}{c|}{\multirow{2}[2]{*}{2146 subjects}} &
          \multicolumn{1}{p{16.8em}}{MRI(Region-level: HC) + MMSE}
          \\
         &
           &
          \multicolumn{1}{c|}{\multirow{2}[2]{*}{AIBL dataset}} &
          \multicolumn{1}{c|}{\multirow{2}[2]{*}{(1711 from ADNI, 435 from AIBL)}} &
          \multicolumn{1}{p{16.8em}}{+ ADAS-Cog13 + APOE4 + RAVLT +}
          \\
         &
           &
           &
          \multicolumn{1}{r|}{} &
          \multicolumn{1}{p{16.8em}}{ FAQ + Amyloid-positive status + Age + Gender}
          \\
        \midrule
        \multicolumn{1}{c|}{\multirow{2}[2]{*}{\cite{cui2018hippocampus}}} &
          \multirow{2}[2]{*}{2019} &
          \multicolumn{1}{c|}{\multirow{2}[2]{*}{ADNI dataset}} &
          \multicolumn{1}{c|}{811 subjects} &
          \multicolumn{1}{c}{\multirow{2}[2]{*}{MRI (Patch-level: HC)}}
          \\
         &
           &
           &
          (192 AD,165 pMCI, 231 sMCI, 223 CN) &
          
          \\
        \midrule
        \multicolumn{1}{c|}{\cite{liu2020multi}} &
          2020 &
          \multicolumn{1}{c|}{ADNI dataset} &
          449 subjects (97 AD, 233 MCI, 119 CN) &
          \multicolumn{1}{c}{MRI (Patch-level: HC)}
          \\
        \midrule
        \multicolumn{1}{c|}{\multirow{2}[2]{*}{\cite{ieracitano2019convolutional}}} &
          \multirow{2}[2]{*}{2018} &
          \multicolumn{1}{p{11.5em}|}{IRCCS Centro Neurolesi} &
          \multicolumn{1}{c|}{\multirow{2}[2]{*}{189 EEG recordings}} &
          \multicolumn{1}{c}{\multirow{3}[2]{*}{EEG}}
          \\
         &
           &
          \multicolumn{1}{p{11.5em}|}{Bonino-Pulejo of Messina (Italy)} &
          \multicolumn{1}{c|}{\multirow{2}[2]{*}{(63 AD, 63 MCI, 63 CN)}} &
          
          \\
        \midrule
        \multicolumn{1}{c|}{\multirow{2}[2]{*}{\cite{li2020detecting}}} &
          \multirow{2}[2]{*}{2020} &
          \multicolumn{1}{c|}{\multirow{2}[2]{*}{ADNI dataset}} &
          \multicolumn{1}{c|}{389 subjects} &
          \multicolumn{1}{c}{\multirow{2}[2]{*}{fMRI}}
          \\
         &
           &
           &
          \multicolumn{1}{c|}{(116 AD, 99 MCI, 174 CN)} &
          
          \\
        \midrule
        \multicolumn{1}{c|}{\multirow{2}[2]{*}{\cite{choi2020cognitive}}} &
          \multirow{2}[2]{*}{2020} &
          \multicolumn{1}{c|}{\multirow{2}[2]{*}{ADNI dataset}} &
          \multicolumn{1}{c|}{1302 subjects} &
          \multicolumn{1}{c}{\multirow{2}[2]{*}{FDG PET}}
          \\
         &
           &
           &
          \multicolumn{1}{c|}{(243 AD, 666 MCI, 393 CN)} &
          
          \\
        \bottomrule
        \end{tabular}%
      \label{tab:12}%
    \end{table*}%


    \begin{table*}[htbp]
      \centering
      \caption{(Continued)}
        \begin{tabular}{c|c|p{10em}|p{16.5em}|p{17em}|}
        \toprule
        \multicolumn{1}{c|}{\textbf{Ref.}} &
          \multicolumn{1}{c|}{\textbf{Year}} &
          \multicolumn{1}{c|}{\textbf{Dataset}} &
          \multicolumn{1}{c|}{\textbf{Sample}} &
          \multicolumn{1}{c}{\textbf{Feature}}
          \\
        \midrule
        \multicolumn{1}{c|}{\multirow{2}[2]{*}{\cite{chitradevi2020analysis}}} &
          \multirow{2}[2]{*}{2020} &
          \multicolumn{1}{c|}{Chettinad Health City,} &
          \multicolumn{1}{c|}{200 subjects} &
          \multicolumn{1}{c}{T2 flair weighted MRI}
          \\
         &
           &
           \multicolumn{1}{c|}{Chennai} &
          \multicolumn{1}{c|}{(100 AD, 100 CN)} &
          \multicolumn{1}{c}{(Region-level: WM, GM, CC, HC)}
          \\
        \midrule
        \multicolumn{1}{c|}{\multirow{2}[2]{*}{\cite{puente2020automatic}}} &
          \multirow{2}[2]{*}{2020} &
          \multicolumn{1}{c|}{ADNI dataset} &
          \multicolumn{1}{c|}{{\multirow{2}[2]{*}{2179 subjects}}} &
          \multicolumn{1}{c}{\multirow{2}[2]{*}{Sagittal MRI + Age + Gender}}
          \\
         &
           &
          \multicolumn{1}{c|}{OASIS dataset} &
          \multicolumn{1}{c|}{{\multirow{2}[2]{*}{(436 from OASIS, 1743 from ADNI)}}}
          \\
          &
           &
          \multicolumn{1}{c|}{ImageNet dataset} &
          \multicolumn{1}{c|}{} 
          \\
        \midrule
        \multicolumn{1}{c|}{\multirow{4}[2]{*}{\cite{spasov2019parameter}}} &
          \multirow{4}[2]{*}{2019} &
          \multicolumn{1}{c|}{\multirow{4}[2]{*}{ADNI dataset}} &
          \multicolumn{1}{c|}{\multirow{3}[2]{*}{785 subjects}} &
          \multicolumn{1}{p{17em}}{CDRSB + ADAS11 + ADAS13 +}
          \\
         &
           &
           &
          \multicolumn{1}{c|}{\multirow{3}[2]{*}{(192 AD,181 pMCI, 228 sMCI, 184 CN)}} &
        \multicolumn{1}{p{17em}}{RAVLT + APOe4 + Demographics + MRI}
          \\
         &
           &
           &
           &
          \multicolumn{1}{p{17em}}{(Region-level: parietal, temporal and}
          \\
         &
           &
           &
           &
          \multicolumn{1}{p{17em}}{frontal lobes)}
          \\
        \midrule
        \multicolumn{1}{c|}{\cite{janghel2021deep}} &
          2021 &
          \multicolumn{1}{c|}{ADNI dataset} &
          \multicolumn{1}{c|}{54 subjects (27 AD, 27 CN)} &
          \multicolumn{1}{c}{fMRI + PET}
          \\
        \midrule
        \multicolumn{1}{c|}{\multirow{2}[2]{*}{\cite{chen2021iterative}}} &
          \multirow{2}[2]{*}{2021} &
          \multicolumn{1}{c|}{\multirow{2}[2]{*}{ADNI dataset}} &
          \multicolumn{1}{c|}{1248 subjects} &
          \multicolumn{1}{c}{MRI}
          \\
         &
           &
           &
          \multicolumn{1}{c|}{(347 AD,179 pMCI, 305 sMCI, 417 CN)} &
          \multicolumn{1}{c}{(Vertex-level: cortical surface)}
          \\
        \midrule
        \multicolumn{1}{c|}{\cite{hazarika2022experimental}} &
          2023 &
          \multicolumn{1}{c|}{ADNI dataset} &
          \multicolumn{1}{c|}{210 subjects (70 AD, 70 MCI, 70 CN)} &
          \multicolumn{1}{c}{MRI (Skull stripped)}
          \\
        \midrule
        \multicolumn{1}{c|}{\multirow{2}[2]{*}{\cite{bae2021transfer}}} &
          \multirow{2}[2]{*}{2021} &
          \multicolumn{1}{c|}{\multirow{2}[2]{*}{ADNI dataset}} &
          \multicolumn{1}{c|}{1080 subjects} &
          \multicolumn{1}{c}{MRI}
          \\
         &
           &
           &
          \multicolumn{1}{c|}{(1406 AD scans, 2084 CN scans)} &
          \multicolumn{1}{c}{(Sagittal + coronal + transverse view)}
          \\
        \midrule
        \multicolumn{1}{c|}{\multirow{2}[2]{*}{\cite{zhang2021explainable}}} &
          \multirow{2}[2]{*}{2022} &
          \multicolumn{1}{|c|}{\multirow{2}[2]{*}{ADNI dataset}} &
          \multicolumn{1}{c|}{1407 subjects} &
          \multicolumn{1}{c}{\multirow{2}[2]{*}{MRI (Skull stripped)}}
          \\
         &
           &
           &
          \multicolumn{1}{p{17em}|}{(353 AD,172 pMCI, 232 sMCI, 650 CN)} &
          \multicolumn{1}{c}{}
          \\
        \midrule
        \multicolumn{1}{c|}{\multirow{4}[2]{*}{\cite{raza2019diagnosis}}} &
          \multirow{4}[2]{*}{2019} &
          \multicolumn{1}{c|}{ADNI dataset} &
          \multicolumn{1}{c|}{ADNI: 432 subjects (232 AD, 200 CN)} &
          \multicolumn{1}{l}{MRI (Coronal slices) +} 
          \\
         &
           &
          \multicolumn{1}{c|}{OASIS dataset} &
          \multicolumn{1}{c|}{OASIS: 416 subjects (100 AD,316 CN)} &
          \multicolumn{1}{l}{3D acceleration signal data +}
          \\
         &
           &
          \multicolumn{1}{c|}{Single-sensor dataset} &
          \multicolumn{1}{c|}{Single-sensor: 30 subjects} &
           \multicolumn{1}{l}{3D angular velocity data}
          \\
         &
           &
          \multicolumn{1}{c|}{Multi-sensor dataset} &
          \multicolumn{1}{c|}{Multi-sensor: 19 subjects} &
          \multicolumn{1}{r}{}
          \\
        \midrule
        \multicolumn{1}{c|}{\multirow{2}[2]{*}{\cite{huang2019diagnosis}}} &
          \multirow{2}[2]{*}{2019} &
          \multicolumn{1}{c|}{\multirow{2}[2]{*}{ADNI dataset}} &
          \multicolumn{1}{c|}{2145 subjects} &
          \multicolumn{1}{l}{MRI (Region-level: HC) +}
          \\
         &
           &
           &
          \multicolumn{1}{c|}{(647 AD,326 pMCI, 441 sMCI, 731 CN)} &
          \multicolumn{1}{l}{PET (Region-level: HC/HC + Cortices)}
          \\
        \midrule
        \multicolumn{1}{c|}{\cite{wang2019ensemble}} &
          2019 &
          \multicolumn{1}{c|}{ADNI dataset} &
          \multicolumn{1}{c|}{624 subjects} &
          \multicolumn{1}{c}{MRI (Voxel-level: whole volume)}
          \\
        \midrule
        \multicolumn{1}{c|}{\multirow{2}[2]{*}{\cite{zhang2019multi}}} &
          \multirow{2}[2]{*}{2019} &
          \multicolumn{1}{c|}{\multirow{2}[2]{*}{ADNI dataset}} &
          \multicolumn{1}{c|}{392 subjects} &
          \multicolumn{1}{l}{MRI (2D slices of different views) + }
          \\
         &
           &
           &
          \multicolumn{1}{c|}{(91 AD, 200 MCI, 101 CN)} &
          \multicolumn{1}{l}{PET + MMSE + CDR}
          \\
        \midrule
        \multicolumn{1}{c|}{\multirow{2}[2]{*}{\cite{liu2019weakly}}} &
          \multirow{2}[2]{*}{2020} &
          \multicolumn{1}{c|}{\multirow{2}[2]{*}{ADNI dataset}} &
          \multicolumn{1}{c|}{\multirow{2}[2]{*}{1469 subjects}} &
          \multicolumn{1}{l}{MRI (Patch-level) + CDR-SB + ADAS-}
          \\
         &
           &
           &
           &
          \multicolumn{1}{l}{Cog11 + ADAS-Cog13 + MMSE}
          \\
        \midrule
        \multicolumn{1}{c|}{\multirow{3}[2]{*}{\cite{abuhmed2021robust}}} &
          \multirow{3}[2]{*}{2021} &
          \multicolumn{1}{c|}{\multirow{3}[2]{*}{ADNI dataset}} &
          \multicolumn{1}{c|}{\multirow{3}[2]{*}{1371 subjects}} &
          \multicolumn{1}{l}{MRI + PET + Cognitive sub-scores +}
          \\
         &
           &
           &
          \multicolumn{1}{c|}{\multirow{3}[2]{*}{(339 AD,140 pMCI, 473 sMCI, 419 CN)}} &
           \multicolumn{1}{l}{APOE4 + Neuropathology data+} 
          \\
         &
           &
           &
           &
          \multicolumn{1}{l}{Neuropsychological battery + CSF}
          \\
        \midrule
        \multicolumn{1}{c|}{\multirow{2}[2]{*}{\cite{lei2022predicting}}} &
          \multirow{2}[2]{*}{2021} &
          \multicolumn{1}{c|}{\multirow{2}[2]{*}{ADNI dataset}} &
          \multicolumn{1}{c|}{\multirow{2}[2]{*}{805 subjects}} &
          \multicolumn{1}{l}{MRI (Region-level: ROI volume) +}
          \\
         &
           &
           &
           &
          \multicolumn{1}{l}{Cognitive sub-scores}
          \\
        \midrule
        \multicolumn{1}{c|}{\multirow{3}[2]{*}{\cite{ghazi2019training}}} &
          \multirow{3}[2]{*}{2019} &
          \multicolumn{1}{c|}{\multirow{3}[2]{*}{ADNI dataset}} &
          \multicolumn{1}{p{16.5em}|}{\multirow{2}[2]{*}{1737 subjects，12741 visits (2745 CN,}} &
          \multicolumn{1}{l}{MRI (Region-level: Ventricles + HC + }
          \\
         &
           &
           &
          \multicolumn{1}{p{17em}|}{\multirow{2}[2]{*}{ 4058 MCI, 2108 AD, 3830 unlabeled)}} &
          \multicolumn{1}{l}{Whole brain + Entorhinal cortex + }
          \\
         &
           &
           &
           &
          \multicolumn{1}{l}{Fusiform + Middle temporal gyrus)}
          \\
        \midrule
        \multicolumn{1}{c|}{\multirow{2}[2]{*}{\cite{ieracitano2020novel}}} &
          \multirow{2}[2]{*}{2019} &
          \multicolumn{1}{p{11.5em}|}{IRCCS Centro Neurolesi} &
          \multicolumn{1}{c|}{\multirow{2}[2]{*}{189 subjects}} &
          \multicolumn{1}{c}{\multirow{2}[2]{*}{EEG}}
          \\
         &
           &
          \multicolumn{1}{p{11.5em}|}{Bonino-Pulejo of Messina (Italy)} &
          \multicolumn{1}{c|}{\multirow{2}[2]{*}{(63 AD, 63 MCI, 63 CN)}} &
          \multicolumn{1}{c}{}
          \\
        \midrule
        \multicolumn{1}{c|}{\cite{bi2019early}} &
          2019 &
          \multicolumn{1}{p{11.5em}|}{Beijing Easy monitor Technology} &
          \multicolumn{1}{c|}{\multirow{2}[2]{*}{12 subjects (4 AD, 4 MCI, 4 CN)}} &
          \multicolumn{1}{c}{\multirow{2}[2]{*}{EEG}}
          \\
        \midrule
        \multicolumn{1}{c|}{\cite{choi2019deep}} &
          2019 &
          \multicolumn{1}{c|}{ADNI dataset} &
          \multicolumn{1}{c|}{1303 subjects (243 AD,667 MCI, 393 CN)} &
          \multicolumn{1}{c}{PET + Age}
          \\
        \midrule
        \multicolumn{1}{c|}{\multirow{2}[2]{*}{\cite{kang2021multi}}} &
          \multirow{2}[2]{*}{2021} &
          \multicolumn{1}{c|}{\multirow{2}[2]{*}{ADNI dataset}} &
          \multicolumn{1}{c|}{\multirow{2}[2]{*}{798 subjects (187 AD, 382 MCI, 229 CN)}} &
          \multicolumn{1}{l}{MRI (Region-level:}
          \\
         &
           &
           &
           &
           \multicolumn{1}{l}{Coronal slices of GM, WM and CSF)}
          \\
        \bottomrule
        \end{tabular}%
      \label{tab:13}%
    \end{table*}%



    \begin{table*}[htbp]
      \centering
      \caption{(Continued)}
        \begin{tabular}{c|c|p{10em}|p{17em}|p{17em}}
        \toprule
        \multicolumn{1}{c|}{\textbf{Ref.}} &
          \multicolumn{1}{c|}{\textbf{Year}} &
          \multicolumn{1}{c|}{\textbf{Dataset}} &
          \multicolumn{1}{c|}{\textbf{Sample}} &
          \multicolumn{1}{c}{\textbf{Feature}}
          \\
        \midrule
        \multirow{2}[2]{*}{\cite{kim2020slice}} &
          \multirow{2}[2]{*}{2020} &
          \multicolumn{1}{c|}{ADNI dataset} &
          ADNI: 486 subjects (139 AD, 347 CN) &
          \multicolumn{1}{c}{\multirow{2}[2]{*}{PET}}
          \\
        \multicolumn{1}{c|}{} &
           &
          \multicolumn{1}{c|}{Severance Hospital} &
          Severance: 141 subjects (73 AD, 68 CN) &
          \multicolumn{1}{c}{}
          \\
        \midrule
        \multirow{1}[1]{*}{\cite{mendoza2020single}} &
          2020 &
          \multicolumn{1}{c|}{OASIS dataset} &
          \multicolumn{1}{c|}{174 subjects (87 AD, 87 CN)} &
          \multicolumn{1}{c}{MRI (Patch-level)}
          \\
        \midrule
        \multirow{2}[2]{*}{\cite{lei2020deep}} &
          \multirow{2}[2]{*}{2020} &
          \multicolumn{1}{c|}{\multirow{2}[2]{*}{ADNI dataset}} &
          \multicolumn{1}{c|}{\multirow{2}[2]{*}{805 subjects}} &
          \multicolumn{1}{c}{(MRI + Cognitive sub-scores)}
          \\
        \multicolumn{1}{c|}{} &
           &
          \multicolumn{1}{c|}{} &
          \multicolumn{1}{c|}{} &
          \multicolumn{1}{c}{(Longitudinal data)}
          \\
        \midrule
        \multirow{2}[2]{*}{\cite{afzal2019data}} &
          \multirow{2}[2]{*}{2019} &
         \multicolumn{1}{c|}{OASIS dataset} &
          \multicolumn{1}{c|}{OASIS：150 subjects (75 CN, 75 AD)} &
          \multicolumn{1}{c}{\multirow{2}[2]{*}{MRI}}
          \\
        \multicolumn{1}{c|}{} &
           &
          \multicolumn{1}{c|}{ImageNet dataset} &
          \multicolumn{1}{c|}{ImageNet: ＞1.2 billion labeled images} &
          \multicolumn{1}{c}{}
          \\
        \midrule
        \multirow{2}[2]{*}{\cite{jain2019convolutional}} &
          \multirow{2}[2]{*}{2019} &
          \multicolumn{1}{c|}{ADNI dataset} &
          \multicolumn{1}{c|}{ADNI: 150 subjects} &
          \multicolumn{1}{c}{\multirow{2}[2]{*}{MRI (The most informative slices)}}
          \\
        \multicolumn{1}{c|}{} &
           &
          \multicolumn{1}{c|}{ImageNet dataset} &
          \multicolumn{1}{c|}{(50 AD, 50 MCI, 50 CN)} &
          \multicolumn{1}{c}{}
          \\
        \midrule
        \multicolumn{1}{c|}{\cite{2019Volumetrie}} &
          2019 &
          \multicolumn{1}{c|}{ADNI dataset} &
          \multicolumn{1}{c|}{706 subjects (279 AD, 427 CN)} &
          \multicolumn{1}{c}{MRI (Voxel-level)}
          \\
        \midrule
        \multirow{2}[2]{*}{\cite{barbaroux2020encoding}} &
          \multirow{2}[2]{*}{2020} &
          \multicolumn{1}{c|}{\multirow{2}[2]{*}{ADNI dataset}} &
          \multicolumn{1}{c|}{589 subjects} &
          \multicolumn{1}{c}{\multirow{2}[2]{*}{MRI (Vertex-level: cortical surface)}}
          \\
        \multicolumn{1}{c|}{} &
           &
          \multicolumn{1}{c|}{} &
          \multicolumn{1}{c|}{(188 AD, 136 pMCI,114 sMCI, 151 CN)} &
          \multicolumn{1}{c}{}
          \\
        \midrule
        \multirow{3}[2]{*}{\cite{tufail2020binary}} &
          \multirow{3}[2]{*}{2020} &
          \multicolumn{1}{c|}{\multirow{2}[2]{*}{OASIS dataset}} &
          \multicolumn{1}{c|}{Dataset-1:180 subjects (90 AD, 90 CN)} &
          \multicolumn{1}{c}{\multirow{3}[2]{*}{Cross-sectional MRI}}
          \\
        \multicolumn{1}{c|}{} &
           &
          \multicolumn{1}{c|}{\multirow{2}[2]{*}{ImageNet dataset}} &
          Dataset-2:114 subjects (30 AD, 84 CN) &
          \multicolumn{1}{c}{}
          \\
        \multicolumn{1}{c|}{} &
           &
          \multicolumn{1}{r|}{} &
          Dataset-3:200 subjects (100 AD,100 CN) &
          \multicolumn{1}{c}{}
          \\
        \midrule
        \multirow{2}[2]{*}{\cite{2020Diagnosis}} &
          \multirow{2}[2]{*}{2020} &
          \multicolumn{1}{c|}{\multirow{2}[2]{*}{ADNI dataset}} &
          \multicolumn{1}{c|}{1000 subjects} &
          \multicolumn{1}{c}{\multirow{2}[2]{*}{Sagittal MRI}}
          \\
        \multicolumn{1}{c|}{} &
           &
          \multicolumn{1}{c|}{} &
          (250 AD, 250 pMCI, 250 sMCI, 250 CN) &
          \multicolumn{1}{c}{}
          \\
        \midrule
        \multirow{1}[1]{*}{\cite{2020Identifying}} &
          2020 &
          \multicolumn{1}{c|}{ADNI dataset} &
          \multicolumn{1}{c|}{120 subjects (70 sMCI, 50 CN)} &
          \multicolumn{1}{c}{MRI + DTI}
          \\
        \bottomrule
        \end{tabular}%
      \label{tab:14}%
    \end{table*}%


\subsection{Modality and method}
\subsubsection{Modality}

\quad In the compiled literature, the modalities involved include images, voiceprints, posture/gait, cognition, biomarkers (genes, cerebrospinal fluid, etc.), and demographic characteristics. Sometimes these modalities are further divided into different types of images, such as MRI, PET, DTI, EEG, or different biomarkers, such as cerebrospinal fluid, SNP, APOE4, Amyloid-β protein. Therefore, some papers refer to the combination of multiple types of images or biomarkers as multimodal fusion too. MRI, PET images, and cognitive assessment test scores are the most commonly used. Cognitive assessment test scores are often used with a certain cutoff value to determine if a person is an AD patient\cite{MMSEvsMoCA}. They can also be used to predict the change in test scores at future time points through regression, to determine if MCI subjects will convert to AD patients\cite{li2019deep}. Several attempts have been made to combine images and cognition, demonstrating the superiority of both modalities\cite{0Development}\cite{2021An}\cite{zhang2019multi}\cite{liu2019weakly}\cite{lei2022predicting}\cite{lei2020deep}. Since general datasets often include demographic data, this modality is commonly used as an auxiliary modality, combined with other modalities to exchange feature information and achieve comprehensive feature capture\cite{spasov2019parameter}. The fields of voiceprints and posture/gait are currently less explored, with most research focusing on a single modality. However, in \cite{raza2019diagnosis}, the combination of posture/gait and image modalities achieved excellent classification results with accuracy above 95\% on several datasets, providing new ideas for future research to explore unexplored combinations of multiple modalities. There is no strict differentiation between genes and biomarkers, and they are often used to describe the same type of features. Studies that only use gene-level features are relatively scarce, such as \cite{park2020prediction}, and they tend to focus more on clinical perspectives rather than computer-aided approaches, which do not meet the requirements of contemporary automated recognition. Therefore, genes and biomarkers are also often used as one of the modalities in multimodal fusion, adding richness to the data structure. The most common combination is still with image, cognition, and demographic modalities\cite{li2019deep}\cite{spasov2019parameter}\cite{abuhmed2021robust}.

\subsubsection{Method}\par







\quad In the compiled literature, several commonly used deep learning architectures are frequently applied. 

\quad Deep Feedforward Neural Network (DFFNN): DFFNN consists of multiple fully connected layers. Due to its high computational cost, it can be challenging to handle high-dimensional input data. Therefore, preprocessing of the input data is often required for dimensionality reduction. For example, in \cite{park2020prediction}, the Limma package was used for feature extraction, \cite{zhou2019effective} extracting features from predefined regions of interest (ROIs), while \cite{liu2019using} focused on volume features of various ROIs in MRI. \cite{lee2019toward} employed a divide-and-conquer strategy by segmenting the gray matter (GM) into sub-blocks and using the voxels as input features for weak classifiers in DFFNN.

\quad Convolutional Neural Network (CNN): CNN is the most widely used network architecture. In practical research, transfer learning methods are often employed to improve efficiency and reduce costs by using pre-trained 2D CNN models directly for classification tasks. This requires converting the neuroimages into appropriate 2D slices, as mentioned in the "Feature Selection" section. \cite{jain2019convolutional} provided specific guidelines for selecting the most informative slices. Common input information for images includes voxel-level, patch-level, region-level, 2D slice and 1D numerical volume features. When dealing with multimodal information, preprocessing is often applied to concatenate the selected features as input, as done in \cite{li2019deep}, \cite{spasov2019parameter}, \cite{zhang2019multi} and \cite{liu2019weakly}.

\quad Recurrent Neural Network (RNN): RNN is well-suited for utilizing sequential information. For example, \cite{li2020detecting} analyzed fMRI data as a time series of neuroimages, aligning with the concept framework of RNN for processing sequential data. \cite{lei2022predicting} predicted the cognitive test scores at a target moment by inputting the scores of previous time periods, allowing for the prediction of whether MCI patients will convert to AD. \cite{ghazi2019training} used Long Short-Term Memory (LSTM) to handle missing values in incomplete data. \cite{abuhmed2021robust} utilized bidirectional LSTM by inputting images and cognitive modal data from four different time points to predict AD progression. Compared to the convolutional nature of CNN, RNN is less compatible with image data and is more suited for sequences such as speech, text and gait. However, the literature reviewed here does not include applications of RNN in these modalities. When neuroimages are decomposed into slices, patches, regions or other sub-elements, RNN can capture the spatial dependencies among these sub-elements.

\quad Generative Networks (GN): \cite{bi2019early} used a deep Boltzmann machine (DCssCDBM), with integrated clear label information to guide feature extraction. \cite{choi2019deep} employed a Variational Autoencoder (VAE) to define anomaly scores based on the distance between given brain images and normal data. The model was trained using only normal brain PET images to identify abnormal patterns. \cite{kang2021multi} utilized Generative Adversarial Networks (GANs) as one of the weak classifiers in the voting decision. \cite{kim2020slice} employed Boundary Equilibrium GAN (BEGAN) with the encoder part of the discriminator network serving as input for an SVM classifier. \cite{mendoza2020single} applied Supervised Switching Autoencoders (multi-channel reconstruction AEs) to individual slices of MRI patches and made predictions based on majority voting.

\quad Deep Polynomial Network (DPN): DPN requires prior feature selection/extraction to find the most discriminative input features for training the network. For example, \cite{lei2020deep} used LASSO for feature extraction and then employed DPN for feature encoding.

\quad This paper compiles the "Modeling methods" content from various studies in this field in recent years, as shown in Table \ref{tab:21}-\ref{tab:25}.


    \begin{sidewaystable*}[htbp]
      \centering
      \caption{Modality and method}
        \begin{tabular}{c|c|p{8em}|p{26em}|p{25em}}
        \toprule
        \multicolumn{1}{c|}{\textbf{Ref.}} &
          \multicolumn{1}{c|}{\textbf{Year}} &
          \multicolumn{1}{c|}{\textbf{Modality}} &
          \multicolumn{1}{c|}{\textbf{Method}} &
          \multicolumn{1}{c}{\textbf{Comment}}
          \\
        \midrule
        \multicolumn{1}{c|}{\multirow{3}[2]{*}{\cite{MMSEvsMoCA}}} &
          \multirow{3}[2]{*}{2020} &
          \multicolumn{1}{c|}{\multirow{3}[2]{*}{Cognition}} &
          Diagnostic accuracy and optimal cut-off scores were calculated &
          \multicolumn{1}{p{25em}}{The MoCA is more sensitive and accurate. Complementing}
          \\
         &
           &
           &
           for each measure, and a method for converting MoCA to MMSE&
          \multicolumn{1}{p{25em}}{the MMSE or the MoCA with the DSRS significantly}
          \\
         &
           &
           &
          scores was presented. &
          \multicolumn{1}{p{25em}}{improved diagnostic accuracy.}
          \\
        \midrule
        \multicolumn{1}{c|}{\multirow{3}[2]{*}{\cite{2010Clin}}} &
          \multirow{3}[2]{*}{2022} &
          \multicolumn{1}{p{8em}|}{Biomarker + Image} &
          Constructed subject similarity networks for all modalities and &
          \multicolumn{1}{p{25em}}{It modeled each modality as a layer of a multidimensional}
          \\
         &
           &
          \multicolumn{1}{c|}{+ Cognition} &
           applied multilayer community detection to find groups with  &
          \multicolumn{1}{p{25em}}{network, so it could be equipped to handle interactions across}
          \\
         &
           &
          \multicolumn{1}{c|}{+ Genetics} &
          shared similarities across modalities. &
          \multicolumn{1}{p{25em}}{modalities and find relation which might had been overlooked.}
          \\
        \midrule
        \multicolumn{1}{c|}{\multirow{3}[2]{*}{\cite{ghoraani2021detection}}} &
          \multirow{3}[2]{*}{2021} &
          \multicolumn{1}{c|}{\multirow{3}[2]{*}{Gait}} &
          A SVM model in a one-vs-one manner was trained for each  &
          \multicolumn{1}{p{25em}}{Gait-based cognitive screening can be easily adapted into}
          \\
         &
           &
           &
          classification, and the majority vote of the three models was  &
          \multicolumn{1}{p{25em}}{clinical settings.}
          \\
         &
           &
           &
          assigned as CN, MCI, or AD. &
          
          \\
        \midrule
        \multicolumn{1}{c|}{\multirow{2}[2]{*}{\cite{mh_1}}} &
          \multirow{2}[2]{*}{2020} &
          \multicolumn{1}{p{8em}|}{Acoustic/Linguistic } &
          Using the spectrogram features extracted from speech data  &
          \multicolumn{1}{p{25em}}{Patient’s speech data can effectively reduce the medical cost,}
          \\
         &
           &
          \multicolumn{1}{c|}{data} &
          and machine learning model LogisticRegressionCV. &
          \multicolumn{1}{p{25em}}{and the data can be collected in real-time and accurately.}
          \\
        \midrule
        \multicolumn{1}{c|}{\multirow{3}[2]{*}{\cite{haider2019assessment}}} &
          \multirow{3}[2]{*}{2021} &
          \multicolumn{1}{p{8em}|}{\multirow{2}[2]{*}{Acoustic/Linguistic }} &
          Using a new active data representation (ADR) method for  &
          \multicolumn{1}{p{25em}}{A more comprehensive acoustic feature set is able to capture}
          \\
         &
           &
          \multicolumn{1}{c|}{\multirow{2}[2]{*}{data}} &
          feature extraction, together with “hard fusion” of feature sets to  &
          \multicolumn{1}{p{25em}}{a wider range of speech subtleties. Besides, a larger number}
          \\
         &
           &
           &
          identify AD through decision tree. &
          \multicolumn{1}{p{25em}}{of features allows for the use of more sophisticated classifiers.}
          \\
        \midrule
        \multicolumn{1}{c|}{\multirow{2}[2]{*}{\cite{luz2020alzheimer}}} &
          \multirow{2}[2]{*}{2020} &
          \multicolumn{1}{p{8em}|}{Acoustic/Linguistic } &
          Operating linear discriminant analysis (LDA) on automatically  &
          \multicolumn{1}{c}{\multirow{2}[2]{*}{One strength of the method is its speaker independent nature.}}
          \\
         &
           &
          \multicolumn{1}{c|}{data} &
          extracted voice features to identify AD. &
          
          \\
        \midrule
        \multicolumn{1}{c|}{\multirow{3}[2]{*}{\cite{shi2017multimodal}}} &
          \multirow{3}[2]{*}{2018} &
          \multicolumn{1}{c|}{\multirow{3}[2]{*}{Image}} &
          Two SDPNs were first used to learn high-level features of MRI  &
          \multicolumn{1}{p{25em}}{The proposed MM-SDPN can be a powerful representation }
          \\
         &
           &
           &
          and PET, respectively, which were then fed to another SDPN to  &
          \multicolumn{1}{p{25em}}{algorithm for not only multimodal neuroimaging data but also }
          \\
         &
           &
           &
          fuse multimodal neuroimaging information (MM-SDPN). &
          \multicolumn{1}{p{25em}}{other medical data.}
          \\
        \midrule
        \multicolumn{1}{c|}{\multirow{3}[2]{*}{\cite{muti_image_2}}} &
          \multirow{3}[2]{*}{2018} &
          \multicolumn{1}{c|}{\multirow{3}[2]{*}{Image}} &
          Using a novel DNN based method that utilized multi-scale and  &
          \multicolumn{1}{p{25em}}{The approach preserved the structural and metabolism }
          \\
         &
           &
           &
          multi-modal information (MMDNN) combining metabolism  &
          \multicolumn{1}{p{25em}}{information at multiple scales.}
          \\
         &
           &
           &
          (FDG-PET) and regional volume (T1-MRI) &
          
          \\
        \midrule
        \multicolumn{1}{c|}{\multirow{3}[2]{*}{\cite{qiu2020development}}} &
          \multirow{3}[2]{*}{2020} &
          \multicolumn{1}{c|}{\multirow{3}[2]{*}{Image + Cognition}} &
          Developing a deep learning framework which linked a fully  &
          \multicolumn{1}{p{25em}}{The intuitive local probabilities outputted by the model were }
          \\
         &
           &
           &
          convolutional network (FCN) to a multilayer perceptron (MLP)  &
          \multicolumn{1}{p{25em}}{readily interpretable.}
          \\
         &
           &
           &
          and generates high resolution disease probability map. &
          
          \\
        \midrule
        \multicolumn{1}{c|}{\multirow{2}[2]{*}{\cite{2018Hierarchical}}} &
          \multirow{2}[2]{*}{2020} &
          \multicolumn{1}{c|}{\multirow{2}[2]{*}{Image}} &
          Proposing a hierarchical fully convolutional network (H-FCN)&
          \multicolumn{1}{p{25em}}{The method had demonstrated better or at least comparable }
          \\
         &
           &
           &
          to identify discriminative patches and regions in sMRI. &
          \multicolumn{1}{p{25em}}{performance, especially in task of pMCI vs sMCI.}
          \\
        \midrule
        \multicolumn{1}{c|}{\multirow{3}[2]{*}{\cite{ebrahimi2021deep}}} &
          \multirow{3}[2]{*}{2021} &
          \multicolumn{1}{c|}{\multirow{3}[2]{*}{Image}} &
          The CNN was ResNet-18 pre-trained on an ImageNet dataset.  &
          \multicolumn{1}{p{25em}}{The proposed models can be applied to any other image-based }
          \\
         &
           &
           &
          The employed sequence-based models were the temporal  &
          \multicolumn{1}{p{25em}}{tasks}
          \\
         &
           &
           &
          convolutional network (TCN) and different types of RNN. &
          
          \\
        \midrule
        \multicolumn{1}{c|}{\multirow{2}[2]{*}{\cite{colliot2008discrimination}}} &
          \multirow{2}[2]{*}{2008} &
          \multicolumn{1}{c|}{\multirow{2}[2]{*}{Image}} &
          Automatically segmenting the Hippocampus with high spatial  &
          \multicolumn{1}{p{25em}}{This automated method can serve as an alternative to manual }
          \\
         &
           &
           &
          resolution, then adapting bootstrap method. &
          \multicolumn{1}{p{25em}}{tracing.}
          \\
        \midrule
        \multicolumn{1}{c|}{\multirow{3}[2]{*}{\cite{multimodal_e}}} &
          \multirow{3}[2]{*}{2022} &
          \multicolumn{1}{c|}{\multirow{3}[2]{*}{Image + Genetics}} &
          Using stacked denoising auto-encoders for EHR and SNP, and  &
          \multicolumn{1}{p{25em}}{Novel perturbation and clustering-based feature extraction }
          \\
         &
           &
           &
          3D CNNs for MRI. Identifying top-performing features learned  &
          \multicolumn{1}{p{25em}}{assisting DL model interpretations were capable of AD stage }
          \\
         &
           &
           &
          by the deep-models with clustering and perturbation analysis &
          \multicolumn{1}{p{25em}}{prediction.}
          \\
        \midrule
        \multicolumn{1}{c|}{\multirow{4}[2]{*}{\cite{zeng2021new}}} &
          \multirow{4}[2]{*}{2021} &
          \multicolumn{1}{c|}{\multirow{4}[2]{*}{Image + Cognition}} &
          Proposing a new framework based on the developed DBN- &
          \multicolumn{1}{p{25em}}{MTL strategy could further improve the performance by}
          \\
         &
           &
           &
          based MTL algorithm. The PCA and MTFS were introduced for  &
          \multicolumn{1}{p{25em}}{taking into account the correlation and difference among}
          \\
         &
           &
           &
          the feature selection, and the dropout technique together with  &
          \multicolumn{1}{p{25em}}{multiple tasks.}
          \\
         &
           &
           &
          the zero-masking technology was utilized. &
          
          \\
        \bottomrule
        \end{tabular}%
      \label{tab:21}%
    \end{sidewaystable*}%


    \begin{sidewaystable*}[htbp]
      \centering
      \caption{(Continued)}
        \begin{tabular}{c|c|p{5em}|p{26em}|p{25em}}
        \toprule
        \multicolumn{1}{c|}{\textbf{Ref.}} &
          \multicolumn{1}{c|}{\textbf{Year}} &
          \multicolumn{1}{c|}{\textbf{Modality}} &
          \multicolumn{1}{c|}{\textbf{Method}} &
          \multicolumn{1}{c}{\textbf{Comment}}
          \\
        \midrule
        \multicolumn{1}{c|}{\multirow{4}[2]{*}{\cite{2021ADVIAN}}} &
          \multirow{4}[2]{*}{2021} &
          \multicolumn{1}{c|}{\multirow{4}[2]{*}{Image}} &
          Proposing a novel VGG-inspired network as the mainstay and  &
          \multicolumn{1}{c}{\multirow{4}[2]{*}{The model did not go through strict clinical environment tests.}}
          \\
         &
           &
           &
          combining the attention mechanism with VIN to produce a  &
          
          \\
         &
           &
           &
          new ADVIAN deep-learning model. The 18-way DA was  &
          
          \\
         &
           &
           &
          harnessed to prevent overfitting in the training set. &
          
          \\
        \midrule
        \multicolumn{1}{c|}{\multirow{4}[2]{*}{\cite{2021Dual}}} &
          \multirow{4}[2]{*}{2021} &
          \multicolumn{1}{c|}{\multirow{4}[2]{*}{Image}} &
          Proposing a dual attention multi-instance deep learning network &
          \multicolumn{1}{p{25em}}{The size of input patches was fixed.However, the structural}
          \\
         &
           &
           &
          (DA-MIDL) including: 1) PatchNets with spatial attention &
          \multicolumn{1}{p{25em}}{changed in the cerebrum caused by brain atrophy may occur}
          \\
         &
           &
           &
           blocks,2) an attention MIL pooling operation, and 3) an &
          \multicolumn{1}{p{25em}}{across multiple regions with different scales.Using the fixed}
          \\
         &
           &
           &
          attention-aware global classifier. &
          \multicolumn{1}{p{25em}}{size could not represent various local features.}
          \\
        \midrule
        \multicolumn{1}{c|}{\multirow{3}[2]{*}{\cite{martinez2019studying}}} &
          \multirow{3}[2]{*}{2020} &
          \multicolumn{1}{c|}{\multirow{3}[2]{*}{Image}} &
          Proposing a deep convolutional autoencoder (CAE) architecture, &
          \multicolumn{1}{p{25em}}{It was a complete data-driven approach, self-supervised so}
          \\
         &
           &
           &
           a tool that could perform an automatic non-linear decomposition &
          \multicolumn{1}{p{25em}}{that the final result was guided only by the particularities of}
          \\
         &
           &
           &
        of a very large dataset (over 2000 images). &
          \multicolumn{1}{p{25em}}{the dataset, and very little pre-processing was required.}
          \\
        \midrule
        \multicolumn{1}{c|}{\multirow{3}[2]{*}{\cite{lin2021bidirectional}}} &
          \multirow{3}[2]{*}{2021} &
          \multicolumn{1}{c|}{\multirow{3}[2]{*}{Image}} &
          Adopting the Reversible Generative Adversarial Network  &
          \multicolumn{1}{p{25em}}{When the data was missing, the performance of AD diagnosis}
          \\
         &
           &
           &
          (RevGAN) to reconstruct the missing data, the 3D CNN with  &
          \multicolumn{1}{p{25em}}{and MCI conversion prediction could be significantly}
          \\
         &
           &
           &
          multi-modality input to classify AD. &
          \multicolumn{1}{p{25em}}{improved using this method.}
          \\
        \midrule
        \multicolumn{1}{c|}{\multirow{3}[2]{*}{\cite{bi2020multimodal}}} &
          \multirow{3}[2]{*}{2020} &
          \multicolumn{1}{c|}{\multirow{2}[2]{*}{Image + Genetics}} &
          Applying the correlation analysis to detect the associations  &
          \multicolumn{1}{p{25em}}{The proposed model found out the abnormal brain regions}
          \\
         &
           &
           &
          between brain regions and genes, then proposing CERF to  &
          \multicolumn{1}{p{25em}}{and pathogenic genes of AD such as thalamic, lingual gyrus,}
          \\
         &
           &
           &
          extract the discriminative fusion features between CN and AD. &
          \multicolumn{1}{p{25em}}{angular gyrus, precuneus, insula DAB1 and LRP1B gene.}
          \\
        \midrule
        \multicolumn{1}{c|}{\multirow{4}[2]{*}{\cite{matsuura2019statistical}}} &
          \multirow{4}[2]{*}{2019} &
          \multicolumn{1}{c|}{\multirow{4}[2]{*}{Gait}} &
          Utilizing Kinect device for collecting gait data to extract 12  &
          \multicolumn{1}{p{25em}}{The combination of single and dual-task performance was}
          \\
         &
           &
           &
          features of single and dual-task performance, then developing  &
          \multicolumn{1}{p{25em}}{useful for MMSE score classification,}
          \\
         &
           &
           &
          a method for automatic dementia score estimation to  &
          
          \\
         &
           &
           &
          investigate which characteristics are the most important. &
          
          \\
        \midrule
        \multicolumn{1}{c|}{\multirow{4}[2]{*}{\cite{li2015robust}}} &
          \multirow{4}[2]{*}{2015} &
          \multicolumn{1}{c|}{\multirow{2}[2]{*}{Image}} &
          Utilizing the dropout technique to to prevent weight co- &
          \multicolumn{1}{p{25em}}{Classical deep learning did not perform well on this small}
          \\
         &
           &
          \multicolumn{1}{c|}{\multirow{2}[2]{*}{+ Cognition}} &
          adaptation. In addition, incorporating stability selection, an  &
          \multicolumn{1}{p{25em}}{dataset, but with the dropout technique, the average accuracy}
          \\
         &
           &
          \multicolumn{1}{c|}{\multirow{2}[2]{*}{+ Biomarker}} &
          adaptive learning factor, and a multi-task learning strategy  &
          \multicolumn{1}{p{25em}}{was improved by 5.9\% on average.}
          \\
         &
           &
           &
          into restricted Boltzmann machine (RBM). &
          
          \\
        \midrule
        \multicolumn{1}{c|}{\multirow{3}[2]{*}{\cite{park2020prediction}}} &
          \multirow{3}[2]{*}{2019} &
          \multicolumn{1}{c|}{\multirow{3}[2]{*}{Genetics}} &
          Using Limma package to identify DEGs and DMPs, then  &
          \multicolumn{1}{p{25em}}{It is supposed that such study will receive much attention}
          \\
         &
           &
           &
          integrating them by intersecting, followed by a DFFNN whose  &
          \multicolumn{1}{p{25em}}{owiconnectionng to the possibility of biomedical explanation}
          \\
         &
           &
           &
          hyper-parameters were optimized by Bayesian Optimization &
          \multicolumn{1}{p{25em}}{and connection to the mechanism study.}
          \\
        \midrule
        \multicolumn{1}{c|}{\multirow{4}[2]{*}{\cite{zhou2019effective}}} &
          \multirow{4}[2]{*}{2019} &
          \multicolumn{1}{c|}{\multirow{4}[2]{*}{Image + Genetics}} &
          Learning latent representations for MRI, PET and Genetic data &
          \multicolumn{1}{p{25em}}{The method was focusing on using ROI features as input to}
          \\
         &
           &
           &
           by 3 DFFNNs, then leaning the resultant feature representations &
          \multicolumn{1}{p{25em}}{the deep learning model. However, such handcrafted features}
          \\
         &
           &
           &
           for each pair of modality combination by another 3 DFFNNs&
          \multicolumn{1}{p{25em}}{may limit the richness of structural and functional brain}
          \\
         &
           &
           &
            and finally, feeding all these feature maps into the last DFFNN. &
          \multicolumn{1}{p{25em}}{information from MRI and PET images}
          \\
        \midrule
        \multicolumn{1}{c|}{\multirow{2}[2]{*}{\cite{liu2019using}}} &
          \multirow{2}[2]{*}{2019} &
          \multicolumn{1}{c|}{\multirow{2}[2]{*}{Image}} &
          Siamese structure adopted on DFFNN to quantify the asymmetry &
          \multicolumn{1}{p{25em}}{Siamese networks had the advantage of reduced complexity}
          \\
         &
           &
           &
        of right and left brain hemispheres as a sign of MCI or AD &
          \multicolumn{1}{p{25em}}{and computational time.}
          \\
        \midrule
        \multicolumn{1}{c|}{\multirow{4}[2]{*}{\cite{lee2019toward}}} &
          \multirow{4}[2]{*}{2019} &
          \multicolumn{1}{c|}{\multirow{4}[2]{*}{Image}} &
          Segmenting GM into regions using atlas, then frequently and  &
          \multicolumn{1}{p{25em}}{This region-based abnormality method is visualizable and }
          \\
         &
           &
           &
          randomly selecting some of the voxels in each region and  &
          \multicolumn{1}{p{25em}}{interpretable.}
          \\
         &
           &
           &
          feeding them into DFFNN as some weak classifiers. Then  &
          
          \\
         &
           &
           &
          concatenating all the outputs into a vector. &
          
          \\
        \midrule
        \multicolumn{1}{c|}{\multirow{3}[2]{*}{\cite{cui2019rnn}}} &
          \multirow{3}[2]{*}{2019} &
          \multicolumn{1}{c|}{\multirow{3}[2]{*}{Image}} &
          Applying independent 3D CNNs for each time point and feeding  &
          \multicolumn{1}{p{25em}}{The proposed method is a data-driven method to jointly learn}
          \\
         &
           &
           &
          the output feature maps to 3 stacked bidirectional GRUs,conca- &
          \multicolumn{1}{p{25em}}{the spatial and longitudinal features and disease classifier}
          \\
         &
           &
           &
          -tenating all the outputs in an FC layer followed by Softmax. &
          \multicolumn{1}{p{25em}}{model.}
          \\
        \bottomrule
        \end{tabular}%
      \label{tab:22}%
    \end{sidewaystable*}%

%

    \begin{sidewaystable*}[htbp]
      \centering
      \caption{(Continued)}
        \begin{tabular}{c|c|p{7em}|p{25em}|p{25em}}
        \toprule
        \multicolumn{1}{c|}{\textbf{Ref.}} &
          \multicolumn{1}{c|}{\textbf{Year}} &
          \multicolumn{1}{c|}{\textbf{Modality}} &
          \multicolumn{1}{c|}{\textbf{Method}} &
          \multicolumn{1}{c}{\textbf{Comment}}
          \\
        \midrule
        \multicolumn{1}{c|}{\multirow{4}[2]{*}{\cite{ge2019multi}}} &
          \multirow{4}[2]{*}{2019} &
          \multicolumn{1}{c|}{\multirow{4}[2]{*}{Image}} &
          Independent multi-scale 3D CNN for GM, WM and CSF &
          \multicolumn{1}{l}{The dataset used in experiments was moderate in size,}
          \\
         &
           &
           &
          inputs to extract feature maps of different levels and applying &
          \multicolumn{1}{l}{which led to some overfitting.}
          \\
         &
           &
           &
           XGBoost on the concatenated features for determining the &
          \multicolumn{1}{r}{}
          \\
         &
           &
           &
          most important features followed FC and Softmax layers &
          \multicolumn{1}{r}{}
          \\
        \midrule
        \multicolumn{1}{c|}{\multirow{3}[2]{*}{\cite{li2019deep}}} &
          \multirow{3}[2]{*}{2019} &
          \multicolumn{1}{c|}{Image + Cognition + } &
          CNN to extract image features and combining them with non- &
          \multicolumn{1}{l}{It found that amyloid-positive MCI with molecular}
          \\
         &
           &
          \multicolumn{1}{c|}{Biomarker + Genetics} &
          image biomarkers for time-to-invent prediction by LASSO &
          \multicolumn{1}{l}{evidence of prodromal AD had higher predicted AD}
          \\
         &
           &
           &
           Cox regression for MCI subjects &
           \multicolumn{1}{l}{dementia progression risks than amyloid-negative MCI.}
          \\
        \midrule
        \multicolumn{1}{c|}{\multirow{4}[2]{*}{\cite{cui2018hippocampus}}} &
          \multirow{4}[2]{*}{2019} &
          \multicolumn{1}{c|}{\multirow{4}[2]{*}{Image}} &
          Applying Paralleled 3D DenseNet on the central patch and  &
          \multicolumn{1}{l}{The proposed method can make full use of the visual and}
          \\
         &
           &
           &
          MLPb on the shape description extracted by SPHARM-PDM  &
          \multicolumn{1}{l}{shape features of hippocampus.}
          \\
         &
           &
           &
          tool for bilateral Hippocampi and combining 4 resultant &
          \multicolumn{1}{r}{}
          \\
         &
           &
           &
          feature maps in FC layers followed by Softmax. &
          \multicolumn{1}{r}{}
          \\
        \midrule
        \multicolumn{1}{c|}{\multirow{3}[2]{*}{\cite{liu2020multi}}} &
          \multirow{3}[2]{*}{2020} &
          \multicolumn{1}{c|}{\multirow{3}[2]{*}{Image}} &
          Multi-task CNN with ResNet block to segment hippocampus  &
          \multicolumn{1}{l}{Multi-task deep model and DenseNet learned complemen-}
          \\
         &
           &
           &
          and using different levels of feature maps, concatenating them  &
          \multicolumn{1}{l}{-tary features for disease classification.}
          \\
         &
           &
           &
          with that of hippocampus obtained by another CNN. &
          \multicolumn{1}{r}{}
          \\
        \midrule
        \multicolumn{1}{c|}{\multirow{3}[2]{*}{\cite{ieracitano2019convolutional}}} &
          \multirow{3}[2]{*}{2018} &
          \multicolumn{1}{c|}{\multirow{3}[2]{*}{Image}} &
          Evaluating the power spectral density (PSD) of the 19- &
          \multicolumn{1}{l}{The originality of the proposed method lies in mapping the}
          \\
         &
           &
           &
           channels EEG traces and representing the related spectral &
           \multicolumn{1}{l}{power spectrum of each subject into a 2D gray scale image.}
          \\
         &
           &
           &
           profiles into 2D gray scale images (PSD-images). &
          \multicolumn{1}{r}{}
          \\
        \midrule
        \multicolumn{1}{c|}{\multirow{2}[2]{*}{\cite{li2020detecting}}} &
          \multirow{2}[2]{*}{2020} &
          \multicolumn{1}{c|}{\multirow{2}[2]{*}{Image}} &
          CNN for feature representation of different time nodes of 4D &
          \multicolumn{1}{l}{The proposed method can extract the spatio-temporal}
          \\
         &
           &
           &
          fMRI and feed them to an LSTM layer followed by Softmax. &
          \multicolumn{1}{l}{property of fMRI data fully for diagnosis of AD.}
          \\
        \midrule
        \multicolumn{1}{c|}{\multirow{2}[2]{*}{\cite{choi2020cognitive}}} &
          \multirow{2}[2]{*}{2020} &
          \multicolumn{1}{c|}{\multirow{2}[2]{*}{Image}} &
          3D-CNN to determine the likelihood of progression to AD  &
          \multicolumn{1}{l}{The deep learning-based cognitive function evaluation}
          \\
         &
           &
           &
          for MCI subjects &
          \multicolumn{1}{l}{model could be transferred to multiple disease domains.}
          \\
        \midrule
        \multicolumn{1}{c|}{\multirow{2}[2]{*}{\cite{chitradevi2020analysis}}} &
          \multirow{2}[2]{*}{2020} &
          \multicolumn{1}{c|}{\multirow{2}[2]{*}{Image}} &
          Gray wolf optimization (GWO) for brain segmentation and  &
          \multicolumn{1}{l}{The pipeline of the process adopted in this study can be}
          \\
         &
           &
           &
          AlexNet for classification. &
          \multicolumn{1}{l}{extended to other neuro developmental disorders.}
          \\
        \midrule
        \multicolumn{1}{c|}{\multirow{2}[2]{*}{\cite{puente2020automatic}}} &
          \multirow{2}[2]{*}{2020} &
          \multicolumn{1}{c|}{\multirow{2}[2]{*}{Image}} &
          Using TL technique, sagittal MRI images and ANN ResNet  &
          \multicolumn{1}{l}{TL allows experiments with little data as well as data}
          \\
         &
           &
           &
          feature extractor with the SVM classifier. &
          \multicolumn{1}{l}{augmentation.}
          \\
        \midrule
        \multicolumn{1}{c|}{\multirow{2}[2]{*}{\cite{spasov2019parameter}}} &
          \multirow{2}[2]{*}{2019} &
          \multicolumn{1}{c|}{Image + Demographic} &
          Consider the classification performance of the proposed &
          \multicolumn{1}{l}{The approach is flexible and can in principle integrate other}
          \\
         &
           &
          \multicolumn{1}{c|}{+Genetics + Cognition} &
          network on four different input biomarker combinations. &
          \multicolumn{1}{l}{imaging modalities and diverse other sets of clinical data.}
          \\
        \midrule
        \multicolumn{1}{c|}{\multirow{2}[2]{*}{\cite{janghel2021deep}}} &
          \multirow{2}[2]{*}{2021} &
          \multicolumn{1}{c|}{\multirow{2}[2]{*}{Image}} &
          VGG-16 for extracting features, SVM, Linear Discriminate,  &
          \multicolumn{1}{l}{Main objective of this paper is to include a preprocessing}
          \\
         &
           &
           &
          Kmeans clustering and Decision tree for classification. &
          \multicolumn{1}{l}{method before CNN model to increase accuracy.}
          \\
        \midrule
        \multicolumn{1}{c|}{\multirow{3}[2]{*}{\cite{chen2021iterative}}} &
          \multirow{3}[2]{*}{2021} &
          \multicolumn{1}{c|}{\multirow{3}[2]{*}{Image}} &
          Parallel ResNet-10 networks for each cortical region and  &
          \multicolumn{1}{l}{It incorporated a sparsity mask generated by sparse}
          \\
         &
           &
           &
          integrating them to a sparse regression while these two  &
        \multicolumn{1}{l}{regression module into a deep learning module for simul-}
          \\
         &
           &
           &
          modules are updated alternatively and iteratively &
           \multicolumn{1}{l}{-taneous AD diagnosis and identification of critical regions.}
          \\
        \midrule
        \multicolumn{1}{c|}{\multirow{3}[2]{*}{\cite{hazarika2022experimental}}} &
          \multirow{3}[2]{*}{2023} &
          \multicolumn{1}{c|}{\multirow{3}[2]{*}{Image}} &
          An experimental analysis of twenty popular CNNs was  &
          \multicolumn{1}{l}{To improve the execution time, it had proposed replacing}
          \\
         &
           &
           &
          performed using MRI. It was observed that the overall  &
          \multicolumn{1}{l}{the convolution layers in the original DenseNet-121}
          \\
         &
           &
           &
          classification accuracy of DenseNet-121 is the best. &
          \multicolumn{1}{l}{architecture with depth-wise convolution layers.}
          \\
        \midrule
        \multicolumn{1}{c|}{\multirow{3}[2]{*}{\cite{bae2021transfer}}} &
          \multirow{3}[2]{*}{2021} &
          \multicolumn{1}{c|}{\multirow{3}[2]{*}{Image}} &
          ResNet-29 was trained by transfer learning. CN and AD scans &
          \multicolumn{1}{l}{It avoided problems previously neglected such as data}
          \\
         &
           &
           &
           were used to pretrain the model, then model was retrained on &
          \multicolumn{1}{l}{shortage, high variance and data leakage.}
          \\
         &
           &
           &
           the target task of pMCI vs sMCI. &
          \multicolumn{1}{r}{}
          \\
        \midrule
        \multicolumn{1}{c|}{\multirow{2}[2]{*}{\cite{zhang2021explainable}}} &
          \multirow{2}[2]{*}{2022} &
          \multicolumn{1}{c|}{\multirow{2}[2]{*}{Image}} &
          Proposing an explainable 3D Residual Attention Deep Neural  &
          \multicolumn{1}{l}{The explainable mechanism is able to identify and high-}
          \\
         &
           &
           &
          Network (3D ResAttNet) for end-to-end learning from sMRI. &
          \multicolumn{1}{l}{-light the contribution of the important brain parts.}
          \\
        \bottomrule
        \end{tabular}%
      \label{tab:23}%
    \end{sidewaystable*}%


    \begin{sidewaystable*}[htbp]
      \centering
      \caption{(Continued)}
        \begin{tabular}{c|c|p{8em}|p{25em}|p{26em}}
        \toprule
        \multicolumn{1}{c|}{\textbf{Ref.}} &
          \multicolumn{1}{c|}{\textbf{Year}} &
          \multicolumn{1}{c|}{\textbf{Modality}} &
          \multicolumn{1}{c|}{\textbf{Method}} &
          \multicolumn{1}{c}{\textbf{Comment}}
          \\
        \midrule
        \multicolumn{1}{c|}{\multirow{3}[2]{*}{\cite{raza2019diagnosis}}} &
          \multirow{3}[2]{*}{2019} &
          \multicolumn{1}{c|}{\multirow{3}[2]{*}{Image + Gait}} &
          Monitoring the subjects’ activities of daily living using body &
          This paper correlates the importance of high accuracy AD 
          \\
         &
           &
           &
          worn inertial sensors to get physical activity data, then classify  &
          diagnosis with physical activities and how these activities can 
          \\
         &
           &
           &
          target together with MRI. &
          be logged without human intervention with high accuracy.
          \\
        \midrule
        \multicolumn{1}{c|}{\multirow{3}[2]{*}{\cite{huang2019diagnosis}}} &
          \multirow{3}[2]{*}{2019} &
          \multicolumn{1}{c|}{\multirow{3}[2]{*}{Image}} &
          Two paralleled VGG11 networks for each modality (MRI + &
          It indicated that the hippocampal area with no segmentation can 
          \\
         &
           &
           &
           PET) and concatenating the feature maps to being fed into &
          be chosen as the input.
          \\
         &
           &
           &
           some FC layers followed by Softmax layer. &
          \multicolumn{1}{r}{}
          \\
        \midrule
        \multicolumn{1}{c|}{\multirow{3}[2]{*}{\cite{wang2019ensemble}}} &
          \multirow{3}[2]{*}{2019} &
          \multicolumn{1}{c|}{\multirow{3}[2]{*}{Image}} &
          Dense connections were introduced to maximize the informa- &
          Dense connections improve the information and gradients 
          \\
         &
           &
           &
           -tion flow. And probability- based fusion method was employ-&
          propagation throughout the network, then make the network 
          \\
         &
           &
           &
          -ed to combine 3D-DenseNets with different architectures. &
          easier to train due to less parameters.
          \\
        \midrule
        \multicolumn{1}{c|}{\multirow{3}[2]{*}{\cite{zhang2019multi}}} &
          \multirow{3}[2]{*}{2019} &
          \multicolumn{1}{c|}{\multirow{3}[2]{*}{Image + Cognition}} &
          Two independent CNNs were used to extract features of MRI &
          The method combined the neuroimaging diagnosis with the 
          \\
         &
           &
           &
          and PET, calculating correlation coefficient and combining &
          clinical neuropsychological diagnosis, so the process is closer to 
          \\
         &
           &
           &
          the outputs with the results of clinical psychological diagnosis. &
          the process of clinician’s diagnosis and easy to implement.
          \\
        \midrule
        \multicolumn{1}{c|}{\multirow{4}[2]{*}{\cite{liu2019weakly}}} &
          \multirow{4}[2]{*}{2020} &
          \multicolumn{1}{c|}{\multirow{4}[2]{*}{Image + Cognition}} &
          Using two scales of landmark patches as two channels into a  &
          The current network primarily worked for estimating multiple
          \\
         &
           &
           &
          densely connected CNNs for each landmark and concatenat- &
          types of clinical scores, without considering the underlying
          \\
         &
           &
           &
          -ing these feature maps to being fed into some FC layers &
          association between clinical scores and class labels of subjects.
          \\
         &
           &
           &
        to predict cognitive scores of four follow-up time points. &
          \multicolumn{1}{r}{}
          \\
        \midrule
        \multicolumn{1}{c|}{\multirow{2}[2]{*}{\cite{abuhmed2021robust}}} &
          \multirow{2}[2]{*}{2021} &
          \multicolumn{1}{p{8em}|}{Image+ Cognition+} &
          Applying Bidirectional LSTM on the multi modal time series  &
          Although the DFBL architecture is more accurate than the 
          \\
         &
           &
          \multicolumn{1}{p{8em}|}{Genetics+Biomarker} &
          data for feature representation and using them in two models. &
          MRBL architecture, the latter is more interpretable.
          \\
        \midrule
        \multicolumn{1}{c|}{\multirow{3}[2]{*}{\cite{lei2022predicting}}} &
          \multirow{3}[2]{*}{2021} &
          \multicolumn{1}{c|}{\multirow{3}[2]{*}{Image + Cognition}} &
          Combining group LASSO and correntropy to reduce dimen- &
          By stacking multiple layers IndRNN, it was possible to study 
          \\
         &
           &
           &
          -sion and obtain features, then using features in a stacked  &
          the cross-channel information in time and obtain deep 
          \\
         &
           &
           &
          independent RNN to predict cognitive scores.  &
          characteristic information.
          \\
        \midrule
        \multicolumn{1}{c|}{\multirow{2}[2]{*}{\cite{ghazi2019training}}} &
          \multirow{2}[2]{*}{2019} &
          \multicolumn{1}{c|}{\multirow{2}[2]{*}{Image}} &
          Applying a modified LSTM to work with incomplete data  &
          This study highlighted the potential of RNNs for modeling the 
          \\
         &
           &
           &
          while LDA providing class labels. &
          progression of AD using longitudinal measurements.
          \\
        \midrule
        \multicolumn{1}{c|}{\multirow{2}[2]{*}{\cite{ieracitano2020novel}}} &
          \multirow{2}[2]{*}{2020} &
          \multicolumn{1}{c|}{\multirow{2}[2]{*}{Image}} &
          Exploiting the potential of both CWT and BiS features &
          This study focused on the AD/MCI/CN classification task and 
          \\
         &
           &
           &
          extracted from time–frequency and bispectrum of EEG. &
          did not address prediction of early onset of AD.
          \\
        \midrule
        \multicolumn{1}{c|}{\multirow{3}[2]{*}{\cite{bi2019early}}} &
          \multirow{3}[2]{*}{2019} &
          \multicolumn{1}{c|}{\multirow{3}[2]{*}{Image}} &
          Using multi-task learning strategy based on discriminative  &
          The designed model was well-performed in comparison to other
          \\
         &
           &
           &
          convolutional high-order Boltzmann Machine with hybrid  &
          generative model since it bridged the connection between feature
          \\
         &
           &
           &
          feature maps. &
          extraction and classification.
          \\
        \midrule
        \multicolumn{1}{c|}{\multirow{2}[2]{*}{\cite{choi2019deep}}} &
          \multirow{2}[2]{*}{2019} &
          \multicolumn{1}{c|}{\multirow{2}[2]{*}{Image}} &
          Using variational autoencoder, abnormality score was defined &
          The approach has advantages in the flexible application for 
          \\
         &
           &
           &
           as how far a given brain image is from the normal data. &
          heterogeneous patients, even for uncommon disorders.
          \\
        \midrule
        \multicolumn{1}{c|}{\multirow{3}[2]{*}{\cite{kang2021multi}}} &
          \multirow{3}[2]{*}{2021} &
          \multicolumn{1}{c|}{\multirow{3}[2]{*}{Image}} &
          A majority voting ensemble was used to combine multi-slice &
          This technique may enhance AD diagnostics when the sample
          \\
         &
           &
           &
          predictions with complementary results.VGG16, ResNet50, &
          size is limited.
          \\
         &
           &
           &
          and DCGAN were used to construct an ensemble classifier. &
          \multicolumn{1}{r}{}
          \\
        \midrule
        \multicolumn{1}{c|}{\multirow{3}[2]{*}{\cite{kim2020slice}}} &
          \multirow{3}[2]{*}{2020} &
          \multicolumn{1}{c|}{\multirow{3}[2]{*}{Image}} &
          Applying slice selective learning to reduce computational cost  &
          When there were insufficient datasets to train deep neural
          \\
         &
           &
           &
          and to extract unbiased features. Extracting features by a &
          networks with single-source training datasets, this approach
          \\
         &
           &
           &
          Boundary Equilibrium Generative Adversarial Network. &
           was a feasible alternative to use datasets from various hospitals.
          \\
        \midrule
        \multicolumn{1}{c|}{\multirow{3}[2]{*}{\cite{mendoza2020single}}} &
          \multirow{3}[2]{*}{2020} &
          \multicolumn{1}{c|}{\multirow{3}[2]{*}{Image}} &
          Applying Supervised Switching AEs on extracted patches of  &
          It used only one scan slice per subject, instead of the whole 3D 
          \\
         &
           &
           &
          a single MRI slice while prediction is based on a majority  &
          volume.
          \\
         &
           &
           &
          rule from the ensemble of patch predictions &
          \multicolumn{1}{r}{}
          \\
        \bottomrule
        \end{tabular}%
      \label{tab:24}%
    \end{sidewaystable*}%


    \begin{sidewaystable*}[htbp]
      \centering
      \caption{(Continued)}
        \begin{tabular}{c|c|p{5em}|p{26em}|p{26em}}
        \toprule
        \multicolumn{1}{c|}{\textbf{Ref.}} &
          \multicolumn{1}{c|}{\textbf{Year}} &
          \multicolumn{1}{c|}{\textbf{Modality}} &
          \multicolumn{1}{c|}{\textbf{Method}} &
          \multicolumn{1}{c}{\textbf{Comment}}
          \\
          \midrule
        \multicolumn{1}{c|}{\multirow{2}[2]{*}{\cite{lei2020deep}}} &
          \multirow{2}[2]{*}{2020} &
          \multicolumn{1}{c|}{Image +} &
          LASSO for feature extraction, deep polynomial network for &
          \multicolumn{1}{l}{The proposed framework can effectively reveal the relationship}
          \\
         &
           &
          \multicolumn{1}{c|}{Cognition} &
          feature encoding followed by an SVR to predict cognitive scores. &
          \multicolumn{1}{l}{between clinical scores and MRI data.}
          \\          
        \midrule
        \multicolumn{1}{c|}{\multirow{3}[2]{*}{\cite{afzal2019data}}} &
          \multirow{3}[2]{*}{2019} &
          \multicolumn{1}{c|}{\multirow{3}[2]{*}{Image}} &
          Proposing a pretrained AlexNet model based on TL technique&
          \multicolumn{1}{l}{Extensive augmentation approaches can prevent overfitting}
          \\
         &
           &
           &
          Using the main view of the brain and the MRI with extensive  &
          \multicolumn{1}{l}{issues in class balance dataset.}
          \\
         &
           &
           &
          image augmentation techniques to avoid overfitting issue.  &
          
          \\
        \midrule
        \multicolumn{1}{c|}{\cite{jain2019convolutional}} &
          2019 &
          \multicolumn{1}{c|}{Image} &
          Using a pretrained VGG16 network for transfer learning. &
          \multicolumn{1}{l}{Transfer learning effectively reduces computational cost.}
          \\
        \midrule
        \multicolumn{1}{c|}{\multirow{2}[2]{*}{\cite{2019Volumetrie}}} &
          \multirow{2}[2]{*}{2019} &
          \multicolumn{1}{c|}{\multirow{2}[2]{*}{Image}} &
          Presenting an effective strategy of augmenting volumetric data &
          \multicolumn{1}{l}{It had shown that using a set of augmentation methods postponed}
          \\
         &
           &
           &
          in a 3D-CNN neural network model. &
          \multicolumn{1}{l}{ the overfitting without any other regularization technique.}
          \\
        \midrule
        \multicolumn{1}{c|}{\multirow{3}[2]{*}{\cite{barbaroux2020encoding}}} &
          \multirow{3}[2]{*}{2020} &
          \multicolumn{1}{c|}{\multirow{3}[2]{*}{Image}} &
          \multirow{2}[2]{*}{Using spherical CNN formulation as an effective deep learning}  &
          \multicolumn{1}{l}{This work demonstrated the feasibility and superiority of the }
          \\
         &
           &
           &
          \multirow{2}[2]{*}{framework for modeling human cortex.} &
          \multicolumn{1}{l}{spherical CNN directly applied on the spherical representation}
          \\
         &
           &
           &
          \multicolumn{1}{r|}{} &
          \multicolumn{1}{l}{in the discriminative analysis of the human cortex.}
          \\
        \midrule
        \multicolumn{1}{c|}{\multirow{3}[2]{*}{\cite{tufail2020binary}}} &
          \multirow{3}[2]{*}{2020} &
          \multicolumn{1}{c|}{\multirow{3}[2]{*}{Image}} &
          The whole brain image was passed through two transfer &
          \multicolumn{1}{l}{Experimental results show that the transfer learning approaches}
          \\
         &
           &
           &
          learning models: Inception V3 and Xception, as well as a  &
          \multicolumn{1}{l}{exceed the performance of non-transfer learning based }
          \\
         &
           &
           &
          custom CNN with separable convolutional layers. &
          \multicolumn{1}{l}{approaches.}
          \\
        \midrule
        \multicolumn{1}{c|}{\cite{2020Diagnosis}} &
          2020 &
          \multicolumn{1}{c|}{Image} &
          Using DCNN and VGG-16 inspired CNN (VCNN) models. &
          \multicolumn{1}{l}{VCNN performed better.}
          \\
        \midrule
        \multicolumn{1}{c|}{\multirow{3}[2]{*}{\cite{2020Identifying}}} &
          \multirow{3}[2]{*}{2020} &
          \multicolumn{1}{c|}{\multirow{3}[2]{*}{Image}} &
          A CNN based on transfer learning is developed to extract &
          \multicolumn{1}{l}{DTI data is an effective supplement to sMRI, and it acts as an }
          \\
         &
           &
           &
          features of MRI and DTI, where an L1-norm is introduced &
          \multicolumn{1}{l}{significant biomarker of MCI.}
          \\
         &
           &
           &
        to reduce the feature dimensionality by LASSO algorithm. &
          
          \\
        \bottomrule
        \end{tabular}%
      \label{tab:25}%
    \end{sidewaystable*}%

\subsection{Evaluation index and performance}
\subsubsection{Evaluation index}

\quad In studies on AD classification based on computer-aided techniques, the effectiveness, reliability, and generalizability of the methods can be evaluated using several metrics to assess their applicability for clinical assistance and diagnosis. The main evaluation criteria include the following:\par

1. Accuracy (ACC): It represents the ratio of correctly classified samples to the total number of samples in a given test dataset. ACC reflects the probability of the classifier accurately classifying AD, MCI, and CN. Generally, higher values indicate better classifier performance.

2. Sensitivity (True Positive Rate, TPR): Also known as recall or hit rate, it represents the probability of correctly detecting all true positive cases. TPR measures the rate of correctly identified AD cases and indicates how well the classifier identifies AD. Larger values imply fewer cases of AD being missed.

3. Specificity (True Negative Rate, TNR): TNR, also known as the true negative rate, represents the probability of correctly detecting all true negative cases. It measures the classifier's ability to accurately identify true negatives, which reflects the classifier's capability to correctly classify CN. Generally, higher values indicate a lower probability of misdiagnosing normal individuals as AD.

4. Precision (Positive Predictive Value, PPV): It is the ratio of true positive samples to all samples classified as positive. Precision measures the proportion of correctly identified true positive cases among all samples classified as positive.

5. Area Under the ROC Curve (AUC): AUC represents the area under the curve plotted with TPR (true positive rate) on the y-axis and FPR (false positive rate) on the x-axis. A larger AUC value indicates better classification performance of the classifier.

6. F1-score: F1-score is the harmonic mean of precision and recall (sensitivity). Since precision and recall are conflicting metrics, using a single precision or recall measure may be biased. Therefore, F1-score provides a balanced evaluation metric by considering both precision and recall.

When dealing with multi-class classification tasks, the calculation of accuracy, sensitivity, and specificity needs to consider the corresponding values for each class. Micro and Macro are two different weighting approaches for calculating these metrics, taking sensitivity as an example.


The calculation formulas are as follows:
\begin{equation}
Acc=\frac{TP+TN}{TP+TN+FP+FN}
\end{equation}

\begin{equation}
Sens=Recall=TPR=\frac{TP}{TP+FN}
\end{equation}

\begin{equation}
Spec=TNR=\frac{TN}{TN+FP}
\end{equation}

\begin{equation}
PPV=\frac{TP}{TP+FP}
\end{equation}

\begin{equation}
F1-scrore=\frac{2\times PPV\times TPR}{PPV+TPR}
\end{equation}

\begin{equation}
FPR=\frac{FP}{TN+FP}
\end{equation}

\begin{equation}
Sens_{macro}=\frac{\sum_{i=1}^{n}Sens_{i}}{n}
\end{equation}

\begin{equation}
Sens_{micro}=\frac{\sum_{i=1}^{n}TP_{i}}{\sum_{i=1}^{n}TP_{i}+\sum_{i=1}^{n}FN_{i}}
\end{equation}

\quad Among them, TP (True Positive) represents the number
of samples predicted as positive and are actually positive; FP
(False Positive) represents the number of samples predicted
as positive but are actually negative; FN (False Negative)
represents the number of samples predicted as negative but are
actually positive; TN (True Negative) represents the number
of samples predicted as negative and are actually negative.


\quad In addition to these commonly used performance metrics, some studies have adopted other evaluation metrics to better align with their model performance. 

1. Matthews Correlation Coefficient (MCC):
   MCC is a measure used to evaluate the classification performance of binary classification tasks. It takes into account true positives, true negatives, false positives, and false negatives. MCC is considered a balanced metric that can be applied even when there is a large difference in sample sizes between the two classes. It essentially describes the correlation coefficient between the actual classification and the predicted classification, with values ranging from -1 to 1. A value of 1 indicates perfect prediction, 0 indicates predictions that are no better than random, and -1 indicates complete disagreement between predicted and actual classifications.

2. Fowlkes-Mallows Index (FMI):
   FMI is an external evaluation method used to determine the similarity between two clusters. A higher FMI indicates greater similarity between the clustering and the reference classification. The minimum possible value of FMI is 0, which corresponds to the worst binary classification where all elements are misclassified. The maximum possible value is 1, which corresponds to the best binary classification where all elements are perfectly classified.

3. Mean Absolute Error (MAE):
   MAE measures the average absolute difference between predicted values and true values. It has a range of [0, +∞), where a value of 0 indicates a perfect model with predictions matching the true values exactly, and larger values indicate larger errors.

4. Pearson Correlation Coefficient (R):
   R measures the linear correlation between two sets of data. It has a range of [-1, 1], where 0 represents no correlation, negative values indicate negative correlation, and positive values indicate positive correlation.

5. Balanced Accuracy (BACC):
   BACC is a performance metric that is better than raw accuracy when the sample sizes of each class (normal and symptomatic) are imbalanced. 

6. C-index (Concordance Index):
   C-index is used to evaluate the predictive ability of a model. It represents the proportion of pairs where the predicted results are consistent with the actual results among all pairs of patients. It estimates the probability of concordance between predicted and observed outcomes.


\quad The calculation formulas/methods are as follows:
\begin{equation}
MCC=\frac{TP\times TN-FP\times FN}{\sqrt{(TP+FP)(TP+FN)(TN+FP)(TN+FN)}}
\end{equation}

\begin{equation}
FMI=\sqrt{PPV\cdot TPR}
\end{equation}

\begin{equation}
R(X,Y)=\frac{E[(X-\mu_{X})(Y-\mu_{Y})]}{\sigma_{X}\sigma_{Y}}
\end{equation}

\begin{equation}
BACC=\frac{TPR+TNR}{2}
\end{equation}

The C-index is calculated by randomly pairing all the research subjects. Taking the prediction of AD as an example, in each pair, at least one individual will progress to AD. If the predicted probability of developing AD is higher for the individual who actually progresses to AD compared to the other individual, it is considered a consistent prediction with the actual outcome.

\subsubsection{Performance}




\quad The literature covered in this article mainly involves several types of prediction: binary classification, multiclass classification and regression prediction. The most common classification involves categorizing subjects into AD, sMCI, pMCI and CN (also known as NC/HC). Among them, the binary classification of sMCI and pMCI predicts whether MCI subjects will progress to AD in the future.

\quad Due to the varying evaluation metrics used in different studies and the differences in selected modalities, datasets and target requirements, the performance of the same model can vary significantly across different classification tasks. Therefore, it is not appropriate to use a single metric to evaluate the superiority or inferiority of all models. The selection of a suitable model should be based on specific requirements. For example, although the accuracy of \cite{ghoraani2021detection} is lower than that of similar studies in both classification tasks (less than 87\%), its gait feature collection cost is much lower than medical imaging. Similarly, although the classification accuracy of \cite{mh_1} is not top-notch, speech data has the advantages of low cost and real-time availability. \cite{liu2019using} achieved significant results in reducing problem complexity and computation time, but at the expense of sacrificing some accuracy. \cite{lee2019toward} proposed a method with the advantages of result visualization and interpretability, although its performance metrics are excellent, but not state-of-the-art. On the other hand, although \cite{2021ADVIAN} achieved an accuracy of 97.76\% in the AD vs CN task, with other metrics also being extremely high, the experiment did not undergo rigorous clinical testing.

\quad However, from the experimental results, it can be observed that combining multiple modalities often outperforms using a single modality as features. This approach can balance and improve overall performance, compensating for the limitations of single modality data in capturing comprehensive features. For example, \cite{2021An} combined imaging with cognitive modalities, achieving accuracy higher than 90\% in all binary classification tasks, with CN vs AD and sMCI vs AD reaching accuracy levels of 98.62\% and 99.62\% respectively, although other metric results were not provided. \cite{raza2019diagnosis} combined imaging with gait modalities and achieved classification accuracy above 95\% for AD vs CN on various datasets, as well as sensitivity, specificity, and precision above 90\%.

\quad This article summarizes the "Evaluation index and Performance" of various studies in this field in recent years, as shown in Table \ref{tab:31}-\ref{tab:36}.


    \begin{table*}[htbp]
      \centering
      \caption{Evaluation index and performance}
        \begin{tabular}{|c|c|c|c|c|c|c|c|c|}
        \toprule
        \multicolumn{1}{|c|}{\multirow{2}[4]{*}{\textbf{Ref.}}} &
          \multicolumn{1}{c|}{\multirow{2}[4]{*}{\textbf{Year}}} &
          \multirow{2}[4]{*}{\textbf{Target}} &
          \multicolumn{6}{c|}{\textbf{Performance}}
          \\
\cmidrule{4-9}         &
           &
          \multicolumn{1}{c|}{} &
          \multicolumn{1}{c|}{\textbf{Acc (\%)}} &
          \multicolumn{1}{c|}{\textbf{Sens (\%)}} &
          \multicolumn{1}{c|}{\textbf{Spec (\%)}} &
          \multicolumn{1}{c|}{\textbf{AUC}} &
          \multicolumn{2}{c|}{\textbf{others}}
          \\
        \midrule
        \multicolumn{1}{|c|}{\multirow{4}[2]{*}{\cite{MMSEvsMoCA}}} &
          \multirow{4}[2]{*}{2020} &
          CN vs MCI(MMSE+DSRS) &
          97 &
          \multirow{4}[2]{*}{} &
          \multirow{4}[2]{*}{} &
          0.97 &
          \multicolumn{2}{c|}{\multirow{4}[2]{*}{}}
          \\
         &
           &
          CN vs MCI(MoCA+DSRS) &
          97 &
           &
           &
          0.97 &
          \multicolumn{2}{c|}{}
          \\
         &
           &
          MCI vs AD(MMSE+DSRS) &
          93 &
           &
           &
          0.91 &
          \multicolumn{2}{c|}{}
          \\
         &
           &
          MCI vs AD(MoCA+DSRS) &
          92 &
           &
           &
          0.9 &
          \multicolumn{2}{c|}{}
          \\
        \midrule
        \multicolumn{1}{|c|}{\multirow{3}[2]{*}{\cite{2010Clin}}} &
          \multirow{3}[2]{*}{2022} &
          AD vs non-AD &
          \multicolumn{1}{c|}{≈90} &
           &
           &
          \multirow{3}[2]{*}{} &
          \multicolumn{2}{c|}{Precision (\%): 86.9}
          \\
         &
           &
          MCI-＞CN prediction &
          88.5 &
          \multicolumn{1}{c|}{＞85} &
          \multicolumn{1}{c|}{＞85} &
           &
          \multicolumn{2}{c|}{}
          \\
         &
           &
          MCI-＞AD prediction &
          90.8 &
          \multicolumn{1}{c|}{＞85} &
          \multicolumn{1}{c|}{＞85} &
           &
          \multicolumn{2}{c|}{}
          \\
        \midrule
        \multicolumn{1}{|c|}{\multirow{2}[2]{*}{\cite{ghoraani2021detection}}} &
          \multirow{2}[2]{*}{2021} &
          CN vs MCI vs AD &
          78 &
          \multirow{2}[2]{*}{} &
          \multirow{2}[2]{*}{} &
          \multirow{2}[2]{*}{} &
          \multicolumn{2}{c|}{F1-score: 0.77}
          \\
         &
           &
          CN vs MCI/AD &
          86 &
           &
           &
           &
          \multicolumn{2}{c|}{F1-score: 0.88}
          \\
        \midrule
        \multicolumn{1}{|c|}{\multirow{3}[2]{*}{\cite{mh_1}}} &
          \multirow{3}[2]{*}{2020} &
          CN vs AD： &
           &
           &
          \multirow{3}[2]{*}{} &
          \multirow{3}[2]{*}{} &
          \multicolumn{1}{c|}{Precision (\%)} &
          \multicolumn{1}{c|}{F1-score}
          \\
         &
           &
          VBSD dataset &
          83.3 &
          86.9 &
           &
           &
          86.9 &
          0.869
          \\
         &
           &
          Dem@Care dataset &
          84.4 &
          87.5 &
           &
           &
          91.3 &
          0.894
          \\
        \midrule
        \multicolumn{1}{|c|}{\cite{haider2019assessment}} &
          2021 &
          AD vs non-AD &
          78.7 &
          80.5 &
          76.8 &
           &
          \multicolumn{2}{c|}{}
          \\
        \midrule
        \multicolumn{1}{|c|}{\multirow{3}[2]{*}{\cite{luz2020alzheimer}}} &
          \multirow{3}[2]{*}{2020} &
          AD vs non-AD： &
           &
           &
          \multirow{3}[2]{*}{} &
          \multirow{3}[2]{*}{} &
          \multicolumn{1}{c|}{Precision (\%)} &
          \multicolumn{1}{c|}{F1-score}
          \\
         &
           &
          Acoustic &
          62 &
          75 &
           &
           &
          60 &
          0.67
          \\
         &
           &
          Linguistic &
          75 &
          62 &
           &
           &
          83 &
          0.71
          \\
        \midrule
        \multicolumn{1}{|c|}{\multirow{4}[2]{*}{\cite{shi2017multimodal}}} &
          \multirow{4}[2]{*}{2018} &
          CN vs AD(a) &
          97.13 &
          95.93 &
          98.53 &
          0.972 &
          \multicolumn{2}{c|}{\multirow{4}[2]{*}{}}
          \\
         &
           &
          MCI vs CN(a) &
          87.24 &
          97.91 &
          67.04 &
          0.901 &
          \multicolumn{2}{c|}{}
          \\
         &
           &
          MCI-C vs MCI-CN(a) &
          78.88 &
          68.04 &
          86.81 &
          0.801 &
          \multicolumn{2}{c|}{}
          \\
         &
           &
          AD vs MCI-C vs MCI-NC vs NC(a) &
          57 &
          53.65 &
          85.05 &
           &
          \multicolumn{2}{c|}{}
          \\
        \midrule
        \multicolumn{1}{|c|}{\multirow{4}[2]{*}{\cite{muti_image_2}}} &
          \multirow{4}[2]{*}{2018} &
          sMCI vs pMCI &
          82.93 &
          79.69 &
          83.84 &
          \multirow{4}[2]{*}{} &
          \multicolumn{2}{c|}{\multirow{4}[2]{*}{}}
          \\
         &
           &
          sCN vs AD &
          84.6 &
          80.2 &
          91.8 &
           &
          \multicolumn{2}{c|}{}
          \\
         &
           &
          sCN vs (pMCI \& AD) &
          86 &
          85.7 &
          86.5 &
           &
          \multicolumn{2}{c|}{}
          \\
         &
           &
          sCN vs. (pCN, pMCI and AD) &
          86.4 &
          86.5 &
          86.3 &
           &
          \multicolumn{2}{c|}{}
          \\
        \midrule
        \multicolumn{1}{|c|}{\multirow{5}[2]{*}{\cite{qiu2020development}}} &
          \multirow{5}[2]{*}{2020} &
          CN vs AD： &
           &
           &
           &
           &
          \multicolumn{1}{c|}{F1-score (\%)} &
          \multicolumn{1}{c|}{MCC}
          \\
         &
           &
          ADNI dataset &
          96.8 &
          95.7 &
          97.7 &
          0.996 &
          96.5 &
          0.937
          \\
         &
           &
          AIBL dataset &
          93.2 &
          87.7 &
          94.3 &
          0.974 &
          81.4 &
          0.78
          \\
         &
           &
          FHS dataset &
          79.2 &
          74.2 &
          80.8 &
          0.867 &
          63.3 &
          0.517
          \\
         &
           &
          NACC dataset &
          85.2 &
          92.4 &
          81 &
          0.954 &
          82.4 &
          0.714
          \\
        \midrule
        \multicolumn{1}{|c|}{\multirow{2}[2]{*}{\cite{2018Hierarchical}}} &
          \multirow{2}[2]{*}{2020} &
          CN vs AD &
          90.3 &
          82.4 &
          96.5 &
          0.951 &
          \multicolumn{2}{c|}{\multirow{2}[2]{*}{}}
          \\
         &
           &
          pMCI vs sMCI &
          80.9 &
          52.6 &
          85.4 &
          0.781 &
          \multicolumn{2}{c|}{}
          \\
        \midrule
        \multicolumn{1}{|c|}{\cite{ebrahimi2021deep}} &
          2021 &
          CN vs AD &
          91.78 &
          91.56 &
          92 &
           &
          \multicolumn{2}{c|}{}
          \\
        \midrule
        \multicolumn{1}{|c|}{\multirow{3}[2]{*}{\cite{colliot2008discrimination}}} &
          \multirow{3}[2]{*}{2008} &
          CN vs AD &
          84 &
          84 &
          84 &
          0.913 &
          \multicolumn{2}{c|}{\multirow{3}[2]{*}{}}
          \\
         &
           &
          MCI vs CN &
          73 &
          75 &
          70 &
          0.808 &
          \multicolumn{2}{c|}{}
          \\
         &
           &
          AD vs MCI &
          69 &
          67 &
          71 &
          0.721 &
          \multicolumn{2}{c|}{}
          \\
        \midrule
        \multicolumn{1}{|c|}{\multirow{3}[2]{*}{\cite{multimodal_e}}} &
          \multirow{3}[2]{*}{2022} &
          CN vs MCI vs AD： &
           &
           &
          \multirow{3}[2]{*}{} &
          \multirow{3}[2]{*}{} &
          \multicolumn{1}{c|}{Precision (\%)} &
          \multicolumn{1}{c|}{F1-score}
          \\
         &
           &
          EHR + SNP + Random Forest &
          78 &
          79 &
           &
           &
          78 &
          0.78
          \\
         &
           &
          EHR + Imaging + Random Forest &
          77 &
          77 &
           &
           &
          76 &
          0.77
          \\
        \bottomrule
        \end{tabular}%
      \label{tab:31}%
    \end{table*}%



    \begin{table*}[htbp]
      \centering
      \caption{(Continued)}
        \begin{tabular}{|c|c|c|c|c|c|c|c|}
        \toprule
        \multicolumn{1}{|c|}{\multirow{2}[4]{*}{\textbf{Ref.}}} &
          \multicolumn{1}{c|}{\multirow{2}[4]{*}{\textbf{Year}}} &
          \multicolumn{1}{c|}{\multirow{2}[4]{*}{\textbf{Target}}} &
          \multicolumn{5}{c|}{\textbf{Performance}}
          \\
\cmidrule{4-8}         &
           &
           &
          \multicolumn{1}{c|}{\textbf{Acc (\%)}} &
          \multicolumn{1}{c|}{\textbf{Sens (\%)}} &
          \multicolumn{1}{c|}{\textbf{Spec (\%)}} &
          \multicolumn{1}{c|}{\textbf{AUC}} &
          \textbf{others}
          \\
        \midrule
        \multicolumn{1}{|c|}{\multirow{6}[2]{*}{\cite{zeng2021new}}} &
          \multirow{6}[2]{*}{2021} &
          \multicolumn{1}{c|}{sMCI vs pMCI} &
          87.78 &
          \multirow{6}[2]{*}{} &
          \multirow{6}[2]{*}{} &
          \multirow{6}[2]{*}{} &
          \multicolumn{1}{c|}{\multirow{6}[2]{*}{}}
          \\
         &
           &
          \multicolumn{1}{c|}{CN vs AD} &
          98.62 &
           &
           &
           &
          \multicolumn{1}{c|}{}
          \\
         &
           &
          \multicolumn{1}{c|}{CN vs sMCI} &
          92.31 &
           &
           &
           &
          \multicolumn{1}{c|}{}
          \\
         &
           &
          \multicolumn{1}{c|}{CN vs pMCI} &
          96.67 &
           &
           &
           &
          \multicolumn{1}{c|}{}
          \\
         &
           &
          \multicolumn{1}{c|}{sMCI vs AD} &
          99.62 &
           &
           &
           &
          \multicolumn{1}{c|}{}
          \\
         &
           &
          \multicolumn{1}{c|}{pMCI vs AD} &
          91.89 &
           &
           &
           &
          \multicolumn{1}{c|}{}
          \\
        \midrule
        \multicolumn{1}{|c|}{\multirow{4}[2]{*}{\cite{2021ADVIAN}}} &
          \multirow{4}[2]{*}{2021} &
          \multicolumn{1}{c|}{\multirow{4}[2]{*}{CN vs AD}} &
          \multirow{4}[2]{*}{97.76} &
          \multirow{4}[2]{*}{97.65} &
          \multirow{4}[2]{*}{97.86} &
          \multirow{4}[2]{*}{0.9852} &
          Precision：97.87
          \\
         &
           &
           &
           &
           &
           &
           &
          F1-score：0.9775
          \\
         &
           &
           &
           &
           &
           &
           &
          MCC：0.9553
          \\
         &
           &
           &
           &
           &
           &
           &
          FMI：97.76
          \\
        \midrule
        \multicolumn{1}{|c|}{\multirow{6}[2]{*}{\cite{2021Dual}}} &
          \multirow{6}[2]{*}{2021} &
          \multicolumn{1}{c|}{CN vs AD on ADNI dataset} &
          92.4 &
          91 &
          93.8 &
          0.965 &
          \multicolumn{1}{c|}{\multirow{6}[2]{*}{}}
          \\
         &
           &
          \multicolumn{1}{c|}{pMCI vs sMCI on ADNI dataset} &
          80.2 &
          77.1 &
          82.6 &
          0.851 &
          \multicolumn{1}{c|}{}
          \\
         &
           &
          \multicolumn{1}{c|}{pMCI vs CN on ADNI dataset} &
          89.5 &
          82.4 &
          92.5 &
          0.917 &
          \multicolumn{1}{c|}{}
          \\
         &
           &
          \multicolumn{1}{c|}{sMCI vs CN on ADNI dataset} &
          82.5 &
          80.4 &
          83.8 &
          0.86 &
          \multicolumn{1}{c|}{}
          \\
         &
           &
          \multicolumn{1}{c|}{CN vs AD on AIBL dataset} &
          90.2 &
          84.8 &
          91.5 &
          0.939 &
          \multicolumn{1}{c|}{}
          \\
         &
           &
          \multicolumn{1}{c|}{pMCI vs sMCI on AIBL dataset} &
          80.9 &
          70.6 &
          82.8 &
          0.824 &
          \multicolumn{1}{c|}{}
          \\
        \midrule
        \multicolumn{1}{|c|}{\multirow{3}[2]{*}{\cite{martinez2019studying}}} &
          \multirow{3}[2]{*}{2020} &
          \multicolumn{1}{c|}{Model: CAE-SVC} &
           &
           &
           &
          \multirow{3}[2]{*}{} &
          F1-score
          \\
         &
           &
          \multicolumn{1}{c|}{CN vs AD (Tissue: GM)} &
          77.1 &
          75.9 &
          78.3 &
           &
          \multicolumn{1}{c|}{0.768}
          \\
         &
           &
          \multicolumn{1}{c|}{sMCI vs pMCI (Tissue: WM)} &
          70.4 &
          69.3 &
          72.6 &
           &
          \multicolumn{1}{c|}{0.709}
          \\
        \midrule
        \multicolumn{1}{|c|}{\multirow{5}[2]{*}{\cite{lin2021bidirectional}}} &
          \multirow{5}[2]{*}{2021} &
          \multicolumn{1}{c|}{CN vs AD(a)，sMCI vs pMCI(b)} &
           &
           &
           &
           &
          \multicolumn{1}{c|}{\multirow{5}[2]{*}{}}
          \\
         &
           &
          \multicolumn{1}{c|}{MRI + PET(a)} &
          92.28 &
          90.38 &
          94.37 &
          0.9276 &
          \multicolumn{1}{c|}{}
          \\
         &
           &
          \multicolumn{1}{c|}{PET + 100\% synthetic MRI(a)} &
          90.77 &
          90.58 &
          90.98 &
          0.916 &
          \multicolumn{1}{c|}{}
          \\
         &
           &
          \multicolumn{1}{c|}{MRI + PET(b)} &
          74.1 &
          75 &
          73.08 &
          0.766 &
          \multicolumn{1}{c|}{}
          \\
         &
           &
          \multicolumn{1}{c|}{MRI + 100\% synthetic PET(b)} &
          73.78 &
          64.52 &
          85.92 &
          0.7509 &
          \multicolumn{1}{c|}{}
          \\
        \midrule
        \multicolumn{1}{|c|}{\cite{bi2020multimodal}} &
          2020 &
          \multicolumn{1}{c|}{CN vs AD (Pearson + CERF)} &
          86.2 &
           &
           &
           &
          \multicolumn{1}{c|}{}
          \\
        \midrule
        \multicolumn{1}{|c|}{\multirow{3}[2]{*}{\cite{matsuura2019statistical}}} &
          \multirow{3}[2]{*}{2019} &
          \multicolumn{1}{c|}{MMSE≥24 vs MMSE<24：} &
          \multirow{3}[2]{*}{} &
           &
           &
          \multirow{3}[2]{*}{} &
          \multicolumn{1}{c|}{\multirow{3}[2]{*}{}}
          \\
         &
           &
          \multicolumn{1}{c|}{Single-task features（Physical）} &
           &
          38.4 &
          87.8 &
           &
          \multicolumn{1}{c|}{}
          \\
         &
           &
          \multicolumn{1}{c|}{Combined features} &
           &
          75.3 &
          79.9 &
           &
          \multicolumn{1}{c|}{}
          \\
        \midrule
        \multicolumn{1}{|c|}{\multirow{4}[2]{*}{\cite{li2015robust}}} &
          \multirow{4}[2]{*}{2015} &
          \multicolumn{1}{c|}{AD vs CN} &
          91.4 &
          \multirow{4}[2]{*}{} &
          \multirow{4}[2]{*}{} &
          \multirow{4}[2]{*}{} &
          \multicolumn{1}{c|}{\multirow{4}[2]{*}{}}
          \\
         &
           &
          \multicolumn{1}{c|}{MCI vs CN} &
          77.4 &
           &
           &
           &
          \multicolumn{1}{c|}{}
          \\
         &
           &
          \multicolumn{1}{c|}{AD vs MCI} &
          70.1 &
           &
           &
           &
          \multicolumn{1}{c|}{}
          \\
         &
           &
          \multicolumn{1}{c|}{MCI.C vs MCI.NC} &
          57.4 &
           &
           &
           &
          \multicolumn{1}{c|}{}
          \\
        \midrule
        \multicolumn{1}{|c|}{\cite{park2020prediction}} &
          2019 &
          \multicolumn{1}{c|}{AD vs CN} &
          82.3 &
           &
           &
           &
          \multicolumn{1}{c|}{}
          \\
        \midrule
        \multicolumn{1}{|c|}{\multirow{4}[2]{*}{\cite{zhou2019effective}}} &
          \multirow{4}[2]{*}{2019} &
          \multicolumn{1}{c|}{CN vs MCI vs AD} &
          \multicolumn{1}{c|}{≈65} &
          \multirow{4}[2]{*}{} &
          \multirow{4}[2]{*}{} &
          \multirow{4}[2]{*}{} &
          \multicolumn{1}{c|}{\multirow{4}[2]{*}{}}
          \\
         &
           &
          \multicolumn{1}{c|}{CN vs sMCI vs pMCI vs AD} &
          \multicolumn{1}{c|}{≈55} &
           &
           &
           &
          \multicolumn{1}{c|}{}
          \\
         &
           &
          \multicolumn{1}{c|}{AD vs CN} &
          \multicolumn{1}{c|}{≈92} &
           &
           &
           &
          \multicolumn{1}{c|}{}
          \\
         &
           &
          \multicolumn{1}{c|}{CN vs MCI} &
          \multicolumn{1}{c|}{≈76} &
           &
           &
           &
          \multicolumn{1}{c|}{}
          \\
        \midrule
        \multicolumn{1}{|c|}{\multirow{2}[2]{*}{\cite{liu2019using}}} &
          \multirow{2}[2]{*}{2019} &
          \multicolumn{1}{c|}{\multirow{2}[2]{*}{CN vs MCI/AD}} &
          \multirow{2}[2]{*}{} &
          \multirow{2}[2]{*}{88.39} &
          \multirow{2}[2]{*}{95.84} &
          \multirow{2}[2]{*}{} &
          BACC: 0.9220
          \\
         &
           &
           &
           &
           &
           &
           &
          F1-score: 0.9259
          \\
        \bottomrule
        \end{tabular}%
      \label{tab:32}%
    \end{table*}%


    \begin{table*}[htbp]
      \centering
      \caption{(Continued)}
        \begin{tabular}{|c|c|c|c|c|c|c|cc|}
        \toprule
        \multicolumn{1}{|c|}{\multirow{2}[4]{*}{\textbf{Ref.}}} &
          \multicolumn{1}{c|}{\multirow{2}[4]{*}{\textbf{Year}}} &
          \multirow{2}[4]{*}{\textbf{Target}} &
          \multicolumn{6}{c|}{\textbf{Performance}}
          \\
\cmidrule{4-9}         &
           &
          \multicolumn{1}{c|}{} &
          \multicolumn{1}{c|}{\textbf{Acc (\%)}} &
          \multicolumn{1}{c|}{\textbf{Sens (\%)}} &
          \multicolumn{1}{c|}{\textbf{Spec (\%)}} &
          \multicolumn{1}{c|}{\textbf{AUC}} &
          \multicolumn{2}{c|}{\textbf{others}}
          \\
        \midrule
        \multicolumn{1}{|c|}{\multirow{6}[2]{*}{\cite{lee2019toward}}} &
          \multirow{6}[2]{*}{2019} &
          AD vs CN &
          92.75 &
          91.89 &
          93.47 &
          0.9804 &
          \multicolumn{2}{c|}{\multirow{6}[2]{*}{}}
          \\
         &
           &
          MCI vs CN &
          89.22 &
          93.33 &
          82.55 &
          0.9573 &
          \multicolumn{2}{c|}{}
          \\
         &
           &
          AD vs MCI &
          81.46 &
          68.66 &
          88.24 &
          0.8954 &
          \multicolumn{2}{c|}{}
          \\
         &
           &
          pMCI vs sMCI &
          88.52 &
          87.5 &
          89.22 &
          0.9568 &
          \multicolumn{2}{c|}{}
          \\
         &
           &
          AD vs MCI vs CN(Macro) &
          71.18 &
          54.97 &
          72 &
           &
          \multicolumn{2}{c|}{}
          \\
         &
           &
          AD vs MCI vs CN(Micro) &
          71.18 &
          55.7 &
          71.18 &
           &
          \multicolumn{2}{c|}{}
          \\
        \midrule
        \multicolumn{1}{|c|}{\multirow{2}[2]{*}{\cite{cui2019rnn}}} &
          \multirow{2}[2]{*}{2019} &
          AD vs CN &
          91.33 &
          86.87 &
          95.2 &
          0.9322 &
          \multicolumn{2}{c|}{\multirow{2}[2]{*}{}}
          \\
         &
           &
          pMCI vs sMCI &
          71.71 &
          65.27 &
          76.27 &
          0.7303 &
          \multicolumn{2}{c|}{}
          \\
        \midrule
        \multicolumn{1}{|c|}{\cite{ge2019multi}} &
          2019 &
          AD vs CN &
          94.74 &
           &
           &
           &
          \multicolumn{2}{c|}{}
          \\
        \midrule
        \multicolumn{1}{|c|}{\multirow{3}[2]{*}{\cite{li2019deep}}} &
          \multirow{3}[2]{*}{2019} &
          AD vs CN on ADNI-GO\&2 &
          90 &
          \multirow{3}[2]{*}{} &
          \multirow{3}[2]{*}{} &
          0.956 &
          \multicolumn{2}{c|}{}
          \\
         &
           &
          AD vs CN on AIBL &
          92.9 &
           &
           &
          0.958 &
          \multicolumn{2}{c|}{}
          \\
         &
           &
          From MCI to AD? &
           &
           &
           &
          0.813 &
          \multicolumn{2}{c|}{C-index: 0.864}
          \\
        \midrule
        \multicolumn{1}{|c|}{\multirow{3}[2]{*}{\cite{cui2018hippocampus}}} &
          \multirow{3}[2]{*}{2019} &
          AD vs CN &
          92.29 &
          90.63 &
          93.72 &
          0.9695 &
          \multicolumn{2}{c|}{\multirow{3}[2]{*}{}}
          \\
         &
           &
          pMCI vs sMCI &
          75 &
          73.33 &
          76.19 &
          0.797 &
          \multicolumn{2}{c|}{}
          \\
         &
           &
          MCI vs CN &
          74.64 &
          77.27 &
          69.96 &
          0.777 &
          \multicolumn{2}{c|}{}
          \\
        \midrule
        \multicolumn{1}{|c|}{\multirow{2}[2]{*}{\cite{liu2020multi}}} &
          \multirow{2}[2]{*}{2020} &
          AD vs CN &
          88.9 &
          86.6 &
          90.8 &
          0.925 &
          \multicolumn{2}{c|}{\multirow{2}[2]{*}{}}
          \\
         &
           &
          MCI vs CN &
          76.2 &
          79.5 &
          69.8 &
          0.775 &
          \multicolumn{2}{c|}{}
          \\
        \midrule
        \multicolumn{1}{|c|}{\multirow{5}[2]{*}{\cite{ieracitano2019convolutional}}} &
          \multirow{5}[2]{*}{2018} &
          \multicolumn{1}{c|}{} &
           &
           &
          \multirow{5}[2]{*}{} &
           &
          \multicolumn{1}{c|}{Precision (\%)} &
          \multicolumn{1}{c|}{F1-score}
          \\
         &
           &
          AD vs CN &
          92.95 &
          95.3 &
           &
          0.97 &
          \multicolumn{1}{c|}{91.02} &
          0.9311
          \\
         &
           &
          AD vs MCI &
          84.62 &
          85.04 &
           &
          0.93 &
          \multicolumn{1}{c|}{84.32} &
          0.8468
          \\
         &
           &
          MCI vs CN &
          91.88 &
          92.31 &
           &
          0.97 &
          \multicolumn{1}{c|}{91.53} &
          0.9191
          \\
         &
           &
          AD vs MCI vs CN &
          83.33 &
          82.91 &
           &
          0.94 &
          \multicolumn{1}{c|}{79.51} &
          0.8117
          \\
        \midrule
        \multicolumn{1}{|c|}{\multirow{4}[2]{*}{\cite{li2020detecting}}} &
          \multirow{4}[2]{*}{2020} &
          AD vs MCI &
          92.11 &
          \multirow{4}[2]{*}{} &
          \multirow{4}[2]{*}{} &
          0.92 &
          \multicolumn{2}{c|}{\multirow{4}[2]{*}{}}
          \\
         &
           &
          MCI vs CN &
          88.12 &
           &
           &
          0.89 &
          \multicolumn{2}{c|}{}
          \\
         &
           &
          AD vs CN &
          97.37 &
           &
           &
          1 &
          \multicolumn{2}{c|}{}
          \\
         &
           &
          AD vs CN vs MCI &
          89.47 &
           &
           &
           &
          \multicolumn{2}{c|}{}
          \\
        \midrule
        \multicolumn{1}{|c|}{\multirow{2}[2]{*}{\cite{choi2020cognitive}}} &
          \multirow{2}[2]{*}{2020} &
          AD vs CN &
          \multirow{2}[2]{*}{} &
          \multirow{2}[2]{*}{} &
          \multirow{2}[2]{*}{} &
          0.94 &
          \multicolumn{2}{c|}{\multirow{2}[2]{*}{}}
          \\
         &
           &
          pMCI vs sMCI &
           &
           &
           &
          0.82 &
          \multicolumn{2}{c|}{}
          \\
        \midrule
        \multicolumn{1}{|c|}{\cite{chitradevi2020analysis}} &
          2020 &
          AD vs CN &
          95 &
          95 &
          94 &
           &
          \multicolumn{2}{c|}{}
          \\
        \midrule
        \multicolumn{1}{|c|}{\multirow{3}[2]{*}{\cite{puente2020automatic}}} &
          \multirow{3}[2]{*}{2020} &
          \multicolumn{1}{c|}{} &
           &
           &
           &
          \multirow{3}[2]{*}{} &
          \multicolumn{1}{c|}{Precision (\%)} &
          \multicolumn{1}{c|}{F1-score}
          \\
         &
           &
          OASIS: AD vs pMCI vs sMCI vs CN &
          86.81 &
          37.81 &
          88.25 &
           &
          \multicolumn{1}{c|}{32.08} &
          0.3347
          \\
         &
           &
          ADNI: AD vs MCI vs CN   &
          78.64 &
          58.28 &
          80.06 &
           &
          \multicolumn{1}{c|}{68.87} &
          0.603
          \\
        \midrule
        \multicolumn{1}{|c|}{\cite{spasov2019parameter}} &
          2019 &
          pMCI vs sMCI &
          86 &
          87.5 &
          85 &
          0.925 &
          \multicolumn{2}{c|}{}
          \\
        \midrule
        \multicolumn{1}{|c|}{\multirow{2}[2]{*}{\cite{janghel2021deep}}} &
          \multirow{2}[2]{*}{2021} &
          AD vs CN on average (fMRI) &
          99.95 &
          \multirow{2}[2]{*}{} &
          \multirow{2}[2]{*}{} &
          1 &
          \multicolumn{2}{c|}{\multirow{2}[2]{*}{}}
          \\
         &
           &
          AD vs CN on average (PET) &
          73.46 &
           &
           &
          0.75 &
          \multicolumn{2}{c|}{}
          \\
        \midrule
        \multicolumn{1}{|c|}{\multirow{2}[2]{*}{\cite{chen2021iterative}}} &
          \multirow{2}[2]{*}{2021} &
          AD vs CN &
          95.32 &
          91.18 &
          93.94 &
          \multirow{2}[2]{*}{} &
          \multicolumn{2}{c|}{\multirow{2}[2]{*}{}}
          \\
         &
           &
          pMCI vs sMCI &
          77.6 &
          71.62 &
          75.85 &
           &
          \multicolumn{2}{c|}{}
          \\
        \bottomrule
        \end{tabular}%
      \label{tab:33}%
    \end{table*}%



    \begin{table*}[htbp]
      \centering
      \caption{(Continued))}
        \begin{tabular}{|c|c|c|c|c|c|c|c|c|}
        \toprule
        \multicolumn{1}{|c|}{\multirow{2}[4]{*}{\textbf{Ref.}}} &
          \multicolumn{1}{c|}{\multirow{2}[4]{*}{\textbf{Year}}} &
          \multirow{2}[4]{*}{\textbf{Target}} &
          \multicolumn{6}{c|}{\textbf{Performance}}
          \\
\cmidrule{4-9}         &
           &
          \multicolumn{1}{c|}{} &
          \multicolumn{1}{c|}{\textbf{Acc (\%)}} &
          \multicolumn{1}{c|}{\textbf{Sens (\%)}} &
          \multicolumn{1}{c|}{\textbf{Spec (\%)}} &
          \multicolumn{1}{c|}{\textbf{AUC}} &
          \multicolumn{2}{c|}{\textbf{others}}
          \\
        \midrule
        \multicolumn{1}{|c|}{\multirow{13}[2]{*}{\cite{hazarika2022experimental}}} &
          \multirow{13}[2]{*}{2023} &
          \multicolumn{1}{c|}{} &
           &
           &
          \multirow{13}[2]{*}{} &
          \multirow{13}[2]{*}{} &
          \multicolumn{1}{c|}{Precision (\%)} &
          \multicolumn{1}{c|}{F1-score}
          \\
         &
           &
          CN vs MCI for 60-69 years &
          93 &
          92 &
           &
           &
          93 &
          0.94
          \\
         &
           &
          CN vs MCI for 70-79 years &
          92 &
          92 &
           &
           &
          94 &
          0.93
          \\
         &
           &
          CN vs MCI for 80+ years &
          87 &
          87 &
           &
           &
          90 &
          0.9
          \\
         &
           &
          MCI vs AD for 60-69 years &
          92 &
          92 &
           &
           &
          90 &
          0.93
          \\
         &
           &
          MCI vs AD for 70-79 years &
          89 &
          91 &
           &
           &
          90 &
          0.9
          \\
         &
           &
          MCI vs AD for 80+ years &
          87 &
          86 &
           &
           &
          87 &
          0.86
          \\
         &
           &
          CN vs AD for 60-69 years &
          92 &
          94 &
           &
           &
          94 &
          0.93
          \\
         &
           &
          CN vs AD for 70-79 years &
          90 &
          90 &
           &
           &
          91 &
          0.88
          \\
         &
           &
          CN vs AD for 80+ years &
          86 &
          87 &
           &
           &
          87 &
          0.86
          \\
         &
           &
          CN vs MCI vs AD for 60-69 years &
          90 &
          89 &
           &
           &
          89 &
          0.87
          \\
         &
           &
          CN vs MCI vs AD for 70-79 years &
          87 &
          88 &
           &
           &
          90 &
          0.91
          \\
         &
           &
          CN vs MCI vs AD for 80+ years &
          87 &
          85 &
           &
           &
          89 &
          0.86
          \\
        \midrule
        \multicolumn{1}{|c|}{\cite{bae2021transfer}} &
          2021 &
          pMCI vs sMCI  &
          82.4 &
           &
           &
           &
          \multicolumn{2}{c|}{}
          \\
        \midrule
        \multicolumn{1}{|c|}{\multirow{2}[2]{*}{\cite{zhang2021explainable}}} &
          \multirow{2}[2]{*}{2022} &
          AD vs CN &
          91 &
          91 &
          92 &
          0.984 &
          \multicolumn{2}{c|}{\multirow{2}[2]{*}{}}
          \\
         &
           &
          pMCI vs sMCI &
          82 &
          81 &
          81 &
          0.92 &
          \multicolumn{2}{c|}{}
          \\
        \midrule
        \multicolumn{1}{|c|}{\multirow{5}[2]{*}{\cite{raza2019diagnosis}}} &
          \multirow{5}[2]{*}{2019} &
          AD vs CN： &
           &
           &
           &
          \multirow{5}[2]{*}{} &
          \multicolumn{1}{c|}{Precision (\%)} &
          \multicolumn{1}{c|}{F1-score}
          \\
         &
           &
          OASIS &
          95.93 &
          92.75 &
          92.51 &
           &
          96.94 &
          0.947
          \\
         &
           &
          ADNI &
          98.74 &
          98.5 &
          98.21 &
           &
          98.81 &
          0.9865
          \\
         &
           &
          Single sensor on smartphone &
          99 &
          96.95 &
          96.41 &
           &
          97 &
          0.9697
          \\
         &
           &
          Multi-sensor &
          99.6 &
          98.41 &
          99.71 &
           &
          98.29 &
          0.9832
          \\
        \midrule
        \multicolumn{1}{|c|}{\multirow{3}[2]{*}{\cite{huang2019diagnosis}}} &
          \multirow{3}[2]{*}{2019} &
          AD vs CN &
          90.1 &
          90.85 &
          89.21 &
          0.9084 &
          \multicolumn{2}{c|}{\multirow{3}[2]{*}{}}
          \\
         &
           &
          CN vs pMCI &
          87.46 &
          90.73 &
          80.61 &
          0.8761 &
          \multicolumn{2}{c|}{}
          \\
         &
           &
          sMCI vs pMCI &
          76.9 &
          68.15 &
          83.93 &
          0.7961 &
          \multicolumn{2}{c|}{}
          \\
        \midrule
        \multicolumn{1}{|c|}{\multirow{4}[2]{*}{\cite{wang2019ensemble}}} &
          \multirow{4}[2]{*}{2019} &
          AD vs MCI &
          93.61 &
          92.45 &
          \multirow{4}[2]{*}{} &
          \multirow{4}[2]{*}{} &
          \multicolumn{2}{c|}{Precision(\%)：94.59}
          \\
         &
           &
          MCI vs CN &
          98.42 &
          98.34 &
           &
           &
          \multicolumn{2}{c|}{Precision(\%)：98.37}
          \\
         &
           &
          AD vs CN &
          98.83 &
          98.7 &
           &
           &
          \multicolumn{2}{c|}{Precision(\%)：98.70}
          \\
         &
           &
          AD vs CN vs MCI &
          97.52 &
           &
           &
           &
          \multicolumn{2}{c|}{}
          \\
        \midrule
        \multicolumn{1}{|c|}{\multirow{3}[2]{*}{\cite{zhang2019multi}}} &
          \multirow{3}[2]{*}{2019} &
          AD vs CN &
          98.47 &
          96.58 &
          95.39 &
          0.9861 &
          \multicolumn{2}{c|}{\multirow{3}[2]{*}{}}
          \\
         &
           &
          MCI vs CN &
          85.74 &
          90.11 &
          91.82 &
          0.8815 &
          \multicolumn{2}{c|}{}
          \\
         &
           &
          AD vs MCI &
          88.2 &
          97.43 &
          84.31 &
          0.8801 &
          \multicolumn{2}{c|}{}
          \\
        \midrule
        \multicolumn{1}{|c|}{\multirow{4}[2]{*}{\cite{liu2019weakly}}} &
          \multirow{4}[2]{*}{2020} &
          Correlation Coefficients of  &
          \multirow{4}[2]{*}{} &
          \multirow{4}[2]{*}{} &
          \multirow{4}[2]{*}{} &
          \multirow{4}[2]{*}{} &
          \multicolumn{2}{c|}{SDR-SB：0.527}
          \\
         &
           &
          cognitive scores at M24 &
           &
           &
           &
           &
          \multicolumn{2}{c|}{ADAS-Cog11：0.552}
          \\
         &
           &
          (Model was trained on ADNI-1， &
           &
           &
           &
           &
          \multicolumn{2}{c|}{ADAS-Cog13：0.583}
          \\
         &
           &
          and tested on ADNI-2) &
           &
           &
           &
           &
          \multicolumn{2}{c|}{MMSE：0.537}
          \\
        \midrule
        \multicolumn{1}{|c|}{\multirow{3}[2]{*}{\cite{abuhmed2021robust}}} &
          \multirow{3}[2]{*}{2021} &
          \multicolumn{1}{c|}{} &
           &
           &
          \multirow{3}[2]{*}{} &
          \multirow{3}[2]{*}{} &
          \multicolumn{1}{c|}{Precision (\%)} &
          \multicolumn{1}{c|}{F1-score}
          \\
         &
           &
          AD vs CN vs MCI  &
          84.95 &
          86.5 &
           &
           &
          85.65 &
          0.8607
          \\
         &
           &
          AD vs CN vs pMCI vs sMCI &
          86.08 &
          87.47 &
           &
           &
          87.2 &
          0.8733
          \\
        \bottomrule
        \end{tabular}%
      \label{tab:34}%
    \end{table*}%


    \begin{table*}[htbp]
      \centering
      \caption{(Continued))}
        \begin{tabular}{|c|c|c|c|c|c|c|c|c|}
        \toprule
        \multicolumn{1}{|c|}{\multirow{2}[4]{*}{\textbf{Ref.}}} &
          \multicolumn{1}{c|}{\multirow{2}[4]{*}{\textbf{Year}}} &
          \multirow{2}[4]{*}{\textbf{Target}} &
          \multicolumn{6}{c|}{\textbf{Performance}}
          \\
\cmidrule{4-9}         &
           &
          \multicolumn{1}{c|}{} &
          \multicolumn{1}{c|}{\textbf{Acc (\%)}} &
          \multicolumn{1}{c|}{\textbf{Sens (\%)}} &
          \multicolumn{1}{c|}{\textbf{Spec (\%)}} &
          \multicolumn{1}{c|}{\textbf{AUC}} &
          \multicolumn{2}{c|}{\textbf{others}}
          \\
        \midrule
        \multicolumn{1}{|c|}{\multirow{5}[2]{*}{\cite{lei2022predicting}}} &
          \multirow{5}[2]{*}{2021} &
          Using CLSIndRNN to predict M36: &
          \multirow{5}[2]{*}{} &
          \multirow{5}[2]{*}{} &
          \multirow{5}[2]{*}{} &
          \multirow{5}[2]{*}{} &
          \multicolumn{1}{c|}{MAE} &
          \multicolumn{1}{c|}{R}
          \\
         &
           &
          MMSE &
           &
           &
           &
           &
          1.56 &
          0.76
          \\
         &
           &
          CDR-SOB &
           &
           &
           &
           &
          0.81 &
          0.67
          \\
         &
           &
          CDR-GLOB &
           &
           &
           &
           &
          0.13 &
          0.85
          \\
         &
           &
          ADAS-Cog &
           &
           &
           &
           &
          3.33 &
          0.86
          \\
        \midrule
        \multicolumn{1}{|c|}{\multirow{4}[2]{*}{\cite{ghazi2019training}}} &
          \multirow{4}[2]{*}{2019} &
          CN vs MCI &
          \multirow{4}[2]{*}{} &
          \multirow{4}[2]{*}{} &
          \multirow{4}[2]{*}{} &
          0.5914 &
          \multicolumn{2}{c|}{\multirow{4}[2]{*}{}}
          \\
         &
           &
          CN vs AD &
           &
           &
           &
          0.9029 &
          \multicolumn{2}{c|}{}
          \\
         &
           &
          MCI vs AD &
           &
           &
           &
          0.7844 &
          \multicolumn{2}{c|}{}
          \\
         &
           &
          CN vs MCI vs AD &
           &
           &
           &
          0.7596 &
          \multicolumn{2}{c|}{}
          \\
        \midrule
        \multicolumn{1}{|c|}{\multirow{5}[2]{*}{\cite{ieracitano2020novel}}} &
          \multirow{5}[2]{*}{2019} &
          \multicolumn{1}{c|}{} &
           &
           &
          \multirow{5}[2]{*}{} &
           &
          \multicolumn{1}{c|}{Precision (\%)} &
          \multicolumn{1}{c|}{F1-score}
          \\
         &
           &
          AD vs CN &
          96.95 &
          94.91 &
           &
           &
          94.72 &
          0.9485
          \\
         &
           &
          AD vs MCI &
          90.24 &
          85.71 &
           &
          0.965 &
          87.14 &
          0.8646
          \\
         &
           &
          MCI vs CN &
          96.24 &
          95.86 &
           &
           &
          95.31 &
          0.9558
          \\
         &
           &
          AD vs MCI vs CN &
          89.22 &
          80.99 &
           &
           &
          80.74 &
          80.87
          \\
        \midrule
        \multicolumn{1}{|c|}{\cite{bi2019early}} &
          2019 &
          AD vs MCI vs CN &
          95.04 &
           &
           &
          0.96 &
          \multicolumn{2}{c|}{}
          \\
        \midrule
        \multicolumn{1}{|c|}{\cite{choi2019deep}} &
          2019 &
          AD vs CN &
           &
           &
           &
          0.9 &
          \multicolumn{2}{c|}{}
          \\
        \midrule
        \multicolumn{1}{|c|}{\multirow{4}[2]{*}{\cite{kang2021multi}}} &
          \multirow{4}[2]{*}{2021} &
          AD vs CN &
          90.36 &
          93.94 &
          83.78 &
          0.8972 &
          \multicolumn{2}{c|}{\multirow{4}[2]{*}{}}
          \\
         &
           &
          AD vs MCI &
          77.19 &
          68.97 &
          54.06 &
          0.7118 &
          \multicolumn{2}{c|}{}
          \\
         &
           &
          MCI vs CN &
          72.36 &
          74.71 &
          84.42 &
          0.6829 &
          \multicolumn{2}{c|}{}
          \\
         &
           &
          pMCI vs sMCI &
          66.7 &
           &
           &
           &
          \multicolumn{2}{c|}{}
          \\
        \midrule
        \multicolumn{1}{|c|}{\multirow{2}[2]{*}{\cite{kim2020slice}}} &
          \multirow{2}[2]{*}{2020} &
          AD vs CN on ADNI dataset &
          94.82 &
          91.78 &
          97.06 &
          0.98 &
          \multicolumn{2}{c|}{\multirow{2}[2]{*}{}}
          \\
         &
           &
          AD vs CN on Sev. &
          94.33 &
          92.11 &
          97.45 &
          0.98 &
          \multicolumn{2}{c|}{}
          \\
        \midrule
        \multicolumn{1}{|c|}{\cite{mendoza2020single}} &
          2020 &
          AD vs CN &
          90 &
          95 &
          85 &
          0.92 &
          \multicolumn{2}{c|}{BACC: 0.90}
          \\
        \midrule
        \multicolumn{1}{|c|}{\multirow{5}[2]{*}{\cite{lei2020deep}}} &
          \multirow{5}[2]{*}{2020} &
          Using CTDE to predict M36: &
          \multirow{5}[2]{*}{} &
          \multirow{5}[2]{*}{} &
          \multirow{5}[2]{*}{} &
          \multirow{5}[2]{*}{} &
          \multicolumn{1}{c|}{MAE} &
          \multicolumn{1}{c|}{R}
          \\
         &
           &
          MMSE &
           &
           &
           &
           &
          1.756 &
          0.858
          \\
         &
           &
          CDR-SOB &
           &
           &
           &
           &
          0.816 &
          0.813
          \\
         &
           &
          CDR-GLOB &
           &
           &
           &
           &
          0.13 &
          0.852
          \\
         &
           &
          ADAS-Cog &
           &
           &
           &
           &
          4.981 &
          0.83
          \\
        \midrule
        \multicolumn{1}{|c|}{\cite{afzal2019data}} &
          2019 &
          AD vs CN &
          98.41 &
           &
           &
           &
          \multicolumn{2}{c|}{}
          \\
        \midrule
        \multicolumn{1}{|c|}{\multirow{4}[2]{*}{\cite{jain2019convolutional}}} &
          \multirow{4}[2]{*}{2019} &
          AD vs MCI vs CN &
          95.73 &
           &
          \multirow{4}[2]{*}{} &
          \multirow{4}[2]{*}{} &
          \multicolumn{1}{c|}{Precision (\%)} &
          \multicolumn{1}{c|}{F1-score}
          \\
         &
           &
          AD vs CN &
          99.14 &
          98.75 &
           &
           &
          99.68 &
          0.9921
          \\
         &
           &
          AD vs MCI &
          99.3 &
          98.75 &
           &
           &
          100 &
          0.9937
          \\
         &
           &
          CN vs MCI &
          99.22 &
          99.06 &
           &
           &
          99.37 &
          0.9921
          \\
        \midrule
        \multicolumn{1}{|c|}{\multirow{2}[2]{*}{\cite{2019Volumetrie}}} &
          \multirow{2}[2]{*}{2019} &
          \multicolumn{1}{c|}{} &
           &
           &
          \multirow{2}[2]{*}{} &
          \multirow{2}[2]{*}{} &
          \multicolumn{1}{c|}{Precision (\%)} &
          \multicolumn{1}{c|}{F1-score}
          \\
         &
           &
          AD vs CN &
          88.8 &
          90 &
           &
           &
          91 &
          0.9
          \\
        \midrule
        \multicolumn{1}{|c|}{\multirow{2}[2]{*}{\cite{barbaroux2020encoding}}} &
          \multirow{2}[2]{*}{2020} &
          AD vs CN &
          92.16 &
          89.73 &
          95.05 &
          0.9459 &
          \multicolumn{2}{c|}{\multirow{2}[2]{*}{}}
          \\
         &
           &
          pMCI vs sMCI &
          76.47 &
          82.62 &
          57.39 &
          0.7445 &
          \multicolumn{2}{c|}{}
          \\
        \midrule
        \multicolumn{1}{|c|}{\multirow{4}[2]{*}{\cite{tufail2020binary}}} &
          \multirow{4}[2]{*}{2020} &
          AD vs CN &
           &
           &
           &
          \multirow{4}[2]{*}{} &
          \multicolumn{1}{c|}{MCC} &
          \multicolumn{1}{c|}{F1-score}
          \\
         &
           &
          Dataset-1 (Xception, Fold-9) &
          64.87 &
          64.88 &
          64.86 &
           &
          0.2975 &
          0.6488
          \\
         &
           &
          Dataset-2 (CNN, Fold-5) &
          82.79 &
           &
           &
           &
           &
          
          \\
         &
           &
          Dataset-3 (Inception V3, Fold-5) &
          99.45 &
           &
           &
           &
           &
          
          \\
        \bottomrule
        \end{tabular}%
      \label{tab:35}%
    \end{table*}%

    \begin{table*}[htbp]
      \centering
      \caption{(Continued))}
        \begin{tabular}{|c|c|c|c|c|c|c|c|}
        \toprule
        \multirow{2}[4]{*}{\textbf{Ref.}} &
          \multicolumn{1}{c|}{\multirow{2}[4]{*}{\textbf{Year}}} &
          \multirow{2}[4]{*}{\textbf{Target}} &
          \multicolumn{5}{c|}{\textbf{Performance}}
          \\
\cmidrule{4-8}        \multicolumn{1}{|c|}{} &
           &
          \multicolumn{1}{c|}{} &
          \multicolumn{1}{c|}{\textbf{Acc (\%)}} &
          \multicolumn{1}{c|}{\textbf{Sens (\%)}} &
          \multicolumn{1}{c|}{\textbf{Spec (\%)}} &
          \multicolumn{1}{c|}{\textbf{AUC}} &
          \multicolumn{1}{c|}{\textbf{others}}
          \\
        \midrule
        \multirow{6}[2]{*}{\cite{2020Diagnosis}} &
          \multirow{6}[2]{*}{2020} &
          CN vs sMCI &
          96.19 &
          \multirow{6}[2]{*}{} &
          \multirow{6}[2]{*}{} &
          \multirow{6}[2]{*}{} &
          \multirow{6}[2]{*}{}
          \\
        \multicolumn{1}{|c|}{} &
           &
          AD vs CN &
          93.78 &
           &
           &
           &
          
          \\
        \multicolumn{1}{|c|}{} &
           &
          CN vs pMCI &
          94.87 &
           &
           &
           &
          
          \\
        \multicolumn{1}{|c|}{} &
           &
          AD vs sMCI &
          92.8 &
           &
           &
           &
          
          \\
        \multicolumn{1}{|c|}{} &
           &
          AD vs pMCI &
          94.79 &
           &
           &
           &
          
          \\
        \multicolumn{1}{|c|}{} &
           &
          pMCI vs sMCI &
          92.91 &
           &
           &
           &
          
          \\
        \midrule
        \multirow{1}[1]{*}{\cite{2020Identifying}} &
          2020 &
          CN vs sMCI &
          94.2 &
          97.3 &
          92.9 &
           &
          
          \\
        \bottomrule
        \end{tabular}%
      \label{tab:36}%
    \end{table*}%

\section{Conclusion}

This article discusses the disease mechanism of Alzheimer's disease and an overview of diagnostic methods, analyzes the advantages and disadvantages of each diagnostic method, and summarizes the prospects of multimodal diagnostic methods in the diagnosis of AD.


The study started with ancillary diagnostic techniques for earlier classification studies, such as cognitive assessment, voice detection, and posture method. Each method has its own merits. Cognitive assessment can directly measure a patient's cognitive function and has been widely used in AD diagnosis and monitoring, demonstrating high clinical validation and reliability. Some methods based on blood or cerebrospinal fluid samples have lower invasiveness and risk compared to other imaging techniques. Similarly, posture and sound methods are non-invasive approaches, and their data can be conveniently obtained through devices. Neuroimaging techniques can obtain brain images at high resolution, revealing subtle changes in brain structure and function. Researchers can quantitatively analyze the images using quantitative analysis methods to support AD diagnosis and monitoring.

But these methods also have some obvious limitations. Some methods may encounter interference during the data collection process. For example, cognitive assessment may be influenced by individual factors such as educational level. Sound detection may be disrupted by environmental noise, unclear articulation, and other factors, affecting feature extraction and interpretation. Other methods require expensive equipment and specialized operations, making them difficult to be widely used in clinical and large-scale applications, such as various neuroimaging techniques. Additionally, some methods may not directly provide information about brain structure and function, limiting the comprehensive evaluation of AD, such as posture detection. Using voice, pose, biomarkers, cognitive assessment scores, and their fusion modalities can often bring more stable and robust diagnostic results to the model. Among them, we discuss in detail such as the fusion of PET and MRI, the fusion of pose and text, the fusion of prosodic features  and speech content and so on. These fusion methods offer significant advantages over single-modality approaches. They leverage the complementary information captured by different modalities, leading to more stable and robust diagnostic results. The integration of diverse data sources allows for a more comprehensive assessment of AD, considering both structural and functional changes, as well as cognitive and physical impairments.

In the process of investigating the multi-modal fusion model, we found that there are currently two mainstream fusion solutions, one is the fusion of different data levels of a single modality, and the other is the data fusion of different modalities. The former requires more attention to combine features extracted from different levels of data within a single modality and combine the decisions or predictions made independently at different data levels within a single modality. The latter focus on the integration of data from different modalities to leverage their complementary information. It also need to learn joint representations or features that capture shared information across different modalities. 

In general, multimodal fusion methods have shown a lot of potential, but there are still many problems. We believe that the main challenges in this field focus on (1) how to align different modality data so that they can work better together and improve the accuracy of diagnosis. (2) How the modal data interact with each other, discover the correlation between the modal and the modal, and make more contributions to the interpretability of the diagnosis. (3) Existing medical images, pose data, and sound data are all separate, and there is no set of relatively complete multi-modal data sets for researchers to use.

\bibliographystyle{ieeetr}
\bibliography{reference.bib}
\end{spacing}
\end{CJK}
\end{document}